\documentclass[12pt]{article}

\usepackage{amsfonts,amsmath,amssymb,color,graphicx,slashed,titlesec,xspace,comment}
\usepackage{multirow}
\usepackage[colorlinks=true,urlcolor=blue,anchorcolor=blue,citecolor=blue,filecolor=blue,linkcolor=blue,menucolor=blue,linktocpage=true]{hyperref}
\usepackage{tikzfeynman}
\usepackage[caption=false]{subfig}
\usepackage{physics, braket}
\usepackage{bbold}
\usepackage{enumerate}
\usepackage{placeins}

\usepackage[top=1in, bottom=1in, left=1in, right=1in, footskip=0.5in, headheight=0.5in, footnotesep=0.2in, marginparwidth=0in, marginparsep=0in]{geometry}
\usepackage{multicol}
\usepackage[labelfont=bf]{caption}
\usepackage{authblk}
\usepackage[compress,numbers]{natbib}

\renewcommand{\arraystretch}{1.5}

\newcommand{\figref}[1]{Fig.~\ref{fig:#1}}
\renewcommand{\eqref}[1]{Eq.~(\ref{eq:#1})}
\newcommand{\eqrefp}[1]{(\ref{eq:#1})}

\newcommand{\eqrefs}[2]{Eqs.\ (\ref{eq:#1}) -- (\ref{eq:#2}) }
\newcommand{\aref}[1]{Appendix \ref{a.#1}}
\newcommand{\sref}[1]{Section \ref{s.#1}}
\newcommand{\ssref}[1]{Section \ref{ss.#1}}

\newcommand{\tref}[1]{Table \ref{t:#1}}

\def\gev{\, {\rm GeV}}

\def\tev{\, {\rm TeV}}

\def\fb{\, {\rm fb}}

\newcommand\MadGraph{{\sc\small MadGraph}}

\begin{document}

\begin{titlepage}
	\begin{flushright}
	\small{YITP-SB-19-25}
	\end{flushright}
	
	\vskip1cm
	\begin{center}
	{\huge \bf Higgs bosons with large couplings \\ to light quarks}
	\end{center}
	\vskip0.2cm

	\begin{center}
	{Daniel Egana-Ugrinovic$^{1,2}$, Samuel Homiller$^{1,3}$ and Patrick Meade$^{1}$}
	\end{center}

	\begin{center}
	{$^{1}$\it C. N. Yang Institute for Theoretical Physics,\\ Stony Brook University, Stony Brook, New York 11794, USA}\\
	\vspace*{0.3cm}
	{$^{2}$\it Perimeter Institute  for Theoretical Physics, Waterloo, Ontario N2L 2Y5, Canada}\\
	\vspace*{0.3cm}
	{$^{3}$\it Department of Physics,\\ Brookhaven National Laboratory, Upton, New York 11973, USA}\\
	\vspace*{0.3cm}
	\vspace*{0.1cm}

	\end{center}

\date{\today}

\begin{abstract}

A common lore has arisen that beyond the Standard Model (BSM) particles, 
which can be searched for at current and proposed experiments, 
should have flavorless or mostly third-generation interactions with Standard Model quarks. 
This theoretical bias severely limits the exploration of BSM phenomenology, 
and is especially constraining for extended Higgs sectors.  
Such limitations can be avoided in the context of Spontaneous Flavor Violation (SFV), 
a robust and UV complete framework that allows for significant couplings to any up or down-type quark, 
while suppressing flavor-changing neutral currents via flavor alignment.   
In this work we study the theory and phenomenology of extended SFV Higgs sectors with large couplings to any quark generation. 
We perform a comprehensive analysis of flavor and collider constraints of extended SFV Higgs sectors, 
and demonstrate that new Higgs bosons with large couplings to the light quarks may be found at the electroweak scale. 
In particular, we find that new Higgses as light as 100 GeV with order $\sim$ 0.1 couplings to first- or second-generation quarks,
which are copiously produced at the LHC via quark fusion,
are allowed by current constraints.
Furthermore, the additional SFV Higgses can mix with the SM Higgs, 
providing strong theory motivation for an experimental program looking for deviations in the light quark--Higgs couplings. 
Our work demonstrates the importance of exploring BSM physics coupled preferentially to light quarks, 
and the need to further develop dedicated experimental techniques for the LHC and future colliders.

\end{abstract}

\end{titlepage}

\tableofcontents

%%%%%%%%%%%%%%%%%%%%%%%%%%%%%%%%%%%%%%%%%%%%%%%%%%%%%%%%
%%%%%%%%%%%%%%%%%%%%%%%%%%%%%%%%%%%%%%%%%%%%%%%%%%%%%%%%
%% SECTION BEGINS
%%%%%%%%%%%%%%%%%%%%%%%%%%%%%%%%%%%%%%%%%%%%%%%%%%%%%%%%
%%%%%%%%%%%%%%%%%%%%%%%%%%%%%%%%%%%%%%%%%%%%%%%%%%%%%%%%
\section{Introduction}

The exploration of new physics at the energy frontier relies on theory guidance to maintain consistency across different experiments, 
to motivate specific experimental searches and to select promising signatures.  However, a narrow selection of models, or the use of sufficient but unnecessary assumptions can lead to theory bias.  Theory bias from the beyond the Standard Model (BSM) perspective 
can undermine our efforts to find new physics unless it is built on solid foundations.

Theory input is particularly important in the flavor sector
due to its unknown and possibly complex origin. 
Assumptions on the flavor structure of new physics are needed in order to determine how new physics couples to the Standard Model (SM) fermions. 
If new physics is close to the electroweak (EW) scale, such assumptions must account for the lack of observation of large flavor-changing neutral currents (FCNCs).
The most common assumption, minimal flavor violation (MFV) \cite{DAmbrosio:2002vsn}, 
allows for new physics at the EW scale with small FCNCs, 
but at the same time constrains new states to couple preferentially to third-generation fermions only or in a flavor-universal way.
In this case, flavor considerations do not give us any more intuition than naturalness arguments. 
Proposed models of new physics coupled preferentially to the third-generation fermions are rather ubiquitous, 
even if some models do not strictly satisfy the MFV criterion.   
While third generation theory bias is reasonable in the context of many models, 
the question remains whether one can build successful BSM models where the coupling to light generations is preferred, 
while maintaining consistency with the results from flavor physics experiments.  
This is particularly important for the exploration of the Higgs sector,
as the Higgs boson itself is at the core of the flavor puzzle.

In \cite{Egana-Ugrinovic:2018znw} a general spurion formalism was developed,
which from the low-energy point of view, allowed for preferential BSM couplings to any specific quark flavor, while suppressing FCNCs via flavor alignment.  
Most importantly, it was shown that a subset of flavor-aligned models exists, called Spontaneous Flavor Violation (SFV), 
which has robust UV completions and allows for BSM physics at the $\mathcal{O}(\mathrm{TeV})$ scale consistent with flavor bounds.

In this work, we apply the concept of SFV to build viable theories of electroweak-scale extended Higgs sectors with sizable couplings to any quark generation.
We introduce two new such theories,
the up-type and down-type SFV two-Higgs doublet models (2HDMs). 
These models allow for generation-specific couplings to down-type quarks or up-type quarks correspondingly,
but constrain the couplings to be MFV-like in the opposite quark sector.
In the SFV 2HDMs, 
tree level FCNCs mediated by the extra Higgses are absent due to flavor alignment of the Yukawa matrices of the two Higgs doublets. 
\footnote{
In the context of the 2HDM, flavor alignment has sometimes been used to refer to a generalized case of MFV \cite{Pich:2009sp, Pich:2010ic, Kagan:2009bn}.
Such theories retain the hierarchical couplings of the SM, and couple the Higgs sector mostly to the third generation. 
In this work we return to the original concept of flavor alignment as defined by Nir and Seiberg in \cite{Nir:1993mx,Egana-Ugrinovic:2018znw}, 
which instead refers to simultaneous diagonalizability of flavored spurions, without necessarily retaining the hierarchies of the SM couplings. 
More details are in \sref{2hdm_basics}.}
Moreover, FCNCs in these theories are further suppressed by Cabibbo-Kobayashi-Maskawa (CKM) matrix elements and the Glashow-Iliopoulos-Maiani (GIM) mechanism.

To assess the viability of our theories, 
we perform a comprehensive analysis of flavor and collider bounds of the up-type SFV 2HDM, 
where Higgs bosons can have large couplings to down, strange and/or bottom quarks.
From our analysis, 
we find that extra Higgses at $100$ GeV coupling preferentially to down or strange quarks with Yukawas of order $\sim 0.1$ 
are allowed by all collider and flavor bounds.
This is despite the fact that such extra Higgses are copiously produced at tree level via quark fusion at the LHC
and can be looked for as dijet resonances. 
With the LHC Run 2 luminosity, 
 $\sim 10^8$ Higgses with such Yukawa couplings to down quarks may have been produced, 
a number that exceeds the amount of produced SM Higgses roughly by a factor of 10. 
These extra Higgses, however,  evade discovery due to large QCD backgrounds.
In addition, 
if such extra Higgses mix with the $125$ GeV Higgs,
they lead to dramatic enhancements of its Yukawa couplings to light quarks, 
which could be measured at the HL-LHC or at a future electron-positron collider. 
We find that enhancements to the down- and strange-quark Yukawas by a factor of $\sim 500$ and $\sim 30$ with respect to their SM expectations can be obtained within a realistic construction.
Larger enhancements are challenging to obtain due to collider constraints on the extra Higgses responsible for such enhancements,
and on currently measured Higgs signal strengths.
Overall, our results provide strong motivation for further developing experimental searches and techniques aiming at identifying new physics coupled mostly to light quarks,
such as light-quark taggers \cite{Duarte-Campderros:2018ouv,Fraser:2018ieu}.

In the literature, work has been already performed in the direction of studying extended Higgs sectors with general flavor alignment \cite{Gatto:1978dy,Gatto:1979mr,Sartori:1979gt,Penuelas:2017ikk,Botella:2018gzy, Rodejohann:2019izm}
and with enhanced Higgs Yukawas \cite{Bodwin:2013gca,Kagan:2014ila,Perez:2015lra,Zhou:2015wra,Brivio:2015fxa,Delaunay:2016brc,Bishara:2016jga,Soreq:2016rae,Yu:2016rvv,Aaboud:2017xnb,Alves:2017avw,Coyle:2019hvs}.
However,
a complete and unified analysis of all the aspects of an extended Higgs sector with generation-specific quark couplings,
including a robust flavor prescription and a complete phenomenological analysis of the extended Higgs sector,
has not been carried out.
These elements are part of a single problem, 
and we will find that it is very illustrative to study them in aggregation to understand their complementarity and to evaluate their viability.

We organize this paper as follows. 
In \sref{2hdm_basics} we introduce the up- and down-type spontaneous flavor-violating 2HDMs.
In Secs.~\ref{s.flavor} and \ref{s.collider} we study the flavor and collider phenomenology of the up-type SFV 2HDM,
providing bounds from $\Delta F=1$ and $\Delta F=2$ FCNCs 
and searches for dijet and diphoton resonances at the LHC. 
In \sref{light_higgs} we study how the up- and down-type SFV 2HDMs can lead to dramatic enhancements of the Yukawas of the $125$ GeV Higgs to the down- or up-type quarks correspondingly,
as compared with the Standard Model expectations. 
An important feature of the SFV 2HDMs is that they are motivated by a UV completion, 
but the discussion of the corresponding technical details is independent of the phenomenological analysis in the body of this work. 
For this reason, we leave the presentation of the UV completion to Appendix \ref{s.uv_completion}.
In Appendixes \ref{a.radiative_corrections}-\ref{a.loop_functions} we present other technical details, including a full renormalization group analysis of the SFV structure and a comparison of the SFV 2HDM with other well-known versions of the two-doublet theory.

%%%%%%%%%%%%%%%%%%%%%%%%%%%%%%%%%%%%%%%%%%%%%%%%%%%%%%%%
%%%%%%%%%%%%%%%%%%%%%%%%%%%%%%%%%%%%%%%%%%%%%%%%%%%%%%%%
%% SECTION BEGINS
%%%%%%%%%%%%%%%%%%%%%%%%%%%%%%%%%%%%%%%%%%%%%%%%%%%%%%%%
%%%%%%%%%%%%%%%%%%%%%%%%%%%%%%%%%%%%%%%%%%%%%%%%%%%%%%%%
\section{Spontaneous Flavor Violation in an extended Higgs sector}
\label{s.2hdm_basics}
In this section we present the up- and down-type spontaneous flavor-violating two-Higgs doublet models. 
We start by reviewing the general two-Higgs doublet theory and the flavor alignment conditions for the absence of tree level FCNCs in \ssref{2hdmlagrangian}.
In \ssref{SFV2HDM} we define the up- and down-type SFV 2HDMs, and we discuss their main properties.
In \ssref{couplings} we present the couplings of the physical Higgs bosons to the fermions in the SFV 2HDMs.

%%%%%%%%%%%%%%%%%%%%%%%%%%%%%%%%%%%%%%%%%%%%%%%%%%%%%%%%
\subsection{Two Higgs doublets, the Higgs basis and flavor alignment}
\label{ss.2hdmlagrangian}
%%%%%%%%%%%%%%%%%%%%%%%%%%%%%%%%%%%%%%%%%%%%%%%%%%%%%%%%
A general 2HDM contains two complex scalar fields $H_a$, $a = 1, 2$, with the quantum numbers of the Standard Model Higgs doublet.
The most general renormalizable Lagrangian for a 2HDM is
%boe%
\begin{equation}
 D_{\mu}H_{a}^{\dagger}D^{\mu}H_a  - V(H_1, H_2) 
 - \bigg[\lambda^{u}_{aij}Q_i H_a \bar{u}_j 
- \lambda^{d\dagger}_{aij} Q_i {H_a}^c \bar{d}_j 
- \lambda^{\ell \dagger}_{aij} L_i {H_a}^c \bar{\ell}_j 
+ \mathrm{h.c.} \bigg],
\label{eq:2HDMLagrangian}
\end{equation}
%eoe%
where the matrices $\lambda^{f}_{aij}$, $f=u,d,\ell$ specify the couplings of the two doublets to the SM fermions  and $V(H_1, H_2)$ is the potential for the doublets. 
The potential is given by
%boe%
\begin{eqnarray}
 \nonumber
 V(H_1,H_2) & = &  m_1^2  H_1^\dagger  H_1+ m_2^2  H_2^\dagger  H_2+\Big( m_{12}^2  H_1^\dagger  H_2 +\textrm{h.c.}\Big) 
 \\
 \nonumber
 &+ &
\frac{1}{2} {\lambda}_1 ( H_1^\dagger  H_1)^2+ \frac{1}{2} {\lambda}_2( H_2^\dagger  H_2)^2 
+{\lambda}_3( H_2^\dagger  H_2)( H_1^\dagger  H_1)+ {\lambda}_4 ( H_2^\dagger  H_1)( H_1^\dagger  H_2) 
\\
& + & 
\bigg[ ~ \frac{1}{2}{\lambda}_5( H_1^\dagger  H_2)^2+{\lambda}_6  H_1^\dagger  H_1  H_1^\dagger  H_2 +{\lambda}_7 ( H_2^\dagger  H_2)( H_1^\dagger  H_2)+\textrm{h.c.}~\bigg] ,
\label{eq:2HDMpotential}
\end{eqnarray}
%eoe%
where, in general, $m_{12}^2, \lambda_5, \lambda_6$, and $\lambda_7$ are complex while the remaining quartic couplings are real. 
We require that the potential \eqref{2HDMpotential} leads to the usual spontaneous gauge symmetry-breaking pattern $SU(2)_L \times U(1)_{Y} \to U(1)_{EM}$~\cite{Ferreira:2004yd}.
In this case, by performing a $U(2)$ rotation in the space of the two Higgs doublets $H_{1,2}$,
it is always possible to find a basis in which only $H_{1}$ is responsible for breaking electroweak symmetry and giving mass to the SM fermions and bosons, 
while $H_2$ does not condense.
This basis is called the Higgs basis~\cite{Georgi:1978ri, Botella:1994cs},
and from here on out $H_a$ will always refer to the doublets in the Higgs basis.
The condensates in the Higgs basis are 
\begin{equation}
\langle H_1^{\dagger} H_1 \rangle = \frac{v^2}{2}, \qquad \langle H_2^{\dagger} H_2 \rangle = 0,
\label{eq:higgs_basis}
\end{equation}
where $v = 246\gev$. 
For more details of the Higgs basis and the electroweak symmetry-breaking conditions leading to the condensates \eqref{higgs_basis} we refer the reader to \cite{Egana-Ugrinovic:2015vgy}. 
Note that since the second Higgs doublet does not condense,
there cannot be any spontaneous $CP$ breaking in the Higgs basis:
any $CP$ violation from the Higgs potential must appear explicitly as phases of the potential couplings.

The Higgs basis is particularly useful for discussing flavor prescriptions in the 2HDM. 
This is because in this basis only $H_1$ gives mass to the SM fermions, so
its Yukawa matrices $\lambda^f_{1ij}$ must correspond to the Yukawa matrices of the SM.
All additional sources of flavor breaking
are contained in the Yukawa matrices of the second doublet, $\lambda^f_{2ij}$,
which can be specified by the flavor prescription.
More explicitly, 
in a general flavor basis, the Yukawas for the first doublet can be written in terms of their singular value decomposition
%boe%
\begin{eqnarray}
\label{eq:sm_yukawas}
\begin{split}
\lambda^u_{1,ij} & \equiv U_{Q_u} Y^u U_{\bar{u}}^{\dagger}, \\
\lambda^{d\dagger}_{1,ij} &  \equiv U_{Q_d} Y^d U_{\bar{d}}^{\dagger}, \\
\lambda^{\ell\dagger}_{1,ij} & \equiv U_{L} Y^\ell U_{\bar{\ell}}^{\dagger},
\end{split}
\end{eqnarray}
%eoe%
where $U_{Q_{u,d}}$, $U_{\bar{u}, \bar{d}}$ and $U_{L,\ell}$ are unitary matrices which depend on the choice of flavor basis,
and $Y^{u,d,\ell}$ are the positive-diagonal matrices containing the SM Yukawa couplings, 
which are flavor invariants. 
The real-diagonal matrices $Y^{u,d,\ell}$ are related to the quark masses through
%boe%
\begin{eqnarray}
\label{eq:sm_yukawas_svd}
\begin{split}
Y^u & \equiv \mathrm{diag}(y^{\textrm{SM}}_u, y^{\textrm{SM}}_c, y^{\textrm{SM}}_t) = \frac{\sqrt{2}}{v} \mathrm{diag}(m_u, m_c, m_t) \quad , \\
Y^d & \equiv \mathrm{diag}(y^{\textrm{SM}}_d, y^{\textrm{SM}}_s, y^{\textrm{SM}}_b) = \frac{\sqrt{2}}{v} \mathrm{diag}(m_d, m_s, m_b) \quad , \\
Y^\ell & \equiv \mathrm{diag}(y^{\textrm{SM}}_e, y^{\textrm{SM}}_{\mu}, y^{\textrm{SM}}_{\tau}) = \frac{\sqrt{2}}{v} \mathrm{diag}(m_e, m_{\mu}, m_{\tau}) \quad . \\
\end{split}
\end{eqnarray}
%eoe
The remaining quark-sector observables are contained in the CKM matrix, defined as the flavor-invariant unitary bilinear
%boe%
\begin{equation}\label{eq:VCKM}
V = U_{Q_u}^T U_{Q_d}^*.
\end{equation}
%eoe%
The Yukawa matrices  $\lambda_2^{u,d,\ell}$ for the second doublet, on the other hand, 
are not fixed by measured SM parameters.
While these matrices are in principle arbitrary, 
in their most general form they lead to tree level FCNCs mediated by the second doublet. 
Forbidding these FCNCs at tree level requires that the second-doublet Yukawa matrices must be simultaneously diagonalizable with the Yukawa matrices of the first Higgs doublet \cite{Gatto:1978dy, Gatto:1979mr, Sartori:1979gt, Grimus:1986mh}.
The conditions for simultaneous diagonalizability are:
%boe%
\begin{equation}
\label{eq:flavor_basis_conditions}
\big[ U_{Q_u}^{\dagger} \lambda_{2}^{u} U_{\bar{u}}\big]_{ij} = \delta_{ij} A_i^{u}
\qquad
\big[ U_{Q_d}^{\dagger} \lambda_{2}^{d\dagger} U_{\bar{d}}\big]_{ij} = \delta_{ij} A_i^{d}
\qquad
\big[ U_L^{\dagger} \lambda_{2}^{u} U_{\bar{\ell}}\big]_{ij} = \delta_{ij} A_i^{\ell}
\quad ,
\end{equation}
%eoe%
where the unitary matrices are the same matrices that diagonalize the first-doublet Yukawas in \eqref{sm_yukawas}.
The $A_i^{u,d,\ell}$ ($i=1..3$) are complex couplings
that control the strength of the second Higgs doublet interactions with the first-, second- and third-generation SM fermions, 
and their phases are physical $CP$ violating phases \cite{Egana-Ugrinovic:2018znw}.
We refer to the flavor prescription for the second-doublet Yukawa matrices in \eqref{flavor_basis_conditions} as ``flavor alignment".   
While flavor alignment can be studied in a general flavor basis, 
for practical purposes it is convenient to choose a particular one.  
For the rest of the paper we commit to the commonly used flavor basis in which the SM down-type Yukawa matrix is diagonal, the SM up-type quark Yukawa matrix contains the CKM angles, and the lepton Yukawa matrix is diagonal. 
In this flavor basis, 
the unitary matrices in \eqref{sm_yukawas} are given by
%boe%
\begin{equation}
U_{\bar{u}} = U_{\bar{d}} = U_{Q_d} = U_{L}  = U_{\bar{\ell}} = \mathbb{1} \quad , \quad  U_{Q_u} = V^T  \quad ,
\label{eq:flavorbasis}
\end{equation}
%eoe%
where $V$ is the CKM matrix. Note that in the flavor basis \eqref{flavorbasis} the up quarks are not mass eigenstates.
To find the couplings of the Higgs bosons to quark mass eigenstates,
one must further perform an $SU(2)_W$-breaking redefinition of the up quark in the left-handed doublet $Q=(u \,\, d)$: 
%boe%
\begin{equation}
u \rightarrow u V^* \quad .
\label{eq:masseigenbasis}
\end{equation}
%eoe%
While the flavor alignment conditions \eqref{flavor_basis_conditions} may be chosen as {\em ad hoc} conditions of the 2HDM as an effective theory,
there is no symmetry principle to impose them.
As a consequence, 
flavor-aligned theories are generically extremely tuned theories.
Exceptionally, 
proportionality of the first- and second-Higgs doublet Yukawas $\lambda_2^{u,d,\ell} \propto \lambda_1^{u,d,\ell}$ (which guarantees simultaneous diagonalizability),
may be imposed via discrete symmetries as in the Natural Flavor Conserving (NFC) types I--IV 2HDM \cite{Glashow:1976nt}.
However, requiring proportionality of the two doublet Yukawas has the drawback of restricting the second doublet to be mostly coupled to third-generation fermions,
limiting the phenomenology at colliders.

To address these limitations, 
in the next section we introduce the spontaneous flavor-violating (SFV) two-Higgs-doublet models, 
which are theories in which flavor alignment is ensured in a technically natural way by a UV completion,
and allow for large couplings to any quark generation.

%%%%%%%%%%%%%%%%%%%%%%%%%%%%%%%%%%%%%%%%%%%%%%%%%%%%%%%%
\subsection{The Spontaneous Flavor-Violating two-Higgs-doublet models}
\label{ss.SFV2HDM}
%%%%%%%%%%%%%%%%%%%%%%%%%%%%%%%%%%%%%%%%%%%%%%%%%%%%%%%%
In this section, we introduce the up- and down-type SFV 2HDMs.
As any other type of 2HDM, 
our theories are defined by specifying the Yukawa matrices of the second doublet $H_2$
and the Higgs potential.
In this section we limit ourselves to discussing the defining features of SFV 2HDMs,
and we leave a detailed discussion of a UV completion leading to the SFV structure to Appendix \ref{s.uv_completion}.
The SFV 2HDMs are defined to be $CP$ conserving, 
in the sense that they do not introduce additional phases beyond the CKM phase contained in the Yukawa matrices. 
In particular, the Higgs potential is allowed to be arbitrary as long as it is $CP$ conserving. 
Without loss of generality, 
we may then take all the Higgs potential couplings and masses in \eqref{2HDMpotential} to be real by performing a $U(1)_{\textrm{PQ}}$ rotation of the second Higgs doublet.
We commit to this Peccei-Quinn (PQ) basis in what follows.

\begin{center}\textbf{The second-doublet quark Yukawa matrices} \end{center}
In the \textbf{up-type SFV 2HDM}, the second-doublet up-type quark Yukawa matrix is required to be equal to the corresponding SM Yukawa matrix up to a proportionality constant.
The second doublet down-type quark Yukawa matrix on the other hand, 
is allowed to be a new matrix which is flavor aligned with the down-type quark SM Yukawa matrix,
without necessarily being proportional to it. 
In the flavor spurion language, 
in up-type SFV no new flavor spurions transforming under $U(3)_{Q}\times U(3)_{\bar{u}}$ are allowed besides the up-type SM quark Yukawa,
but a new flavor-aligned spurion $\lambda^d_{2,ij}$, which transforms under $U(3)_{Q} \times U(3)_{\bar{d}}$, is allowed.
Explicitly, the second-doublet quark Yukawa matrices in the up-type SFV 2HDM, in the previously discussed flavor basis, are given by
\begin{eqnarray}
\lambda^u_{2} &=&
 \xi \, V^T Y^u  
 \quad ,
\nonumber \\
\lambda^d_{2} &=& K^d
 \equiv  \mathrm{diag}( \kappa_d , \kappa_s,\, \kappa_b )  \quad ,
 \label{eq:uptypeSFV}
\end{eqnarray}
where
$\xi$ is a proportionality constant.
Furthermore, 
since the SFV 2HDMs are $CP$ conserving, 
the new couplings $\kappa_d\, , \kappa_s$, and $\kappa_b$, and the proportionality constant $\xi$ must be real  in our PQ basis.

In the up-type SFV 2HDM, the Yukawas $\kappa_d\, , \kappa_s$, and $\kappa_b$ independently control the couplings of the second doublet to each down-type quark generation,
while the couplings to up-type quarks are universally proportional to the corresponding SM Yukawas.
As a consequence, 
the second doublet may couple to the different down-type quark generations with arbitrary hierarchies, 
but the couplings to up-type quarks respect the SM hierarchies.

The \textbf{down-type SFV 2HDM} is the same as the up-type, 
but with the roles of up- and down-type quarks exchanged. 
The second doublet Yukawa matrices in the down-type SFV 2HDM are given by
\begin{eqnarray}
\lambda^u_{2} &=& V^T \, K^u \equiv  V^T \, \mathrm{diag}( \kappa_u,\, \kappa_c,\, \kappa_t )
\quad ,
\nonumber \\
\lambda^d_{2} &=& \xi \,  Y^d
 \quad ,
 \label{eq:downtypeSFV}
\end{eqnarray}
where $\xi$ is a real proportionality constant and $\kappa_{u},\kappa_c,\kappa_t$ are real Yukawa couplings in our PQ basis.
Since these Yukawas are free parameters,
in down-type SFV the second-Higgs doublet couplings to up-type quarks do not necessarily respect the SM Yukawa hierarchies.

\begin{center} \textbf{The second-doublet lepton Yukawa matrices} \end{center}
In both the up- and down-type SFV 2HDMs no new spurions transforming as the SM lepton Yukawa matrix are allowed. 
This means that in both types of SFV 2HDMs,  
the lepton Yukawa matrix of the second doublet must be proportional to the corresponding SM one.
The second-doublet Yukawa matrix is then
\begin{eqnarray}
\lambda^\ell_{2} &=& \xi^\ell  Y^\ell  
\quad ,
\label{eq:leptonyukawas}
\end{eqnarray}
where the proportionality constant $\xi^\ell$ is real in our PQ basis.

\begin{comment}
We conclude by stressing a series of important properties of the SFV 2HDMs.
First, note that the only difference between the up and down-type SFV 2HDMs lies on the different second-doublet quark Yukawa matrices in both models, 
the lepton Yukawa and Higgs potential structure being the same.
Second, the up-type (down-type) SFV 2HDM allows for generation specific couplings to down-type (up-type) quarks, 
while the couplings to the opposite quark sector and lepton sector are limited to be MFV-like.
Third,
the SFV 2HDMs are flavor-aligned (c.f. \eqref{flavor_basis_conditions}), 
so they do not lead to tree level FCNCs.
Finally, 
since the SFV 2HDM does not contain any additional sources of $CP$ violation besides the CKM phase, 
it does not suffer from stringent bounds from measurements of the electron or neutron electric dipole moments~\cite{Andreev:2018ayy,Egana-Ugrinovic:2018fpy}.
In addition, in \sref{uv_completion} we will show that the strong-$CP$ problem is automatically solved in our SFV UV completion.

While in this section we have discussed the couplings of the second doublet to the SM fermions,
the physical Higgs fields in the theory are the scalar mass eigenstates,
so we now move on to discussing the couplings of such Higgs bosons to the SM fermions.

\end{comment}

%%%%%%%%%%%%%%%%%%%%%%%%%%%%%%%%%%%%%%%%%%%%%%%%%%%%%%%%
\subsection{Physical Higgs bosons and their couplings to SM fermions}
\label{ss.couplings}
%%%%%%%%%%%%%%%%%%%%%%%%%%%%%%%%%%%%%%%%%%%%%%%%%%%%%%%%
Having defined the SFV 2HDMs, we now review the physical couplings of the Higgs mass eigenstates, 
which are needed for a phenomenological investigation.  
We confine ourselves to the the couplings to fermions, which distinguish the SFV theories.
The couplings to gauge bosons and the self-couplings correspond to the ones of a generic $CP$ conserving 2HDM, 
and can be found elsewhere \cite{Gunion:2002zf}.

In unitary gauge, the $SU(2)$ components of the doublet fields $H_1$ and $H_2$ can be written in terms of three real and neutral Higgs fields $h_{a}$, $a=1..3$ and one charged Higgs boson $H^\pm$
%boe%
\begin{eqnarray}\label{eq:doublet_components}
\begin{split}
H_1 & = \bigg( 
	{\def\arraystretch{0.9}\tabcolsep=10pt
	\begin{array}{c}  0\\ H_1^0  \end{array} }
	\bigg) 
	= \frac{1}{\sqrt{2}} \bigg( 
	{\def\arraystretch{1}\tabcolsep=10pt
	\begin{array}{c} 0 \\ v+ h_1 \end{array} }
	\bigg) \quad , \quad \\
H_2 & = \bigg( 
	{\def\arraystretch{0.9}\tabcolsep=10pt
	\begin{array}{c} H^+ \\ H_2^0 \end{array} }
	\bigg) 
	= \frac{1}{\sqrt{2}}\bigg( 
	{\def\arraystretch{1}\tabcolsep=10pt
	\begin{array}{c} \sqrt{2} H^+ \\ h_2 + i h_3 \end{array} }
	\bigg)  \quad .
\end{split}
\end{eqnarray}
%eoe%
The physical mass eigenstates in the two Higgs doublets are the charged Higgs $H^\pm$, the pseudoscalar Higgs
$h_3$, usually denoted as $h_3 \equiv A$, 
and two $CP$ even scalars $h, H$ which are a combination of the components $h_1,h_2$ above.
The charged Higgs boson $H^{\pm}$ resides entirely in the second Higgs doublet $H_2$, and has mass
%boe%
\begin{equation}
m^2_{H\pm}=m_2^2+\frac{1}{2}\lambda_3v^2
\quad .
\end{equation}
%eoe%
The $CP$-odd higgs has mass
\begin{equation}
m_{A}=m_2^2+ \frac{1}{2}v^2 \Big( \lambda_3+ \lambda_4 - \lambda_5\Big) \quad.
\end{equation}
Finally, the masses of the neutral $CP$-even mass eigenstates $h, H$ can be obtained by diagonalizing the scalar mass matrix
%boe%
\begin{equation}
{\cal M}^2=\left(
{\def\arraystretch{1}\tabcolsep=10pt
\begin{array}{cc} 
v^2 \lambda_1 
&  
v^2 \lambda_6 
\\  
v^2 \lambda_6 &
m_2^2+ \frac{1}{2}v^2 \Big( \lambda_3+ \lambda_4 + \lambda_5\Big) 
\end{array}}\right).
\label{eq:massmatrixHiggsbasis}
\end{equation}
%eoe%
The $CP$-even mass eigenstates $h,H$ are given in terms of the original fields $h_{1,2}$ by the linear combinations 
%boe%
\begin{eqnarray}
\begin{split}
\label{eq:higgs_mass_eigenstates}
h & \equiv \sin(\beta - \alpha) h_1 + \cos(\beta - \alpha)h_2 \quad , \\
H & \equiv -\cos(\beta - \alpha) h_1 + \sin(\beta - \alpha) h_2 \quad , \\
\end{split}
\end{eqnarray}
%eoe%
where $\beta - \alpha$ is traditionally referred as the alignment angle. 
We will refer to the alignment parameter as  $\cos(\beta - \alpha)$.
In terms of the elements of the mass matrix \eqref{massmatrixHiggsbasis}, the alignment angle is given by
%boe%
\begin{eqnarray}
\begin{split}
\tan\big[ 2(\beta - \alpha) \big] & = \frac{-2 \mathcal{M}_{12}^2}{ \mathcal{M}_{22}^2 - \mathcal{M}_{11}^2 } \\
&= \frac{2 \lambda_6 v^2}{\lambda_1 v^2 - \big( m_2^2 +  \frac{1}{2}(\lambda_3 + \lambda_4 + \lambda_5)v^2 \big)}.
\label{eq:alignment_angle}
\end{split}
\end{eqnarray}
%eoe%
In what follows and without loss of generality, 
we associate the mass eigenstate $h$ with the $125$ \textrm{GeV} Higgs boson observed at the LHC.
Note that in the limit where $\cos(\beta-\alpha) = 0$,
usually referred to as the \textit{Higgs alignment limit} (not to be confused with flavor alignment),
the $125$ \textrm{GeV} Higgs boson resides entirely in the Higgs doublet $H_1$ breaking electroweak symmetry, 
as in the SM.
For this reason, 
in the Higgs alignment limit the $125$ \textrm{GeV} Higgs boson of the 2HDM is Standard Model-like, 
with tree level couplings to fermions and gauge bosons that coincide with the SM expectations.
From \eqref{alignment_angle} we see that the alignment limit can be reached in two ways while retaining perturbativity: either by taking the mass of the second doublet $m_2\to \infty$ while holding the renormalizable couplings fixed (the ``decoupling limit"), 
or by taking $\lambda_6 \to 0$ (``alignment without decoupling"). 
We refer the reader to \cite{Egana-Ugrinovic:2015vgy,Gunion:2002zf} for a detailed discussion of the alignment limit.

\begin{center}
\textbf{
Up-type SFV physical Higgs-fermion couplings
}
\end{center}
We first summarize the Yukawa matrices in the up-type SFV 2HDM. 
In our selected flavor basis  \eqref{flavorbasis}, 
and collecting the first- and second-doublet Yukawas from \eqref{sm_yukawas}, \eqref{uptypeSFV} and \eqref{leptonyukawas}, 
the up-type SFV 2HDM Yukawa matrices are
%boe%
\begin{equation}
\label{eq:uptypeflavorbasis}
 \quad
\begin{array}{ccccccccccc}
\lambda_1^u &=& V^T Y^u  
&
\quad
&
 \lambda_1^d &=& Y^d 
&
\quad
&
 \lambda_1^\ell &=& Y^\ell  
 \quad ,
\\
\lambda_2^u &=& \xi V^T Y^u 
&
\quad
&
\lambda_2^d &=& K^d
&
\quad
&
\lambda_2^\ell &=& \xi^\ell Y^\ell \quad 
\, .
\end{array}
\end{equation}
%eoe%
where the real-diagonal SM Yukawa matrices $Y^{u,d,\ell}$ and the real-diagonal SFV Yukawa matrix $K^{d}$ are defined in Eqns.~\eqrefp{sm_yukawas_svd} and \eqrefp{uptypeSFV},
and $\xi, \xi^\ell$ are real proportionality constants.
The couplings of the physical Higgs bosons $h, H, A$ and $H^\pm$ to the SM fermions in up-type SFV can be easily obtained from using the Yukawa matrices \eqref{uptypeflavorbasis} in the 2HDM Lagrangian \eqref{2HDMLagrangian}, 
the definition of the doublet components \eqref{doublet_components} and of the neutral mass eigenstates \eqref{higgs_mass_eigenstates}, 
and by performing the rotation from our flavor basis to the quark mass eigenbasis \eqref{masseigenbasis}.
We summarize the couplings of the physical Higgs bosons to the quark mass eigenstates in Appendix \ref{a.couplings}, 
\tref{yukawaup}. \\

\begin{center}
\textbf{
Down-type SFV 2HDM physical Higgs-fermion couplings
}
\end{center}
In our selected flavor basis \eqref{flavorbasis}, 
and collecting the Yukawa matrices \eqref{sm_yukawas}, \eqref{downtypeSFV} and \eqref{leptonyukawas},
the down-type SFV 2HDM Yukawa matrices are given by

%boe%%
\begin{equation}
\label{eq:downtypeflavorbasis}
 \quad
\begin{array}{ccccccccccc}
\lambda_1^u &=& V^T Y^u  
&
\quad
&
 \lambda_1^d &=& Y^d 
&
\quad
&
 \lambda_1^\ell &=& Y^\ell  
 \quad ,
\\
\lambda_2^u &=& V^T K^u 
&
\quad
&
\lambda_2^d &=& \xi Y^d
&
\quad
&
\lambda_2^\ell &=& \xi^\ell Y^\ell \quad 
\, ,
\end{array}
\end{equation}
%eoe%
where the real-diagonal SM Yukawa matrices $Y^{u,d,\ell}$ and the real-diagonal SFV Yukawa matrix $K^{u}$ is defined in Eqns.~\eqrefp{sm_yukawas_svd} and \eqrefp{downtypeSFV}.
We summarize the corresponding couplings of the physical Higgs bosons to the quark mass eigenstates in in Appendix \ref{a.couplings}, \tref{yukawadown}.

%%%%%%%%%%%%%%%%%%%%%%%%%%%%%%%%%%%%%%%%%%%%%%%%%%%%%%%%
%%%%%%%%%%%%%%%%%%%%%%%%%%%%%%%%%%%%%%%%%%%%%%%%%%%%%%%%
%% SECTION BEGINS
%%%%%%%%%%%%%%%%%%%%%%%%%%%%%%%%%%%%%%%%%%%%%%%%%%%%%%%%
%%%%%%%%%%%%%%%%%%%%%%%%%%%%%%%%%%%%%%%%%%%%%%%%%%%%%%%%
\section{Flavor phenomenology of the up-type SFV 2HDM}
\label{s.flavor}
While the SFV 2HDM is free from FCNCs at tree level, 
contributions to FCNCs arise at loop level.
In this section we obtain bounds from FCNCs on the SFV 2HDM. 
For brevity, we concentrate on the up-type SFV 2HDM 
and leave a study of flavor constraints on down-type SFV for future work.

In theories with generic Yukawa structure for the second Higgs doublet, 
loop suppression of FCNCs is generically not enough to allow for extra Higgs states close to the electroweak scale to be consistent with stringent experimental bounds from the absence
of FCNCs \cite{Crivellin:2013wna}. 
However,
FCNCs in SFV theories are further suppressed by CKM elements and SM Yukawas \cite{Egana-Ugrinovic:2018znw}.
In our up-type SFV 2HDM, this feature may be seen by writing down all the quark bilinears leading to FCNCs together with the appropriate Yukawa spurions required for consistency with the flavor symmetries. 
Using our second-Higgs-doublet Yukawa matrices  \eqref{uptypeflavorbasis},
the FCNC quark bilinears at leading order in a spurion expansion are
%boe
\begin{eqnarray}
 &
 d\,\, (V^T Y_{u}^2 V^*) \, \,K^d\, \bar{d}  \,\, , \,\,
\bar{d}^\dagger  K^d\, (V^T Y_{u}^2 V^*) \,K^d\,\bar{d} \,\, , \,\,
 \bar{d}^\dagger Y^d\, (V^T Y_{u}^2 V^*) \,K^d\, \bar{d} 
 &  
 \quad
 \text{down-sector FCNC} 
 \nonumber 
 \\
\label{eq:SFVbilinearsdown}
 \\
&
u\,(\, V^* K_d^2 V^T \,)u^\dagger \,\, , \,\,
u\,(\, V^* K_d^2 V^T \,) \,Y^{u} \bar{u}  &  
 \quad
\text{up-sector FCNC} 
\nonumber 
\\
\label{eq:SFVbilinearsup}
\end{eqnarray}
%% eoe
along with the same bilinears where $K^d$ is replaced by $Y^d$, which exist also in MFV theories.
Here we remind that $u$ and $d$ are the components of the left-handed $SU(2)$ doublet while $\bar{u}$ and $\bar{d}$ are the right-handed $SU(2)$ singlet quarks.
From \eqref{SFVbilinearsdown} we see that all down-type FCNCs
are strongly suppressed by the off-diagonal elements of the matrix combination $(V^T Y_{u}^2 V^*)_{ij}  \simeq y_t^2 V_{3i} V^{*}_{3j}$. 
Contributions proportional to other CKM matrix elements are suppressed by the GIM mechanism and the smallness of the up and charm SM Yukawas. 
Up-type FCNCs, on the other hand, \eqref{SFVbilinearsup},
are only suppressed by factors of $V^* K_d^2 V^T\equiv V^* \textrm{diag}(\kappa_d^2, \kappa_s^2 , \kappa_b^2) V^T$.
If we take $\kappa_d$ to be large,
we expect loop-induced $D-\bar{D}$ mixing to be suppressed only by factors of $\kappa_d^2 \, V_{11} V^{*}_{12} \sim 0.22\, \kappa_d^2$.
This indicates that in up-type SFV, 
up-type meson mixing phenomenology is particularly relevant,
as we will see in detail in the following sections. 
\footnote{On the other hand and by similar arguments, we expect that in the down-type SFV, down-type FCNCs will lead to the most stringent flavor constraints instead. Due to strong limits especially from mixing of $K$ mesons, we expect
the flavor constraints in the down-type SFV 2HDM to be more severe than the ones studied in this section for the up-type SFV 2HDM.}

Loop-level FCNCs in the SFV 2HDM can be divided in FCNCs induced by direct contributions of one loop diagrams and those due to flavor misalignment between the Yukawa matrices of the two Higgs doublets due to RGE running.
We dedicate the rest of this section to study bounds from direct contributions to $\Delta F=1$ and $\Delta F=2$ processes, 
and we leave a dedicated study of radiative corrections to the SFV Yukawas for \aref{radiative_corrections}.

To simplify the study of flavor violation, 
and motivated by the proximity of the 125 \textrm{GeV} Higgs coupling measurements to the SM expectations \cite{Craig:2013hca}, 
for the rest of this section and in \sref{collider} we work in the Higgs alignment limit. 
In this limit, the alignment parameter in \eqref{higgs_mass_eigenstates} is equal to zero,
$\cos(\beta - \alpha) = 0$, which is obtained by setting $\lambda_6 \rightarrow 0$ in the Higgs potential.
Also for simplicity, we take the Higgs mass eigenstates belonging to the second Higgs doublet to be degenerate,
$m_H = m_{H^{\pm}} = m_A$, 
by further setting $\lambda_4, \lambda_5 \to 0$.
For the purposes of flavor bounds, nonzero values of $\lambda_4$ and $\lambda_5$ only introduce mass splittings between the different heavy Higgs states, 
which do not significantly affect our discussion.
In addition, since we are mostly interested in quark phenomenology,
for the rest of this work we set the lepton Yukawas of the second doublet to zero
by choosing $\xi^\ell=0$ in \eqref{uptypeflavorbasis}.
Including nonzero lepton Yukawas would only complicate our presentation and does not significantly affect our conclusions unless 
$\xi^\ell \gg 1$.
Nonetheless,
we have checked that for $\xi^\ell \leq 1$, the flavor bounds that we present in this section on the SFV quark-sector parameters are stronger than bounds from semileptonic B- and D-meson decays that arise at one loop when couplings to leptons are allowed.

With these simplifications, the up-type SFV 2HDM is described by five new parameters: the mass of the extra Higgs bosons $m_H$, the MFV-like proportionality factor for the up-type quark Yukawa matrices, $\xi$, and the three new Yukawa couplings, $\kappa_d, \kappa_s$, and $\kappa_b$.

%%%%%%%%%%%%%%%%%%%%%%%%%%%%%%%%%%%%%%%%%%%%%%%%%%%%%%%%
\subsection{Constraints from $B \to X_{s,d}\gamma$ transitions}
\label{ss.btosgamma}
%%%%%%%%%%%%%%%%%%%%%%%%%%%%%%%%%%%%%%%%%%%%%%%%%%%%%%%%
\begin{figure}
  \centering
  \includegraphics[width=0.4\linewidth]{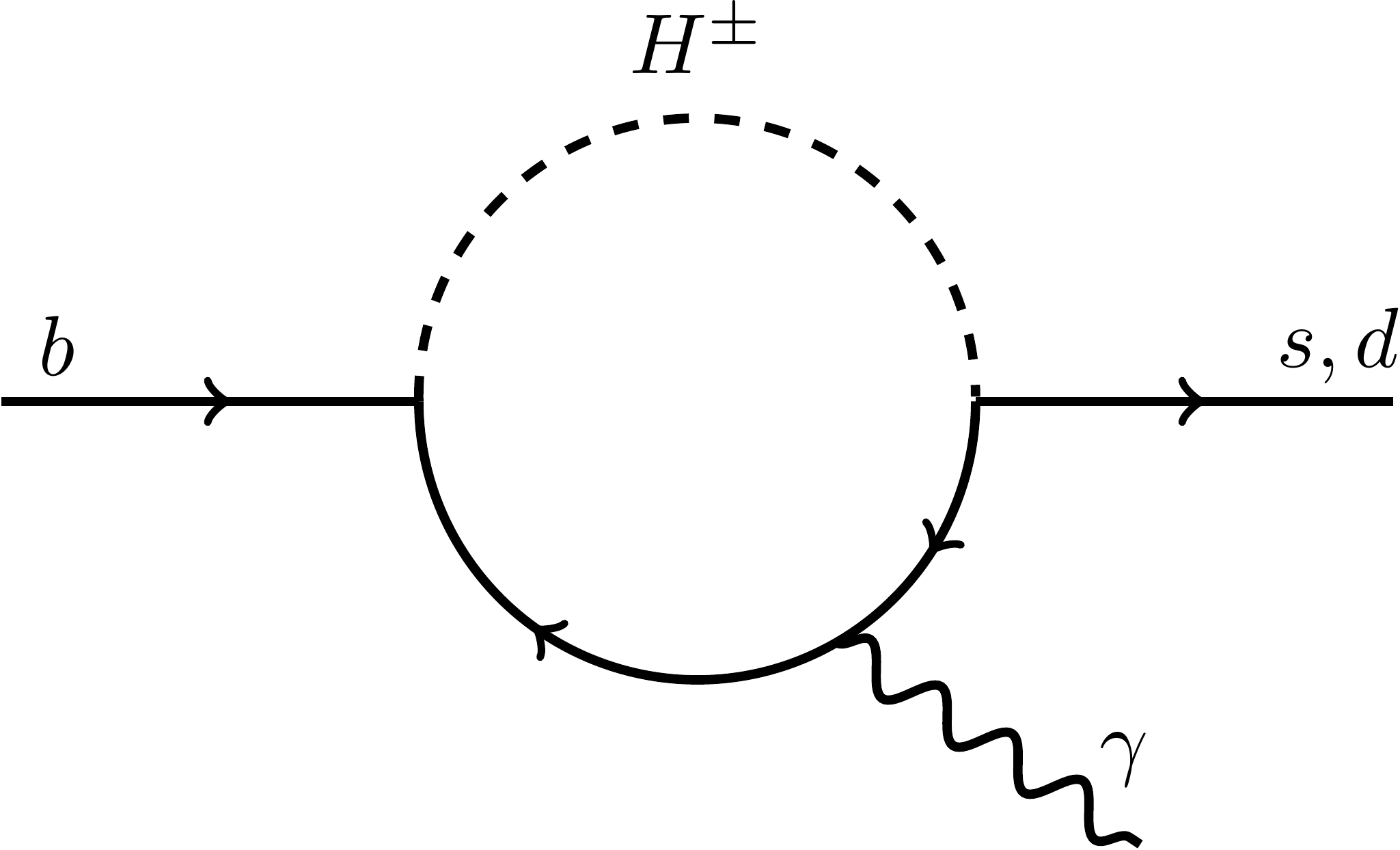}
  \caption{Charged Higgs contribution to the $B\rightarrow X_{s,d} \gamma$ amplitude.}
  \label{fig:b_sgamma}
\end{figure}
%%%%%%%%%%%%%%%%%%%%%%%%%
Contributions from the charged Higgs to B-meson radiative decays into a photon arise at one loop from penguin diagrams such as, e.g., \figref{b_sgamma}.
Such diagrams contribute to the coefficients $C^{bs}_{7},C'^{bs}_{7}$ of the  $b\to s\gamma$ transition operators
%boe%
\begin{eqnarray}
\begin{split}
\mathcal{O}^{bs}_7 & = i\frac{e}{8\pi^2} m_b \big( Q_2 \sigma_{\mu\nu} \bar{d}_3 \big)^{\dagger} F^{\mu\nu},\\
\mathcal{O}^{bs}_{7'} & = i \frac{e}{8\pi^2} m_b \big( Q_3 \sigma_{\mu\nu} \bar{d}_2 \big) F^{\mu\nu} \quad .
\label{eq:btosgamma_ops}
\end{split}
\end{eqnarray}
%eoe%
The Wilson coefficients $C_7$ and $C_7'$ for the 2HDM have been computed in~\cite{Crivellin:2013wna,Hou:1987kf,Borzumati:1998tg} in terms of generic charged Higgs-fermion couplings. 
Using these results and the charged Higgs couplings in \tref{yukawaup}, 
we obtain
\begin{align}
\begin{split}
  C^{bs}_7 & = \frac{v^2}{V_{tb}V^*_{ts}} \sum_{j=u,c,t} V_{jb}V_{js}^* \left(
  -\xi\, \kappa_b\, y_j  \frac{C^0_{7,XY}(z_j)}{m_b\, m_{u_j}}
  + \xi^2\, y_j^2\, \frac{C^0_{7,YY}(z_j)}{m_{u_j}^2}
  \right) \\
  C'^{bs}_7 & = \frac{v^2}{V_{tb}V^*_{ts}} \sum_{j=u,c,t} V_{jb}V_{js}^* \left(
  -\xi\, \kappa_s\, y_j \frac{C^0_{7,XY}(z_j)}{m_b\, m_{u_j}}
  + \kappa_s\, \kappa_b \frac{C^0_{7,YY}(z_j)}{m_{u_j}^2}
  \right)
  \quad ,
 \end{split}
  \label{eq:bsgamma}
\end{align}
where $z_j \equiv m_{j}^2/m_{H}^2$, and the functions $C^0_{7,XY}, C^0_{7,YY}, C^0_{8,XY}$, and $C^0_{8,YY}$ are given in \aref{loop_functions}.
The coefficients for the $b\to d\gamma$ transitions, $C^{bd}_{7}$ and $C'^{bd}_{7}$, are obtained by simply replacing all the indices $s \to d$ in \eqref{bsgamma}. 

Limits on $C'^{bd}_{7}$ were presented in ref.~\cite{Crivellin:2011ba}, and on $C^{bs}_{7}, C'^{bs}_{7}$ have been presented in ref.~\cite{Capdevila:2017bsm}.
We summarize the constraints on these operators coefficients in  \tref{flavor_measurements}.
We translate these limits into constraints on the quark-sector parameters of the SFV quark Yukawa couplings using \eqref{bsgamma}. 
The results in the $m_H$ v.s $\kappa_j$ ($j=d,s,b$) plane are shown in Figs. \ref{fig:flavor_kappad} -- \ref{fig:flavor_kappab} in green. 
In each figure, 
in the left panel we show the constraints for $\xi = 0.1$ and in the right panel for  $\xi =1.0$.

In principle there are also constraints on $\Delta F = 1$ transitions in the up-sector which could be important given that up-sector FCNCs need not  be GIM and Yukawa suppressed. We have checked explicitly for our model, however, that the bounds on $c\to u\gamma$ transitions presented in refs.~\cite{deBoer:2017que} and \cite{Cerri:2018ypt} are always weaker than the $D-\bar{D}$ mixing constraints discussed below.

 %%%%%%%%%%%%%%%%%%%%%%%%%
\begin{table*}[htbp!]
\centering
\begin{tabular}{|c|c|c|c|}
\hline
Process & Parameters &  $95\%$ C.L. range & Reference \\ \hline
$b \to d\gamma$ & ${C'}^{bd}_{7}$ & $ \leq 0.85$ & \cite{Crivellin:2011ba} 
\\ \hline
\multirow{2}{*}{$b\to s\gamma (\ell^+\ell^-)$} 
& ${C}^{bs}_{7}$ & $\in [-3.0, 7.0]  \times 10^{-2} $ 
& \multirow{2}{*}{\cite{Capdevila:2017bsm}} 
\\
& ${C'}^{bs}_{7}$ & $ \in [-3.0, 8.0] \times 10^{-2}  $ &  
 \\ \hline
\multirow{2}{*}{$K-\bar{K}$ mixing} & $\Im C_{K,1}$ & $\in [-4.4, 2.8] \times 10^{-15}$ & \multirow{2}{*}{\cite{Bona:2007vi}} \\
 & $\Im C_{K,2}$ & $\in [-5.1, 9.3] \times 10^{-17}$ &  \\ \hline
$B_d-\bar{B}_d$ mixing & $C_{B_d}$, $(\Phi_{B_d} = 0)$ & $\in [0.8, 1.3]$ & \cite{Bona:2016bvr} \\ \hline
$B_s-\bar{B}_s$ mixing & $C_{B_s}$, $(\Phi_{B_s} = 0)$ & $\in [0.82, 1.3]$ & \cite{Bona:2016bvr} \\ \hline
$D-\bar{D}$ mixing & $|M_{12}|$ & $ \leq 7.4 \times 10^{-3}$ ps & %\cite{Bevan:2014tha, Bona:2017gut} \\ \hline
\cite{Aaij:2019jot} \\ \hline
\end{tabular}
\caption{Flavor-changing processes that arise at one loop in our SFV 2HDM, 
and the 95\% C.L. experimental range on the parameters used to set limits. 
For the limit from $D-\bar{D}$ mixing, 
we have converted the updated global fit on the value $x$ presented in ref.~\cite{Aaij:2019jot} into a limit on $\abs{M_{12}}$ using the relations in \cite{Bevan:2014tha, Bona:2017gut}.}
\label{t:flavor_measurements}
\end{table*}
%%%%%%%%%%%%%%%%%%%%%%%%%

%%%%%%%%%%%%%%%%%%%%%%%%%%%%%%%%%%%%%%%%%%%%%%%%%%%%%%%%
\subsection{Constraints from neutral meson mixing}
\label{ss.box_mixing}
%%%%%%%%%%%%%%%%%%%%%%%%%%%%%%%%%%%%%%%%%%%%%%%%%%%%%%%%
We next consider the short-distance contributions to neutral meson mixing in our 2HDM.
These contributions can be matched onto the $\Delta F = 2$ Hamiltonian
%boe%
\begin{equation}
\mathcal{H}_{\textrm{eff}}^{\Delta F = 2} = \sum_{j = 1}^5 C_j \mathcal{O}_j + \sum_{j = 1}^3 C_{j'} \mathcal{O}_{j'} 
+ \textrm{h.c.}
\label{eq:box_mixing_Heff}
\end{equation}
%eoe%
In the case of $B_s-\bar{B_s}$ mixing, the effective operators are
%boe%
\begin{equation}
\begin{aligned}
\mathcal{O}_1^{bd} & = \big( \bar{d}_{2a}^{\dagger} \bar{\sigma}^{\mu} d_{3a}\big) \big( \bar{d}_{2a}^{\dagger} \bar{\sigma}^{\mu} d_{3a}\big),\\% \quad &
\mathcal{O}_2^{bd} & = \big( \bar{d}_{2a} d_{3a} \big) \big(\bar{d}_{2b} d_{3b}\big), \\
\mathcal{O}_3^{bd} & = \big( \bar{d}_{2a} d_{3b} \big) \big(\bar{d}_{2b} d_{3a}\big),\\% \quad &
\mathcal{O}_4^{bd} & = \big( \bar{d}_{2a} d_{3a} \big) \big( Q_{2b}^{\dagger} \bar{d}_{3b}^{\dagger}\big), \\
\mathcal{O}_5^{bd} & = \big( \bar{d}_{2a} d_{3b} \big) \big( d_{2b}^{\dagger} \bar{d}_{3a}^{\dagger}\big), \\%&
\end{aligned}
\label{eq:mixing_operators}
\end{equation}
%eoe%
where $a$ and $b$ represent color indices. 
The primed operators are related to $O_{1,2,3}$ by interchanging the left- and right-handed fields.
The operators for $B_d-\bar{B_d}$, $K-\bar{K}$ and $D-\bar{D}$ mixing are obtained by simply replacing the appropriate quark fields in the expressions above. 

At one loop, 
the second Higgs doublet contributes to the four-fermion operator coefficients via box diagrams with two charged Higgs bosons, 
and one charged Higgs boson and one $W$ or Goldstone boson.
We calculate these coefficients in Feynman gauge.
For $B_s-\bar{B_s}$ mixing, 
the charged Higgs boxes give
%boe%
\begin{equation}\label{eq:SFVboxcoefficients1}
\begin{aligned}
C_1 & = \frac{-1}{128\pi^2} \sum_{j,k = 1}^3 
	\lambda^{H^+ *}_{d_2 \bar{u}_j}\, \lambda^{H^+}_{d_3 \bar{u}_j}\,
	\lambda^{H^+ *}_{d_2 \bar{u}_k}\, \lambda^{H^+}_{d_3 \bar{u}_k}\,
	D_2\left( m_{u_j}^2, m_{u_k}^2, m_H^2, m_H^2 \right) \\
C_2 & = \frac{-1}{32 \pi^2} \sum_{j,k = 1}^3 m_{u_j}\, m_{u_k}\,
	\lambda^{H^- *}_{u_j \bar{d}_2}\, \lambda^{H^+}_{d_3 \bar{u}_j}\,
	\lambda^{H^- *}_{u_k \bar{d}_2}\, \lambda^{H^+}_{d_3 \bar{u}_k}\,
	D_0\left( m_{u_j}^2, m_{u_k}^2, m_H^2, m_H^2\right) \\
C_4 & = \frac{-1}{16\pi^2} \sum_{j, k = 1}^3 m_{u_j}\, m_{u_k}\,
	\lambda^{H^- *}_{u_j \bar{d}_2}\, \lambda^{H^+}_{d_3 \bar{u}_j}\,
	\lambda^{H^+ *}_{d_2 \bar{u}_k}\, \lambda^{H^-}_{u_k \bar{d}_3}\,
	D_0\left( m_{u_j}^2, m_{u_k}^2, m_H^2, m_H^2\right) \\
C_5 & = \frac{1}{32\pi^2} \sum_{j,k = 1}^3
	\lambda^{H^- *}_{u_j \bar{d}_2}\, \lambda^{H^-}_{u_j \bar{d}_3}\,
	\lambda^{H^+ *}_{d_2 \bar{u}_k}\, \lambda^{H^+}_{d_3 \bar{u}_k}\,
	D_2\left( m_{u_j}^2, m_{u_k}^2, m_H^2, m_H^2\right)
\end{aligned}
\end{equation}
%eoe%
The coefficients for the primed operators in \eqref{SFVboxcoefficients1} are obtained by replacing $\lambda^{H^+}_{d_i \bar{u}_j} \leftrightarrow \lambda^{H^-}_{u_j \bar{d}_i}$.
For the diagrams with a charged Higgs and a $W$ or Goldstone boson we find
%boe%
\begin{equation}\label{eq:SFVboxcoefficients2}
\begin{aligned}
C_1 & = \frac{-1}{128 \pi^2}\frac{g^2}{m_W^2} \sum_{j, k = 1}^3
	V_{j2}^*\, V^{\phantom{*}}_{k3}\, m_{u_j}\, m_{u_k}\,
	\lambda^{H^+}_{d_3 \bar{u}_j}\, \lambda^{H^+ *}_{d_2 \bar{u}_k} \\
	& \qquad \times \left[ D_2\left(m_{u_j}^2, m_{u_k}^2, m_W, m_H\right) - 4 m_W^2\, D_0\left(m_{u_j}^2, m_{u_k}^2, m_W^2, m_H^2\right) \right]\\
{C}'_1 & = \frac{-1}{128 \pi^2} \frac{g^2}{m_W^2} \sum_{j, k = 1}^3\, 
	V_{j2}^* V^{\phantom{*}}_{k3}\, m_b\, m_s\,
	\lambda^{H^-}_{u_j \bar{d}_3}\, \lambda^{H^- *}_{u_k \bar{d}_2}\, 
	 D_2\left(m_{u_j}^2, m_{u_k}^2, m_W^2, m_H^2\right) \\
C_2 & = \frac{-1}{32 \pi^2} \frac{g^2}{m_W^2} \sum_{j, k = 1}^3 
	V^{\phantom{*}}_{j3}\, V_{k2}^*\, m_s\, m_{u_j}^2 m_{u_k} 
	\lambda^{H^- *}_{u_j \bar{d}_2}\, \lambda^{H^+}_{d_3 \bar{u}_k}\,
	D_0\left(m_{u_j}^2, m_{u_k}^2, m_W^2, m_H^2\right) \\
{C}'_2 & = \frac{-1}{32\pi^2} \frac{g^2}{m_W^2} \sum_{j, k = 1}^3 
	V_{j2}^*\, V^{\phantom{*}}_{k3}\, m_b\, m_{u_j}^2\, m_{u_k}\,
	\lambda^{H^-}_{u_j \bar{d}_3}\, \lambda^{H^+ *}_{d_2 \bar{u}_k}\,
	D_0\left(m_{u_j}^2, m_{u_k}^2, m_W^2, m_H^2\right) \\
C_4 & = \frac{-1}{32\pi^2} \frac{g^2}{m_W^2} \sum_{j, k = 1}^3 
	\bigg[ 
	\big( V_{j2}^*\, V^{\phantom{*}}_{k3}\, m_{u_j}\, m_{u_k}\, m_s m_b\, 
	\lambda^{H^+ *}_{d_2 \bar{u}_j}\, \lambda^{H^+}_{d_3 \bar{u}_j} \\
	& \qquad \qquad 
	+ V_{j2}^*\, V^{\phantom{*}}_{k3} m_{u_j}^2 m_{u_k}^2\, 
	\lambda^{H^-}_{u_j \bar{d}_3}\, \lambda^{H^- *}_{u_k \bar{d}_2} \big) 
	D_0 \left(m_{u_j}^2, m_{u_k}^2, m_W^2, m_H^2 \right) \\
	& \qquad - m_W^2\, V_{k2}^*\, V^{\phantom{*}}_{j3}\, 
	\lambda^{H^- *}_{u_j \bar{d}_2}\, \lambda^{H^-}_{u_k \bar{d}_3}\,
	D_2\left(m_{u_j}^2, m_{u_k}^2, m_W^2, m_H^2 \right)
	\bigg] \\
C_5 & = \frac{1}{64\pi^2} \frac{g^2}{m_W^2} \sum_{j, k = 1}^3 \big( 
	V_{j2}^*\, V^{\phantom{*}}_{k3}\, m_{u_j}\, m_b\, 
	\lambda^{H^+}_{d_3 \bar{u}_j} \lambda^{H^- *}_{u_k \bar{d}_2}
	+ V^{\phantom{*}}_{j3}\, V_{k2}^*\, m_{u_j}\, m_s\, 
	\lambda^{H^+ *}_{d_2 \bar{u}_j} \lambda^{H^-}_{u_k \bar{d}_3} \big)\\ 
	& \qquad \times 
	D_2 \left(m_{u_j}^2, m_{u_k}^2, m_W^2, m_H^2 \right) \quad .
\end{aligned}
\end{equation}
%eoe%
The charged Higgs Yukawas are given by (see \tref{yukawaup})
%boe%
\begin{equation}
\lambda^{H^+}_{d_j \bar{u}_k} = -\xi \left( V^T Y^u \right)_{jk}, \qquad
\lambda^{H^-}_{u_j \bar{d}_k} = \left( V^* K^d \right)_{jk} \quad ,
\end{equation}
%eoe% 
and the loop functions $D_0$ and $D_2$ are given in \aref{loop_functions}.
The total contribution to the four-fermion operator coefficients is obtained by summing \eqrefs{SFVboxcoefficients1}{SFVboxcoefficients2}.
The operator coefficients for  $B_d-\bar{B_d}$, $K-\bar{K}$ are obtained by replacing the corresponding down-type quarks in the expressions above,
while the coefficients for $D-\bar{D}$ mixing can be obtained by interchanging $\lambda^{H^-}_{u_i \bar{d}_j} \leftrightarrow \lambda^{H^+ *}_{d_i \bar{u}_j}$, substituting the appropriate quark masses, and replacing all CKM matrix insertions with their conjugate transpose.
We have checked that our results are consistent with the results in \cite{Buras:2001mb,Altmannshofer:2007cs}.
\footnote{The charged Higgs-Goldstone and charged-Higgs $W$ boxes quoted in \cite{Crivellin:2013wna}, however, disagree with both our results and the results in \cite{Buras:2001mb,Altmannshofer:2007cs}.}

To set constraints on the SFV 2HDM from down-type meson mixing we proceed as follows. 
For $K-\bar{K}$ mixing, 
we use the limits on the real and imaginary parts of the coefficients $C_{i}(m_H)$ given in~\cite{Bona:2007vi}.
The strongest constraints in our model are set by the limits on the imaginary part of the coefficients $C_1$ and $C_2$, 
summarized in \tref{flavor_measurements}.
\footnote{We have also checked that using instead limits on $C_{\epsilon_K}$ and $C_{\Delta M_K}$ from~\cite{Bona:2016bvr}, and requiring them to lie in their 95\% C.L. range leads to similar constraints to the ones we present here.}
The constraints in the $m_H$ vs. $\kappa_j$ ($j=d,s,b$) plane are shown in Figs. \ref{fig:flavor_kappad} -- \ref{fig:flavor_kappab} in dashed-red contours.

For $B-\bar{B}$ mixing we use the the latest limits from \cite{Bona:2016bvr}.
In  \cite{Bona:2016bvr}, constraints are reported in terms of the coefficient $C_{B_q}$ and phase $\varphi_{B_q}$, $q=s,d$, defined by 
%boe%
\begin{equation}
C_{B_q}e^{2i\varphi_{B_q}} 
\equiv 
1 + 
\frac{\bra{B_q } \mathcal{H}_{\textrm{eff}}^{\Delta F = 2, \textrm{NP}} \ket{\bar{B}_q}}
{\bra{B_q } \mathcal{H}_{\textrm{eff}}^{\Delta F = 2, \textrm{SM}} \ket{\bar{B}_q}} \quad .
\label{eq:Bq_mixing_params}
\end{equation}
%eoe%
The Standard Model matrix elements in \eqref{Bq_mixing_params} are given by \cite{Crivellin:2013wna,Lenz:2010gu}
%boe%
\begin{align}
\label{eq:SMvalues}
  \bra{B_d^0} \mathcal{H}^{\Delta F = 2, \textrm{SM}}\ket{\bar{B}_d^0} & = \left( 1.1 + 1.3i \right)\times 10^{-13} \gev \quad ,\\
  \bra{B_s^0} \mathcal{H}_{SM}^{\Delta F = 2, \textrm{SM}}\ket{\bar{B}_s^0} & = \left( 59 - 2.2i \right)\times 10^{-13} \gev \quad .
\end{align}
%eoe%
The new physics matrix elements at the hadronic scale in \eqref{Bq_mixing_params} may be computed using the four-fermion operator coefficients \eqref{SFVboxcoefficients1} and \eqref{SFVboxcoefficients2} at the heavy Higgs mass scale, 
together with the expression for the Hamiltonian matrix element at the hadronic scale~\cite{Bona:2007vi}
%boe%
\begin{equation}
\label{eq:bs_magic_numbers}
\bra{B_{q}} \mathcal{H}_{\textrm{eff}}^{\Delta F = 2} \ket{\bar{B}_{q}} 
= 
\sum_{i,l,r= 1}^5 \left( b_l^{(r,i)} + \eta c_{l}^{(r,i)}\right) \eta^{a_l} C_i \bra{B_{q}} \mathcal{O}_r^{bq} \ket{\bar{B_q}}
\quad ,
\end{equation}
%eoe%
where $\eta = \alpha_s(m_H) / \alpha_s (m_t)$ and the magic numbers $a_l$, $b_l^{(r,j)}$, $c_l^{(r,j)}$ account for the RGE evolution. 
The operator matrix elements $\bra{B_{q}} \mathcal{O}_r^{bq} \ket{\bar{B_q}}$ at the hadronic scale are given in~\cite{Becirevic:2001jj}.
The expressions \eqref{bs_magic_numbers} also hold for the primed operators, with the same magic numbers and matrix elements.
Using Eqns. \eqrefp{SFVboxcoefficients1}, \eqrefp{SFVboxcoefficients2} and \eqrefp{Bq_mixing_params} -- \eqrefp{bs_magic_numbers}, we may now compute the parameters $C_{B_q}$, $\varphi_{B_q}$ in the SFV 2HDM.
In all parameter space of interest, we find that $\varphi_{B_q}$ is negligible. 
Therefore, we set limits on the SFV 2HDM parameters by requiring $C_{B_q}$ to lie within the $95\%$ C.L. constraint computed in refs.~\cite{Bona:2016bvr} (shown in \tref{flavor_measurements}).
The resulting $95 \%$ C.L.  bounds in the $m_H$ vs. $\kappa_j$ ($j=d,s,b$) plane are shown in Figs. \ref{fig:flavor_kappad} -- \ref{fig:flavor_kappab} in solid and dotted red contours for $B_d$ and $B_s$ mixing respectively.

Finally, 
to set bounds from $D-\bar{D}$ mixing we make use of the $95\%$ C.L. limits on the dispersive part of the mixing hamiltonian $M_{12}$, 
given in \tref{flavor_measurements}.
Within our model, the $D-\bar{D}$ mixing hamiltonian matrix element $\bra{D} \mathcal{H}_{\textrm{eff}}^{\Delta F = 2} \ket{\bar{D}}$ is obtained using Eqns. \eqref{SFVboxcoefficients1}, \eqrefp{SFVboxcoefficients2}, \eqrefp{bs_magic_numbers} and the magic numbers in~\cite{Bona:2007vi}.
In our normalization, 
the dispersive part of the mixing hamiltonian $M_{12}$ is equal to the short-distance hamiltonian matrix element \cite{Kagan:2009gb}
%boe%
\begin{equation}
M_{12}
=
\bra{D} \mathcal{H}_{\textrm{eff}}^{\Delta F = 2} \ket{\bar{D}} \quad .
\label{eq:dispersivepart}
\end{equation}
%eoe%
Since the long-distance SM contributions to D-meson mixing are currently unknown, \footnote{For a recent review of the status of the SM predictions for $D-\bar{D}$ mixing, see for instance \cite{Bazavov:2017weg}}
we set limits by demanding that no fine cancellations must occur between the calculable 
charged Higgs contributions and the unknown SM contribution to explain the measured value of $M_{12}$.
Explicitly, we require
%boe%
\begin{equation}
\abs{ \bra{D} \mathcal{H}_{\textrm{eff}}^{\Delta F = 2} \ket{\bar{D}} }
  \leq 
  7.4 \times 10^{-3}~\text{ps}
\label{eq:DD_mixing_bound}
\end{equation}
%eoe%
where $\mathcal{H}_{\textrm{eff}}$ includes the contributions from all the operators induced by the 2HDM. 
With this caveat, 
constraints from $D-\bar{D}$ mixing are shown in purple contours in Figs. \ref{fig:flavor_kappad}--\ref{fig:flavor_kappab}.

%%%%%%%%%%%%%%%%%%%%%%%%%%%%%%%%%%%%%%%%%%%%%%%%%%%%%%%%
\subsection{Summary and discussion of flavor constraints}
\label{s.summary}
\begin{figure*}[htbp!]
\centering
\subfloat{{\includegraphics[width=0.47\linewidth]{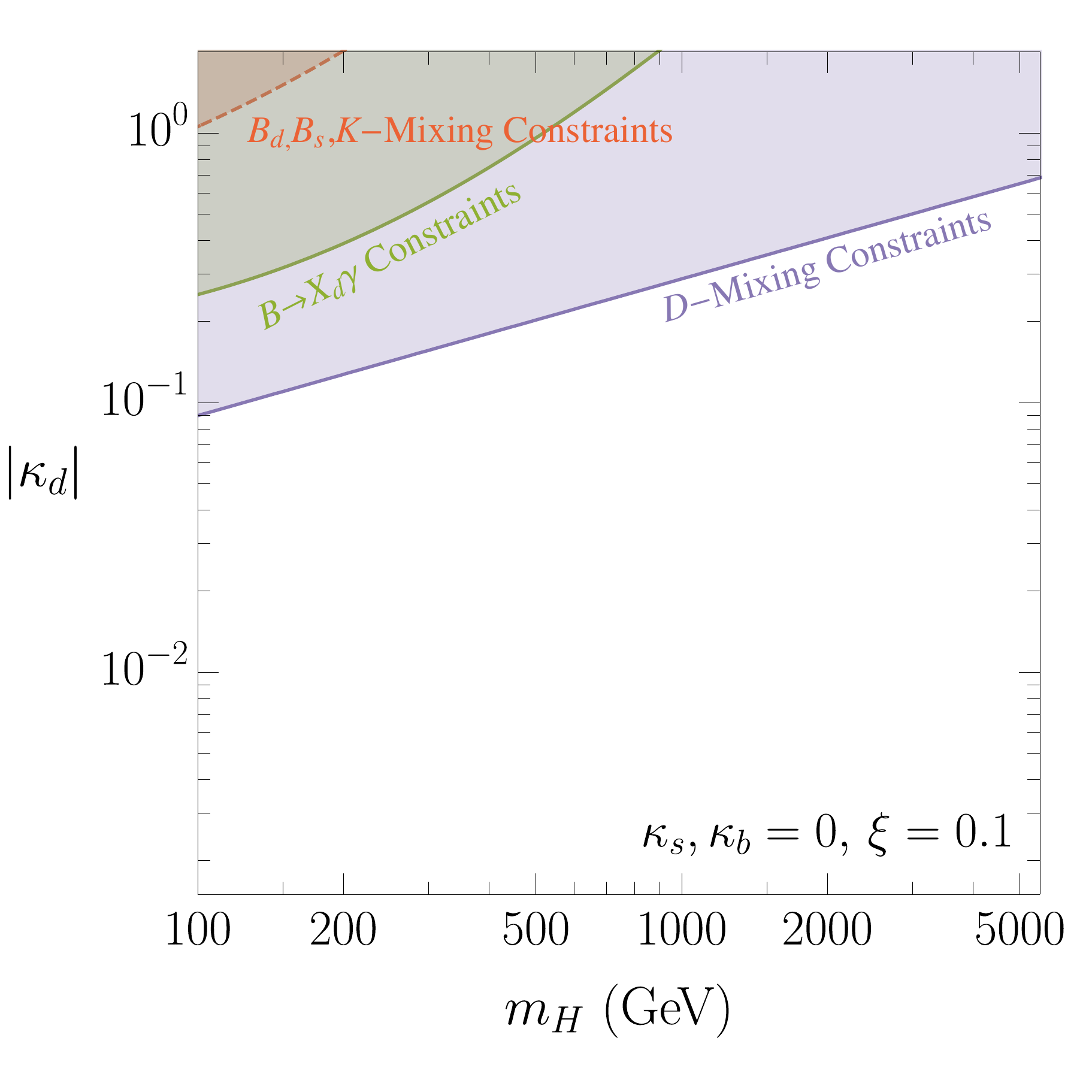} }}
\quad
\subfloat{{\includegraphics[width=0.47\linewidth]{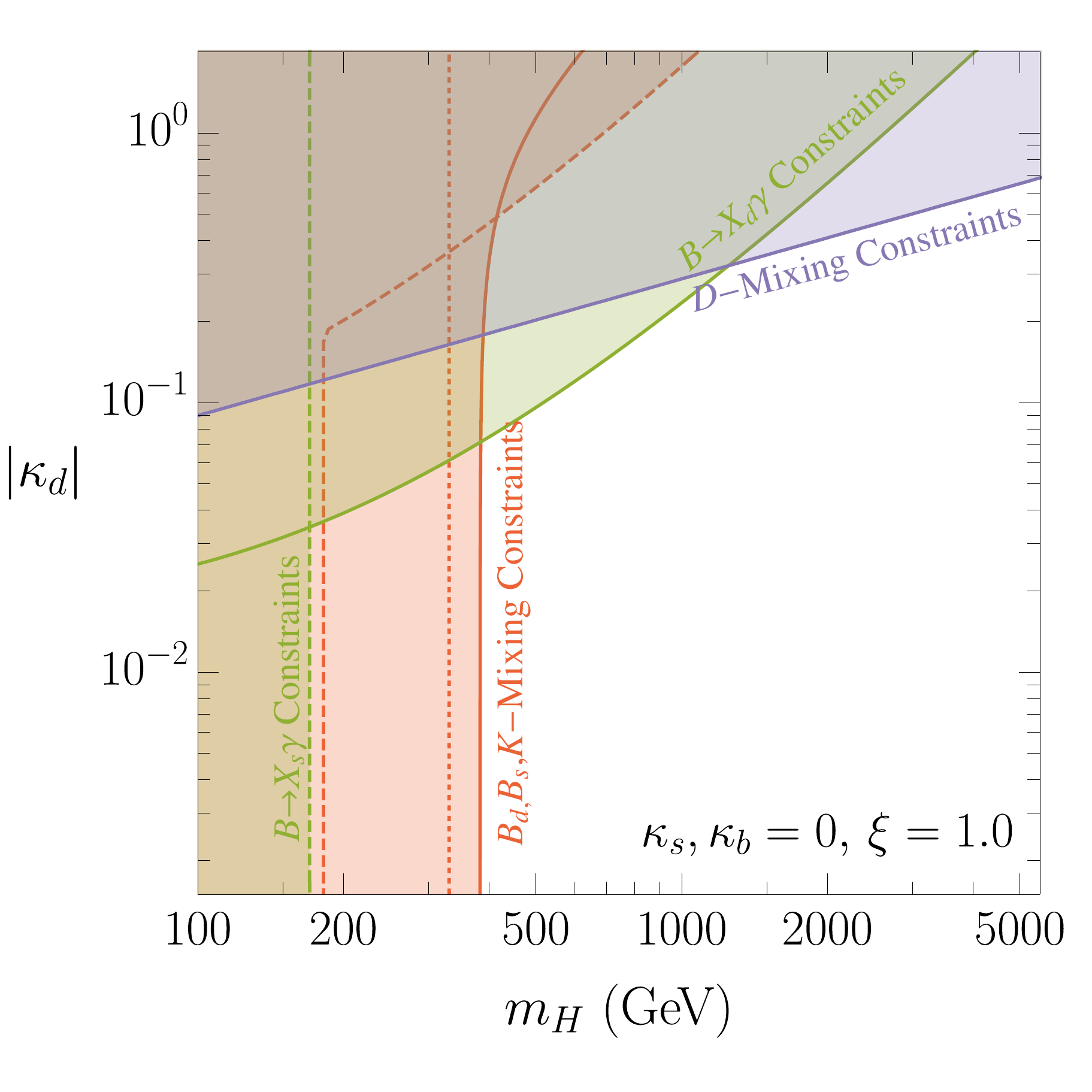} }}
\caption{Constraints on the up-type SFV 2HDM from one loop FCNC measurements in the plane of the second-Higgs doublet mass scale $m_H$ vs. its Yukawa coupling to down quarks $\kappa_d$, 
assuming $\kappa_s = \kappa_b = 0$. 
The couplings of the second Higgs doublet to up-type quarks in SFV are universally proportional to the Standard Model ones, 
with proportionality constant $\xi = 0.1$ (left panel) and $\xi = 1.0$ (right panel). 
Couplings of the second doublet to leptons have been set to zero.
All the Higgses in the second doublet, $H,A,H^\pm$ are taken to be mass degenerate. 
Constraints from $b \to s\gamma$ and $b \to d\gamma$ transitions are shown in green, 
with the constraints on $C_{7'}$ ($C_{7}$) indicated by solid and dashed lines, respectively, and the particular transition as indicated on the figure.
%with the constraint on $C^{bd}_{7'}$ ($C^{bs}_{7}$) indicated by the solid (dashed) line, respectively. 
Constraints from $B_d$, $B_s$ and $K$ mixing are shown as solid, dotted and dashed red lines respectively. 
The constraint from requiring the absence of fine-tuning in $D-\bar{D}$ mixing is shown in purple.} 
\label{fig:flavor_kappad}
\end{figure*}

\begin{figure*}[h]
\centering
\subfloat{{\includegraphics[width=0.47\linewidth]{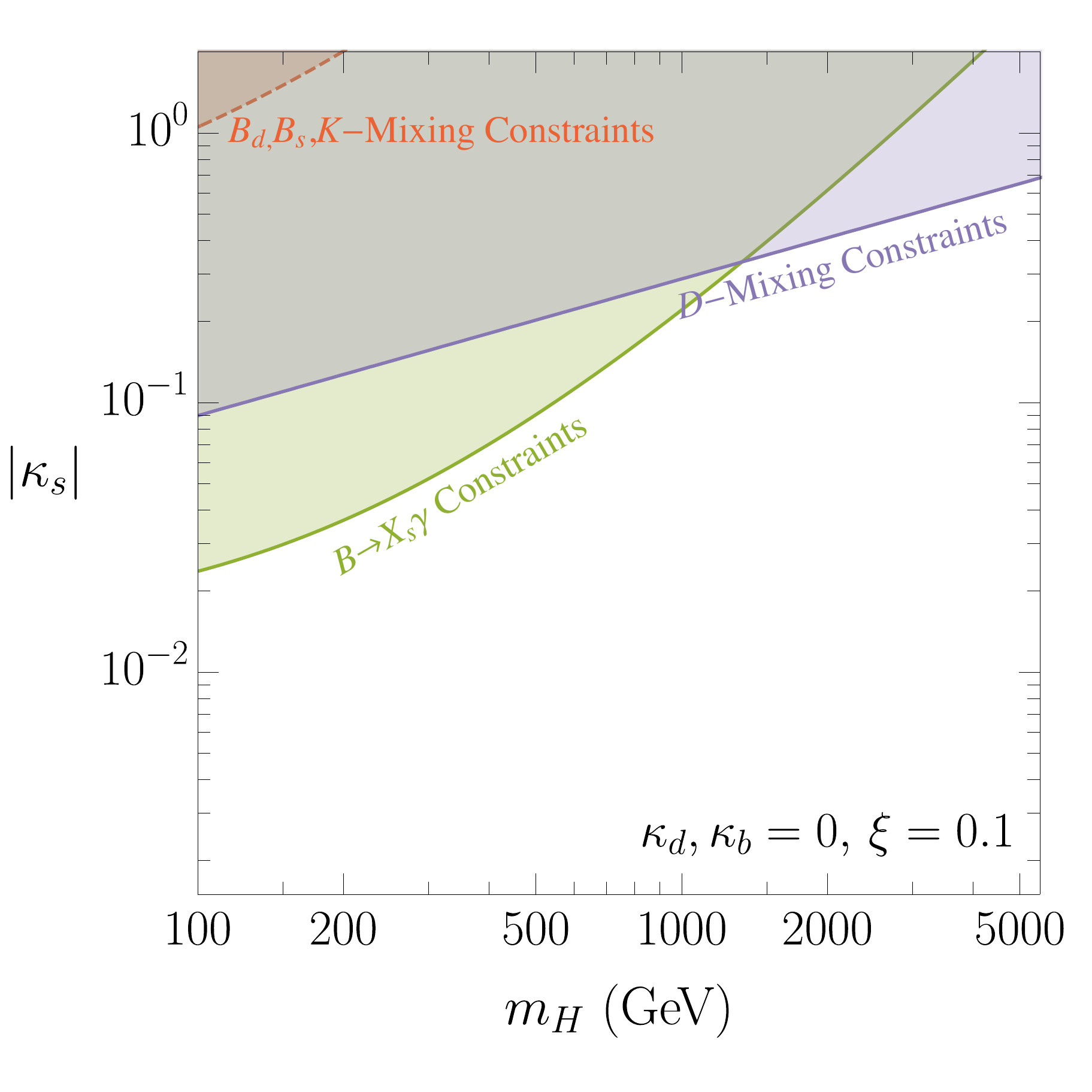} }}
\quad
\subfloat{{\includegraphics[width=0.47\linewidth]{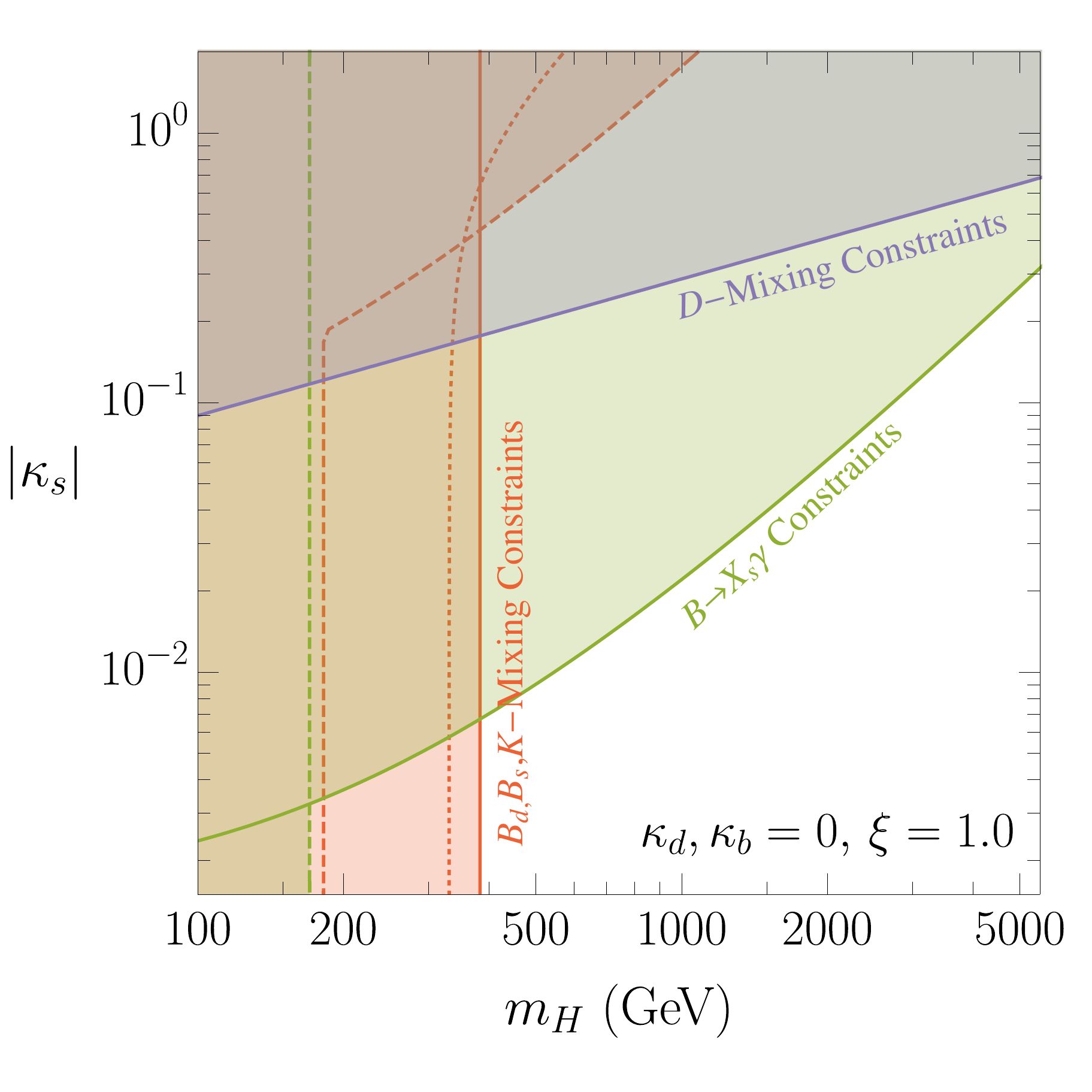} }}
\caption{The same as \figref{flavor_kappad}, but for $\kappa_s$, with $\kappa_d = \kappa_b = 0$. Both the solid and dashed green contours arise from $B\to X_s\,\gamma$ transitions.}
\label{fig:flavor_kappas}
\end{figure*}

\begin{figure*}[h]
\centering
\subfloat{{\includegraphics[width=0.47\linewidth]{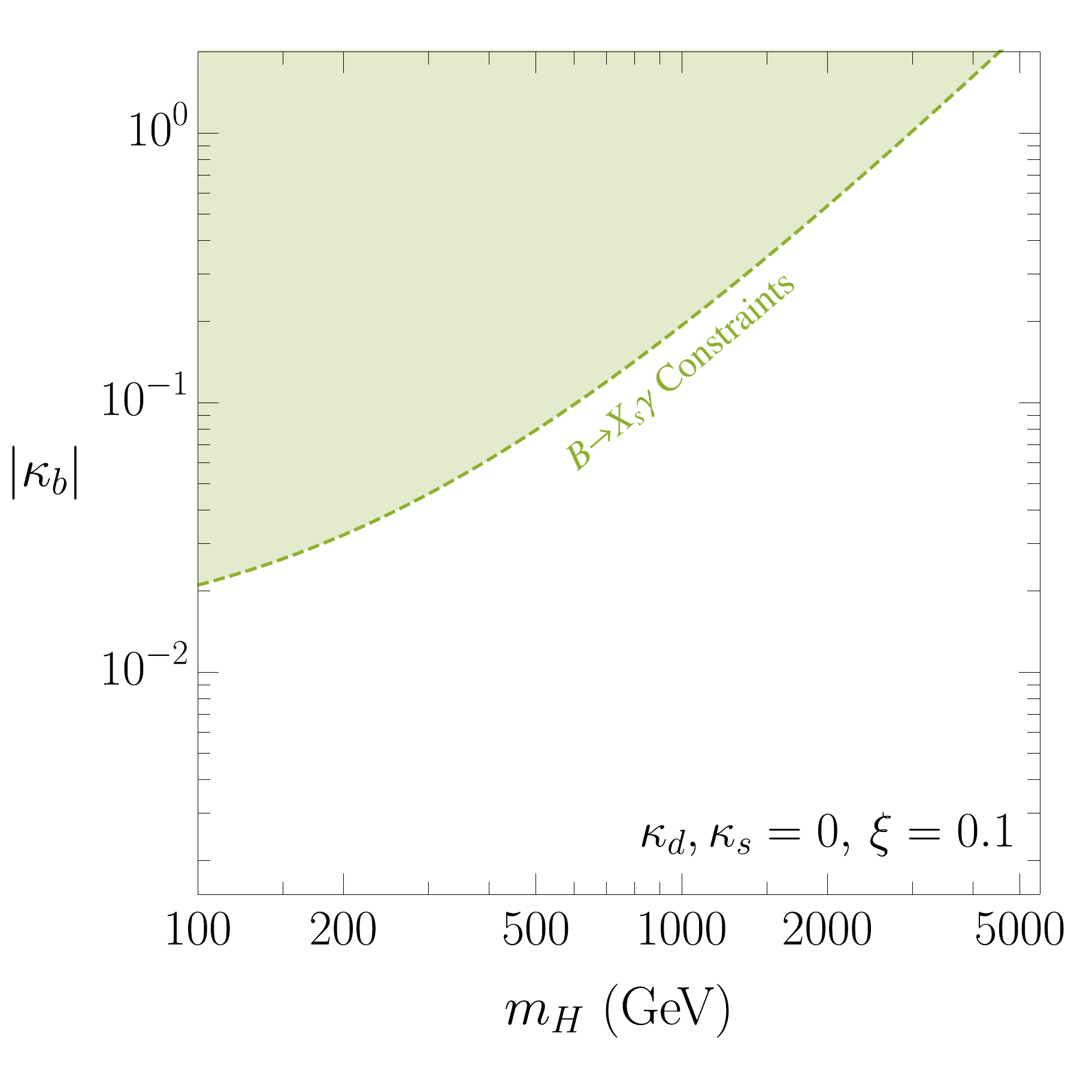} }}
\quad
\subfloat{{\includegraphics[width=0.47\linewidth]{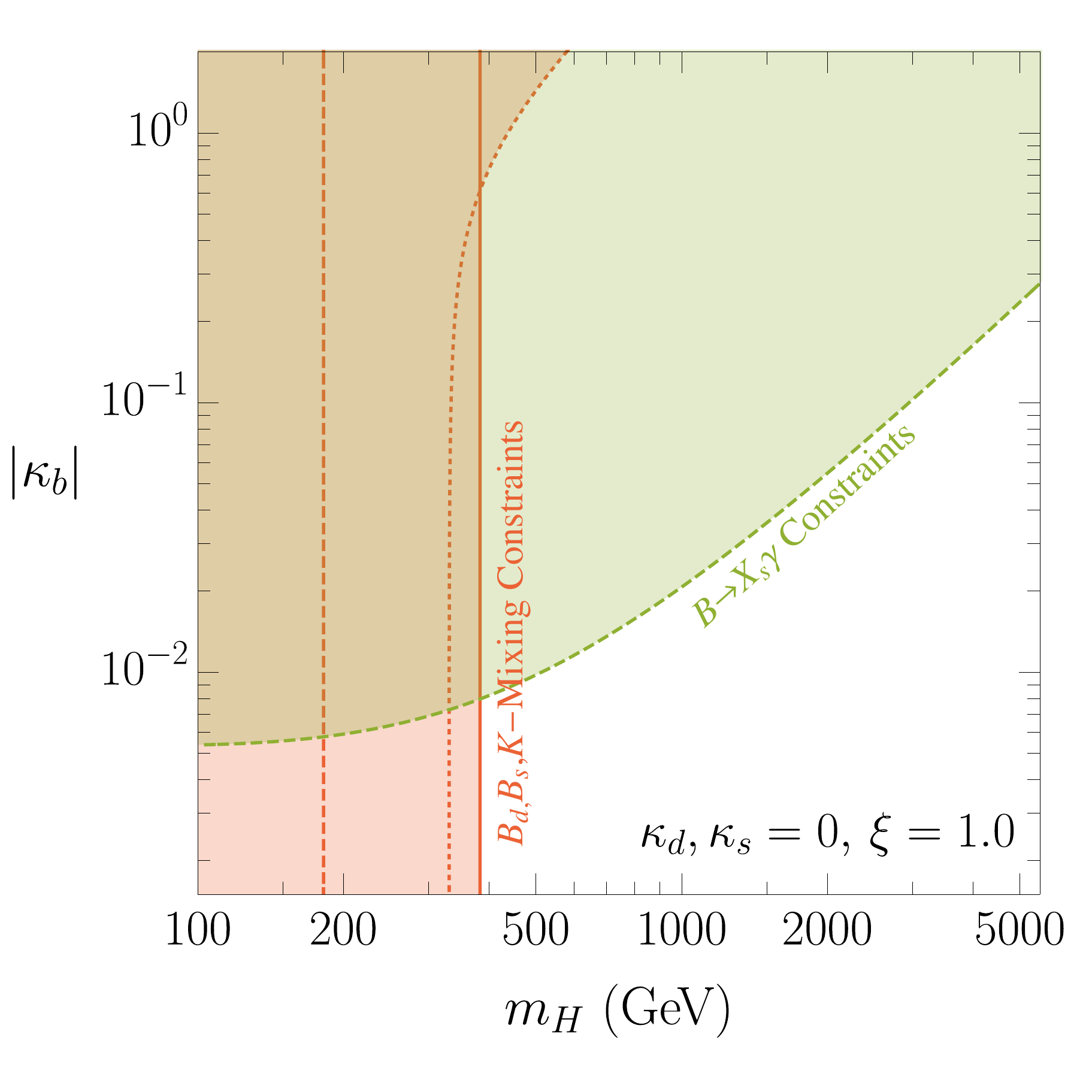} }}
\caption{The same as \figref{flavor_kappad} but for $\kappa_b$, with $\kappa_d = \kappa_s = 0$.}
\label{fig:flavor_kappab}
\end{figure*}

%%%%%%%%%%%%%%%%%%%%%%%%%%%%%%%%%%%%%%%%%%%%%%%%%%%%%%%%
We summarize our flavor bounds on the up-type SFV 2HDM in Figs.~\ref{fig:flavor_kappad}--\ref{fig:flavor_kappab}. 
We present bounds by turning on one down-type SFV Yukawa coupling $\kappa_d$, $\kappa_s$ or $\kappa_b$ at a time, 
in the corresponding $\kappa_j-m_H$ plane, 
where $m_H$ is the mass of the second Higgs doublet. 
In the left panel of each figure, 
we present the bounds for $\xi=0.1$ and in the right panel for $\xi=1$, 
where $\xi$ is the universal proportionality constant between the up-type Yukawas of the two Higgs doublets; cf. \eqref{uptypeSFV} 
(so for instance $\xi=0.1$ means that the second doublet couples to top quarks with strength $0.1 \, y_t$).

The most striking feature in the up-type SFV 2HDM,
is that an electroweak-scale second Higgs doublet may specifically couple to first- or second-generation down-type quarks with large Yukawa couplings while retaining consistency with flavor bounds.
This effect is most evident in the case of couplings to first-generation quarks.
From Fig. \ref{fig:flavor_kappad} (left),
we see that a second Higgs doublet with a mass of order $\sim100 \, \textrm{GeV}$ and down-quark coupling $\kappa_d \sim 0.1 $ remains consistent with all flavor bounds.
Note that this corresponds to a Yukawa coupling that is four orders of magnitude larger than the Standard Model down-quark Yukawa.
Interestingly, 
the most constrained couplings in the up-type SFV are not to first- or second-generation quarks as in the case of flavor-anarchic theories (see e.g. \cite{Crivellin:2013wna})
but to third generation quarks,
as can be seen by comparing the bounds in Fig. \ref{fig:flavor_kappab} and Figs. \ref{fig:flavor_kappad}, \ref{fig:flavor_kappas}.
This provides strong motivation to study and try to set limits on new physics with preferential couplings to light quarks at high-energy colliders.

Up-type SFV allows for such large generation-specific couplings by strongly suppressing down-type FCNCs via CKM matrix and small up and charm SM Yukawa insertions, and also via the GIM mechanism. 
In order to see explicitly how SFV works to suppress flavor bounds, 
in \tref{2hdm_operators} we present the coupling dependence (scaling) of the leading penguin and box diagrams contributing to FCNC operators in the $B_s$ meson system.
For simplicity, in the table we omit numerical prefactors and loop functions that depend only on the top quark and second Higgs doublet masses and on the top quark mass.
First, from the table we see that all diagrams are strongly suppressed by the CKM matrix combination $V^{\phantom{*}}_{tb}V^*_{ts}$.
Second, 
note that all the diagrams that we present in the table include insertions of top-quark Yukawas,
since they correspond to diagrams with internal top-quarks in the loops. 
Contributions from diagrams with internal up or charm quarks are suppressed via the GIM mechanism and small light-quark SM Yukawas, 
and are not shown in the table.
Finally, 
we see that some of the diagrams are further suppressed by down-quark SM Yukawa insertions. 
The combination of all these factors, 
anticipated at the beginning of this section using flavor symmetries,
leads to the strong suppression of flavor bounds in up-type SFV.

From \tref{2hdm_operators} we can also understand the generic features of bounds from down-type FCNCs in Figs. \ref{fig:flavor_kappad}-\ref{fig:flavor_kappab}.
In the figures, 
we observe a series of bounds that are independent on the value of the down-type Yukawas $\kappa_j$ ($j=d,s,b$),
and depend only on the proportionality factor $\xi$ between the first- and second-doublet up-type Yukawas. 
These bounds come from the limits on the operators $\mathcal{O}_1$ for down-type meson mixing and $\mathcal{O}_7$ for radiative $B$-meson decay,
which get contributions from box and penguin diagrams that depend only on up-type Yukawas.
For $\xi=1$, the strongest $\kappa_j$-independent limit comes from the operator $\mathcal{O}_1^{bd}$ contributing to $B_d-\bar{B}_d$ mixing,
which sets a limit $m_H \gtrsim 420\gev$, 
as can be seen from Figs. \ref{fig:flavor_kappad}-\ref{fig:flavor_kappab} (right panels).
This limit becomes irrelevant for for $\xi = 0.1$,
(Figs. \ref{fig:flavor_kappad}-\ref{fig:flavor_kappab}, left panels),
in which case constraints from radiative B-meson decays and $D-\bar{D}$ mixing are dominant.
For $\xi=0$ (a second Higgs doublet that \textit{does not} couple to up-type quarks) and allowing only for $\kappa_s$ \textit{or} $\kappa_d$ to be nonzero,
the only relevant limits on $\kappa_d$ and $\kappa_s$ come from $D-\bar{D}$ mixing.
In this scenario,
$\kappa_b$ is essentially unconstrained. 

Finally, limits from $D-\bar{D}$ are throughout significant, 
since in up-type SFV some of the meson mixing operators are not suppressed by the Standard Model GIM mechanism or by any small SM Yukawas, 
as discussed at the very beginning of this section. 
In particular, the coefficient of the operator $\mathcal{O}_1^{cu}$ is only suppressed by CKM matrix insertions (see \eqref{SFVbilinearsup}). 

We conclude that in the up-type SFV 2HDM, 
large and preferential couplings to down or (to a lesser extent) strange quarks of a second Higgs doublet with a mass $\mathcal{O}(100) \, \textrm{GeV}$ are allowed by flavor constraints. 
While complementary measurements of the $B$ system at Belle-II will improve these bounds ~\cite{Kou:2018nap},
in such scenarios flavored BSM physics might be more efficiently probed via direct production at colliders.
We explore this possibility in detail in the next two sections.

 %%%%%%%%%%%%%%%%%%%%%%%%%%%%%%%%%%%%%%%%%%%%%%%%%%%%%%%%
 %%%% Figures
%% three two-panel figures showing flavor constraints in the \kappa_i vs. m_H plane for \xi = 1.0 and 0.1 

%%%%%%%%%%%%%%%%%%%%%%%%%%%%%%%%%%%%%%%%%%%%%%%%%%%%%%%%

\begin{table*}[htbp!]
\centering
\begin{tabular}{l|cccccl}
\multicolumn{1}{c|}{Operator} & \multicolumn{6}{c}{Scaling of leading diagrams in SFV}
\\ \hline
\multicolumn{1}{l|}{$B \to X_s \gamma$ Transitions}  & & \multicolumn{3}{c}{$H^\pm$ penguin} \\ 
$\mathcal{O}_7 = \left(Q_2 \sigma_{\mu\nu} \bar{q}_3\right)^{\dagger} F^{\mu\nu}$	
	&
	\multirow{2}{*}{$y_{t}^2 \, \big(V_{tb}{V_{ts}}^*\big)\, \times\,$} 
	&
	$\xi\, \kappa_b$ 
	&
	&
	$y_{b}\, \xi^2$ 
	&
	&
\\
$\mathcal{O}'_7 = \left(Q_3 \sigma_{\mu\nu} \bar{q}_2\right) F^{\mu\nu}$	& & 
	$\xi\, \kappa_s$
	&
	& 
	$\kappa_b\, \kappa_s$
	&
	&
	\\ 
\hline
\multicolumn{1}{l|}{$B_s - \bar{B}_s$ Mixing}      & &  \multicolumn{1}{c}{$H^\pm$ box} & &  \multicolumn{1}{c}{$H^\pm\,\text{-}\,W^\pm$ box} & & \multicolumn{1}{c}{$H^\pm\,\text{-}\,G^\pm$ box} \\
$\mathcal{O}_1 = \left(Q_3 \bar{\sigma}^{\mu} Q_2^{\dagger}\right)^2$ 	& 	
\multirow{6}{*}{$y_{t}^4\big(V_{tb}{V_{ts}}^*\big)^2\, \times\, $}	
  	& 
	$\xi^4$ 
	&
	
	&
	$g_2^2\, \xi^2$ 
	&
	
	&
	$g_2^2\, \xi^2$
	\\
$\mathcal{O}'_1 = \left(\bar{d}_2^{\dagger} \bar{\sigma}^{\mu} \bar{d}_3\right)^2$	& &
	$\kappa_s^2\, \kappa_b^2$ 
	&
	
	&
	 
	&
	&
	$g_2^2\, y_{s}\, y_{b}\, \kappa_s\, \kappa_b$
	\\
$\mathcal{O}_2 = \left(\bar{d}_2 Q_3\right)^2$	& &
	$\xi^2\, \kappa_s^2$
	&
	
	&
	
	&
	&
	$g_2^2\, y_{s}\, \xi\, \kappa_s$
	\\	
$\mathcal{O}'_2 = \left(\bar{d}_3 Q_2\right)^2$	
	& & 
	$\xi^2\, \kappa_b^2$ 
	&
	
	&
	
	&
	&
	$g_2^2\, y_{b}\, \xi\, \kappa_b$
	\\
$\mathcal{O}_4 = \left(\bar{d}_2 Q_3\right) \left(\bar{d}_3^{\dagger} Q_2^{\dagger}\right)$	& & 
	$\xi^2\, \kappa_s\, \kappa_b$
	&
	
	&
	$g_2^2\, \kappa_s\, \kappa_b$
	&
	&
	$g_2^2\, \kappa_s\,\kappa_b$
	\\
$\mathcal{O}_5 = \left(\bar{d}_2 Q_2^{\dagger}\right) \left(\bar{d}_3^{\dagger} Q_3\right)$	& & 
	$\xi^2\, \kappa_s\, \kappa_b$ 
	&
	
	&
	
	&
	
	&
	$g_2^2\, y_{s}\, \xi\, \kappa_b$,~$g_2^2\, y_{b}\, \xi\, \kappa_s$ 
\end{tabular}
\caption{
Operators induced by the SFV 2HDM leading to $b \to s\gamma$ transitions (top) and $B_s - \bar{B}_s$ mixing (bottom).
For each operator, we show the scaling of the leading one loop contribution mediated by the top and dictated by the SFV ansatz. 
The operators arising in $B_d - \bar{B}_d$ mixing, $K-\bar{K}$ mixing and $b \to d\gamma$ transitions can be obtained by a simple replacement of the indices in those shown above, though the diagrams with top quarks may no longer dominate due to the smallness of $V_{td}$. 
In $\mathcal{O}_5$ the parentheses indicate color index contraction, while for the other operators the spinor and color contraction is the same.
}
\label{t:2hdm_operators}
\end{table*}

%%%%%%%%%%%%%%%%%%%%%%%%%%%%%%%%%%%%%%%%%%%%%%%%%%%%%%%%
%%%%%%%%%%%%%%%%%%%%%%%%%%%%%%%%%%%%%%%%%%%%%%%%%%%%%%%%
%% SECTION BEGINS
%%%%%%%%%%%%%%%%%%%%%%%%%%%%%%%%%%%%%%%%%%%%%%%%%%%%%%%%
%%%%%%%%%%%%%%%%%%%%%%%%%%%%%%%%%%%%%%%%%%%%%%%%%%%%%%%%
\section{Collider phenomenology of the up-type SFV 2HDM}
\label{s.collider}

The collider phenomenology of the SFV extra Higgs bosons differs significantly from the one of the Standard Model Higgs,
and from the one of more popular 2HDMs as the MFV or types I-IV 2HDM,
since the SFV Higgses may couple preferentially to light quarks. 
In this case, the main differences between the SM Higgs and the extra SFV Higgs bosons are twofold. 
First, 
while the SM Higgs is produced mostly via gluon fusion at hadron colliders,
extra SFV Higgs bosons are mostly produced at tree level via quark fusion.
And second, 
while the SM Higgs decays are most easily detected in diphoton or multilepton channels,
extra SFV Higgs bosons decay almost entirely to quarks and are most efficiently probed by dijet searches.

We organize the discussion of the collider phenomenology as follows.
In section \ref{s.production} we study the main production and decay modes for the SFV Higgs bosons.
In section \ref{s.dijets} we study constraints from resonance searches in the dijet final state.
We dedicate section \ref{s.othersearches} to study constraints from diphoton and other final states. 
Finally, in section \ref{s.discussioncollider} we summarize and discuss collider constraints and their interplay with flavor constraints.
As in the previous section, 
for brevity we focus only in the up-type SFV 2HDM with Higgs potential parameters $\lambda_4 = \lambda_5 = \lambda_6 = 0$, 
and leave a study of down-type SFV for future work.
In this case the 2HDM is in the alignment limit, 
and the Higgs bosons $H,A,H^\pm$ are all degenerate.
We also continue to assume throughout that the couplings of extra Higgs states to leptons vanish, $\xi^\ell = 0$, 
to concentrate on the quark phenomenology. 

%%%%%%%%%%%%%%%%%%%%%%%%%%%%%%%%%%%%%%%%%%%%%%%%%%%%%%%%
\subsection{Production and decay modes of the extra Higgs states}
\label{s.production}
%%%%%%%%%%%%%%%%%%%%%%%%%%%%%%%%%%%%%%%%%%%%%%%%%%%%%%%%
\subsubsection{Production of neutral Higgses}

%%%%%%%%%%%%%%%%%%%%%%%%%
\begin{figure} [h!]
\begin{center}
\includegraphics[width=0.3\linewidth]{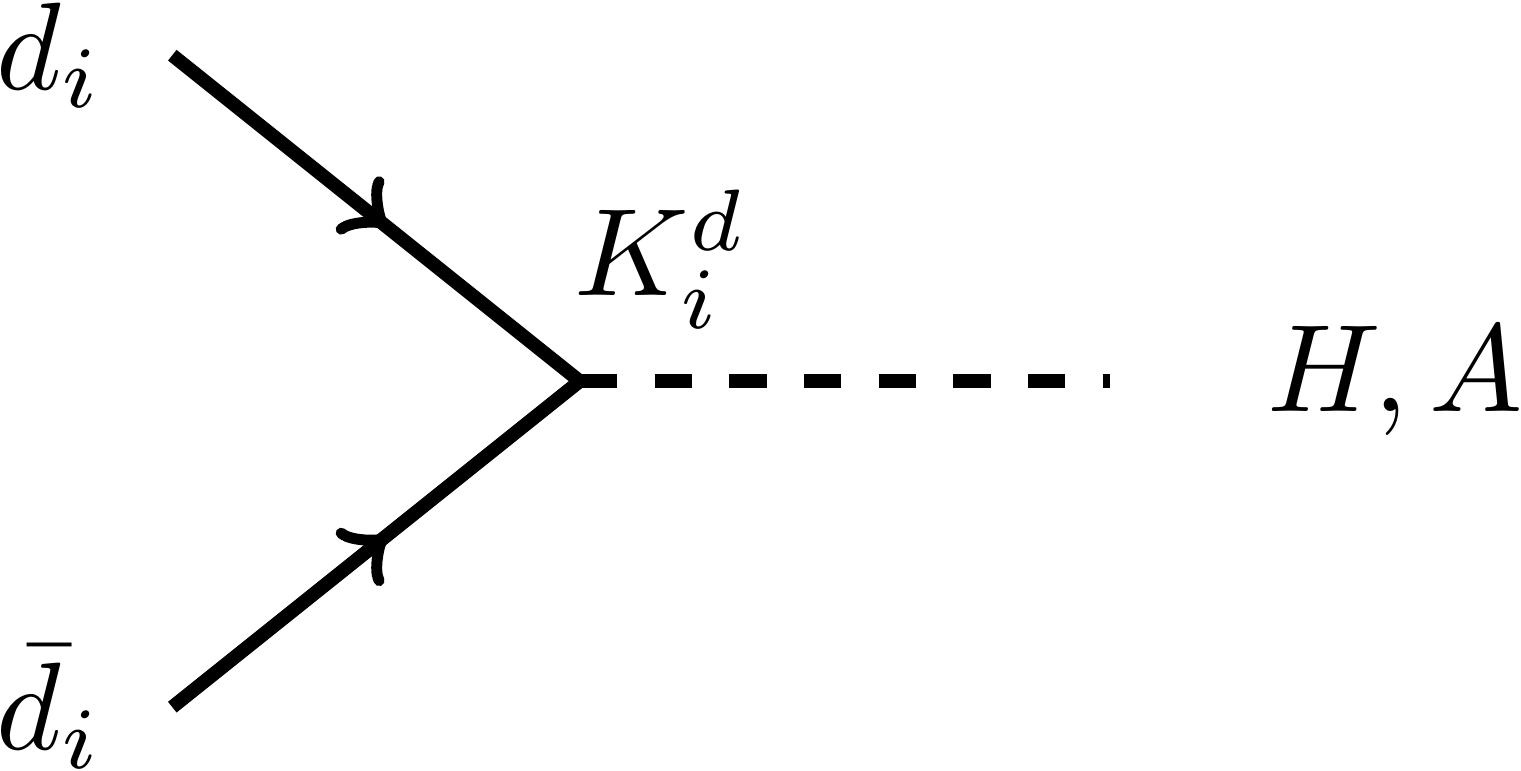}
~
\includegraphics[width=0.3\linewidth]{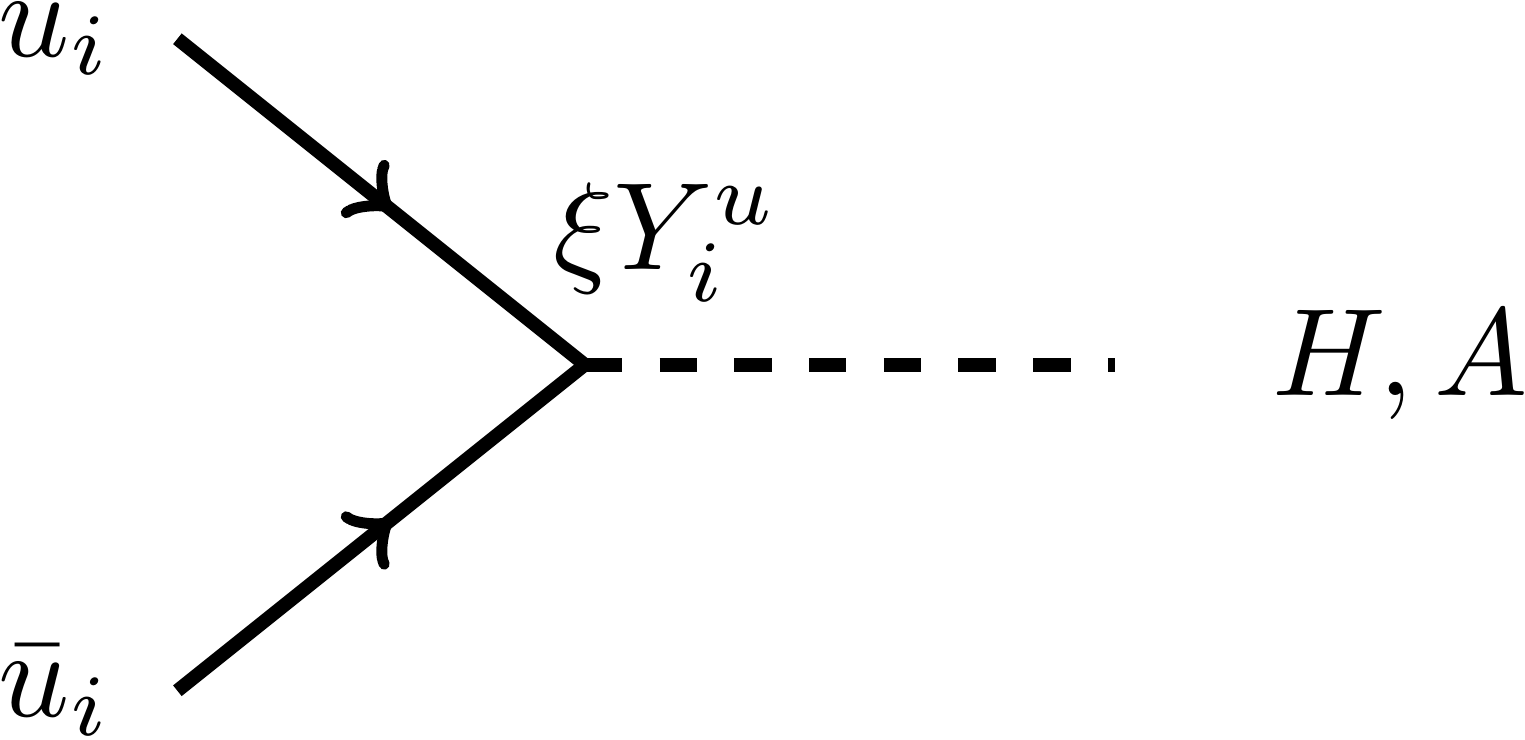}
\end{center}
\begin{center}
\includegraphics[width=0.3\linewidth]{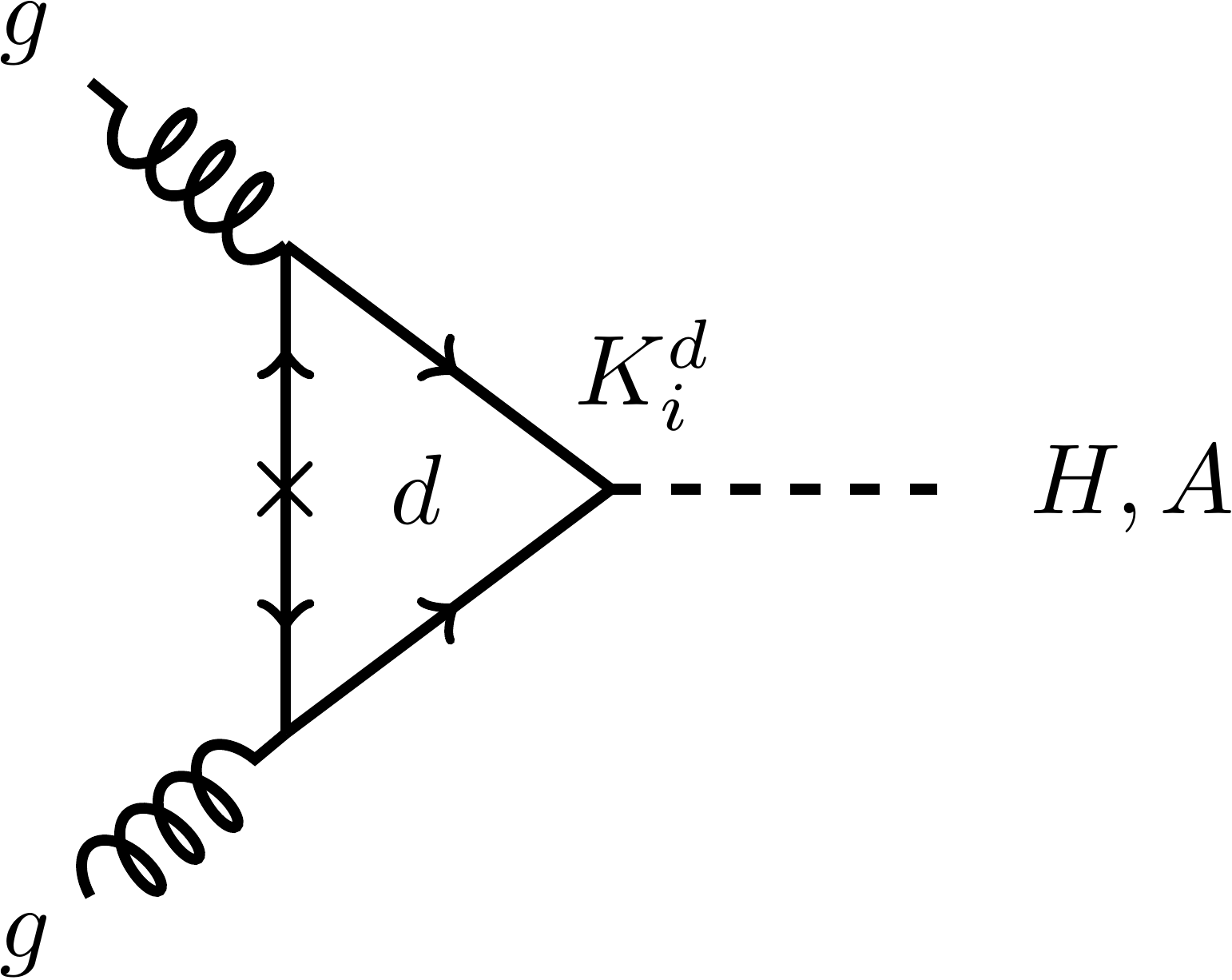}
~
\includegraphics[width=0.3\linewidth]{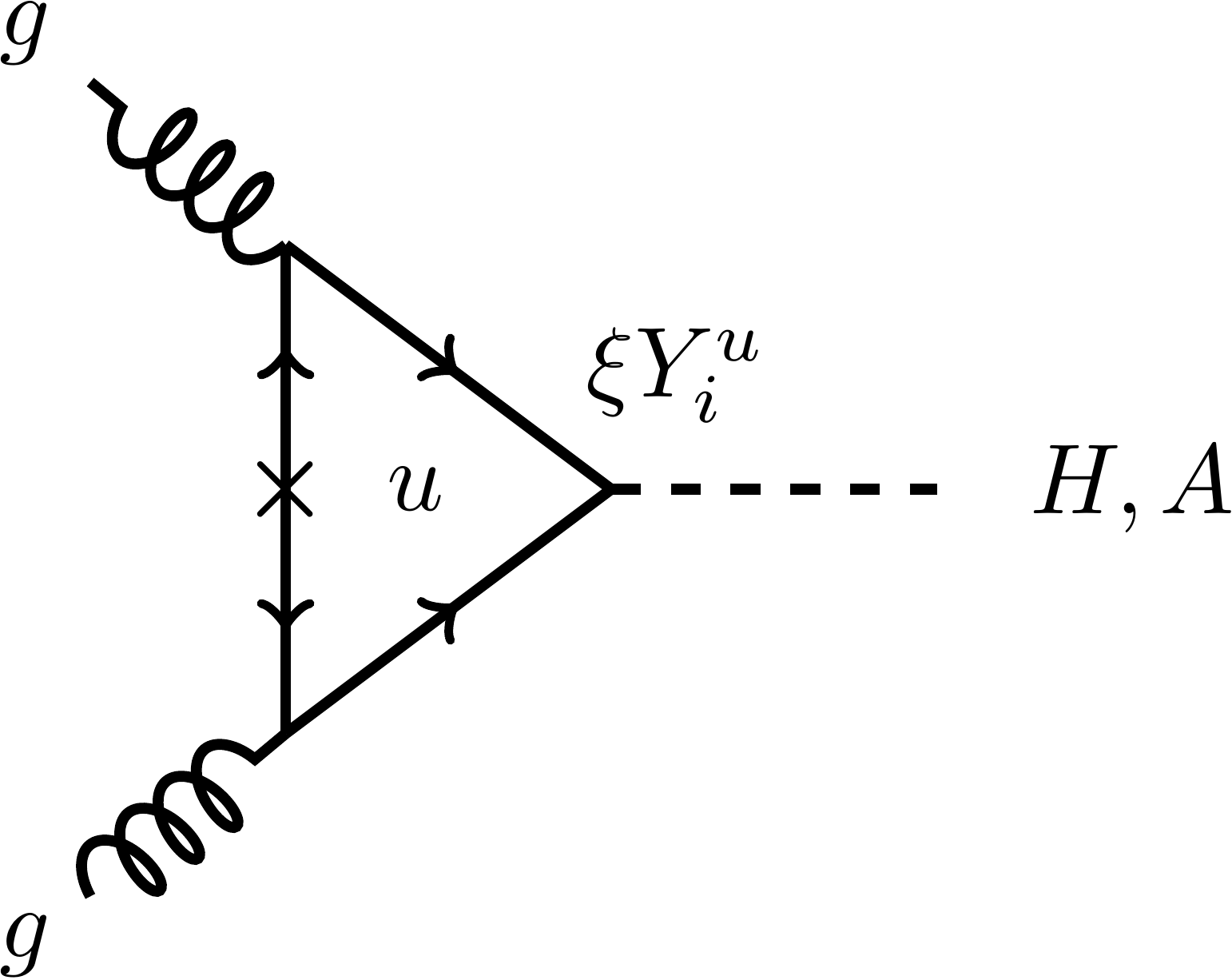}
  \end{center}
\caption{Diagrams leading to $s$-channel production of neutral Higgses in the alignment limit in the up-type SFV 2HDM. 
The couplings of the Higgs bosons to the fermions are given in \tref{yukawaup}.
Here, $Y^u = \textrm{diag}(y^{\textrm{SM}}_u, y^{\textrm{SM}}_c, y^{\textrm{SM}}_t)$ and $K^{d} = \textrm{diag}(\kappa_d, \kappa_s, \kappa_b)$.
}
\label{fig:neutral_higgs_diagrams}
\end{figure} 
%%%%%%%%%%%%%%%%%%%%%%%%%

In the SFV 2HDM, if the neutral Higgs states $H,A$ are below the LHC energy threshold, they can be produced both at tree level from quark fusion and at loop level from gluon fusion, as shown in \figref{neutral_higgs_diagrams}. 
In particular, 
neutral Higgses with large couplings to down or strange quarks may be copiously produced thanks to the large parton distribution functions (PDFs) of such quarks.
While large Yukawas for the light down-type quarks also enhance the gluon fusion diagram \figref{neutral_higgs_diagrams} (lower left),
such diagrams are still chirally suppressed by one small SM quark Yukawa insertion.
As a consequence, 
production via gluon fusion is mostly due to diagrams with top-quarks in the loop as for the SM Higgs, 
\figref{neutral_higgs_diagrams} (lower-right).

\begin{figure}%[ht]
\centering
\includegraphics[width=0.8\linewidth]{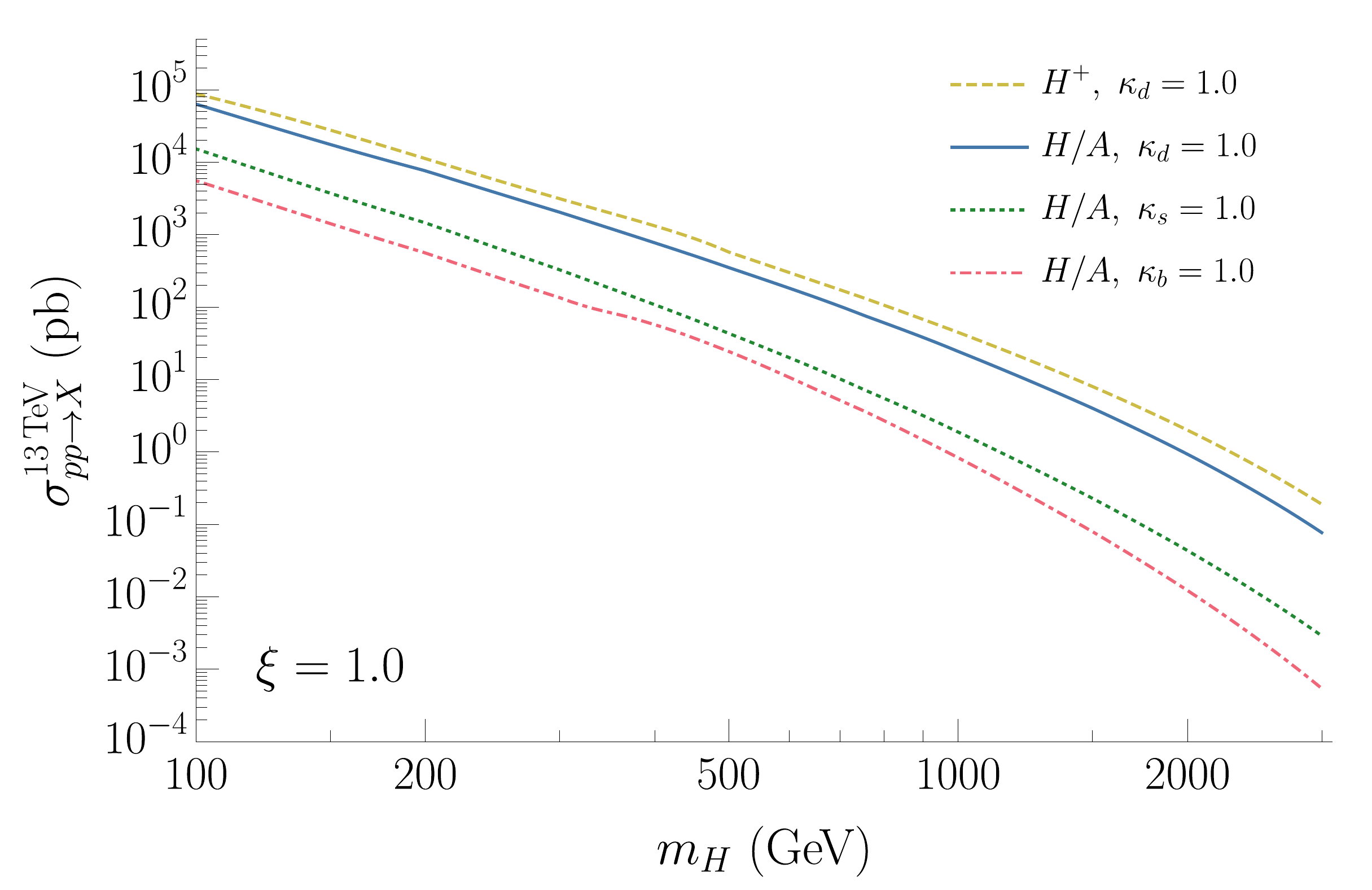}
\caption{
LHC production cross sections for the heavy neutral Higgses, $H/A$, 
for the three benchmark cases: 
coupling exclusively to the down, strange and bottom quarks.
We also show the production cross section for the charged Higgs, $H^+$, but coupling only to down quarks.
The corresponding cross section for $H^-$ is somewhat smaller due to the charge asymmetry in $pp$ collisions.
In each case the couplings to the rest of the quarks are set to zero.
Here we show only the leading-order cross section, though QCD corrections are expected to be large, as in Drell-Yan.
}
\label{fig:total_csx}
\end{figure}

In \figref{total_csx} we show the neutral Higgs production cross sections at the $13$ TeV LHC for three benchmark cases with neutral Higgses coupling to each individual down-type quark generation. 
The benchmark cases are defined by $\kappa_d=1, \kappa_{s,b}=0\,$; $\kappa_s=1, \kappa_{d,b}=0$ and $\kappa_b=1, \kappa_{d,s}=0$.
In all three scenarios for simplicity we set the Yukawa couplings to up-type quarks to zero, $\xi=0$, 
so that gluon fusion production is negligible and Higgses are produced exlusively via quark fusion.
Note that in this case, 
the production cross section scales simply as $\kappa_j^2$, ($j=d,s\,\mathrm{{ or }}\,b$).

The significance of the large down quark Yukawa coupling is clear from the figure: for the case with $\kappa_d = 1$, the cross section for a $150\gev$ neutral Higgs is $1.9 \times 10^4 \, \textrm{pb}$, 
a factor of $\sim 400$ larger than the SM gluon fusion Higgs production cross section $\sigma_{h,\textrm{SM}}^{\textrm{ggf}}=49 \, \textrm{pb}$,
while for $\kappa_d = 0.1, \xi=0$, it is $\sim 4$ times larger.
\footnote{Even if we allow for couplings to up-type quarks by setting $\xi$ to be nonzero, 
down quark fusion remains being one of the dominant production mechanisms at the $13$ TeV LHC, 
over gluon fusion.
For reference, 
with $\kappa_d=10^{-3}$, $\kappa_{s,b}=0$ and $\xi=1$, down-quark fusion is already the largest production mechanism.}
For SFV Higgses coupling mostly to the strange or bottom quarks, 
the contribution to Higgs production from quark fusion is smaller due to the smaller parton luminosities for $s$ and $b$ quarks,
but remains important. 
As a reference, for a $150\gev$ neutral Higgs $H$ or $A$, 
the (leading-order) down-type quark fusion production cross section exceeds the Standard Model Higgs production cross section for strange Yukawa $\kappa_s \geq 0.11$,
or for bottom Yukawa $\kappa_b\geq 0.18$.

%%%%%%%%%%%%%%%%%%%%%%%%%
%%%%%%%%%%%%%%%%%%%%%%%%%

%%%%%%%%%%%%%%%%%%%%%%%%%%%%%%%%%%%%%%%%
\subsubsection{Charged Higgs production}
%%%%%%%%%%%%%%%%%%%%%%%%%
\begin{figure}% [h!]
\begin{center}
\includegraphics[width=0.3\linewidth]{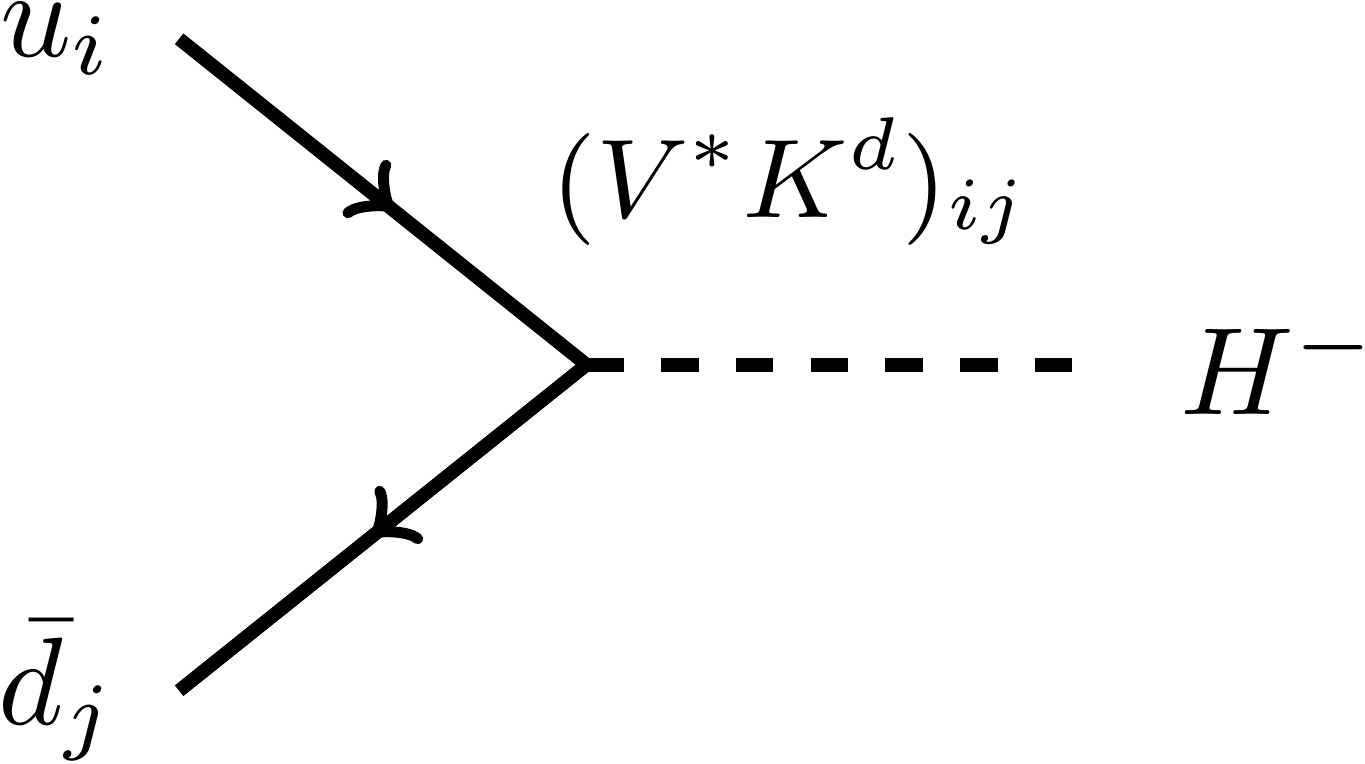}
~~
\includegraphics[width=0.3\linewidth]{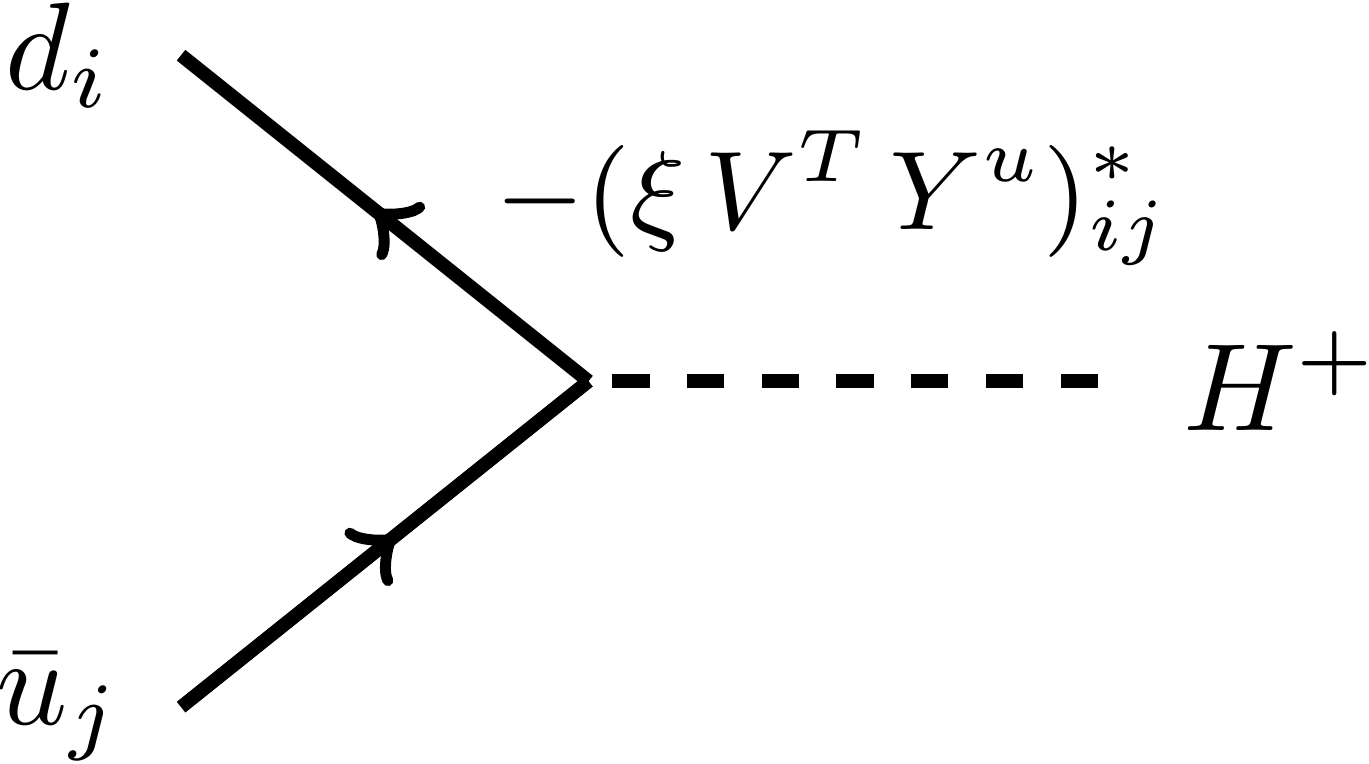}
\end{center}
\caption{
Quark fusion production diagrams for the charged Higgs in the alignment limit, in the up-type SFV 2HDM.
The couplings of the charged Higgs to the fermions are given in \tref{yukawaup}.
Here $V$ is the CKM matrix, 
and $Y^u = \textrm{diag}(y^{\textrm{SM}}_u, y^{\textrm{SM}}_c, y^{\textrm{SM}}_t)$ and $K^{d} = \textrm{diag}(\kappa_d, \kappa_s, \kappa_b)$.
}
\label{fig:charged_higgs_diagrams}
\end{figure} 
%%%%%%%%%%%%%%%%%%%%%%%%%

In the most popular versions of the 2HDM, e.g., the MFV or types I-IV 2HDMs, 
the most significant charged-Higgs production mode for $m_H < m_t - m_b$ near the alignment limit is $tb$ associated production~\cite{DiazCruz:1992gg, Alwall:2004xw,Craig:2015jba}. 
For larger charged-Higgs masses this mode is suppressed, and the dominant process is instead $g b \to t H^{\pm}$~\cite{Plehn:2002vy,Berger:2003sm}.
Quark-fusion production of the charged Higgs is also possible, but in MFV or in the types I-IV models (including the 2HDM in the minimal supersymmetric Standard Model), this relies on the $b$-quark Yukawa, and is suppressed by the $b$-quark PDF or $V_{ub}$~\cite{Gunion:1986pe, Moretti:1996ra, He:1998ie, Dittmaier:2007uw}. Associated $W^{\pm}H^{\mp}$ production vanishes in the alignment limit.

In contrast, in the up-type SFV 2HDM, 
the enhanced Yukawa couplings to first- and second-generation down-type quarks entirely change the dominant production modes for charged Higgs bosons, 
and allow for comparatively much larger cross section via quark fusion of first- or second-generation quarks,
shown in Fig.~\ref{fig:charged_higgs_diagrams} (left).
Quark fusion via the SFV up-type Yukawas in  Fig.~\ref{fig:charged_higgs_diagrams} (right) is suppressed by the up and charm SM Yukawas or CKM elements.
We show the charged-Higgs production cross section for $\kappa_d=1$, $\kappa_s=\kappa_b=0$ in \figref{total_csx}. 
Quite differently from the case of the types I-IV or MFV 2HDMs, 
the charged-Higgs production cross section is in this case the largest among all SFV Higgs bosons.  

%%%%%%%%%%%%%%%%%%%%%%%%%%%%%%%%%%%%%%%%%%%%%%%%%%%%%%%%
\subsubsection{Total width and branching ratios}

We now discuss the decays of the extra Higgs states.
With our choice of Higgs potential parameters $\lambda_4 = \lambda_5 = \lambda_6 = 0$ the extra Higgses are degenerate and decays among these states are forbidden~\cite{Craig:2013hca,Davidson:2005cw}.
Decays to gauge bosons are also forbidden in the alignment limit, $\lambda_6=0$,
while decays to leptons are not allowed in our simplified scenario where we have set the second-doublet lepton Yukawas to zero. 

\begin{figure*}[h]
\centering
\centering
\subfloat{{\includegraphics[width=0.48\linewidth]{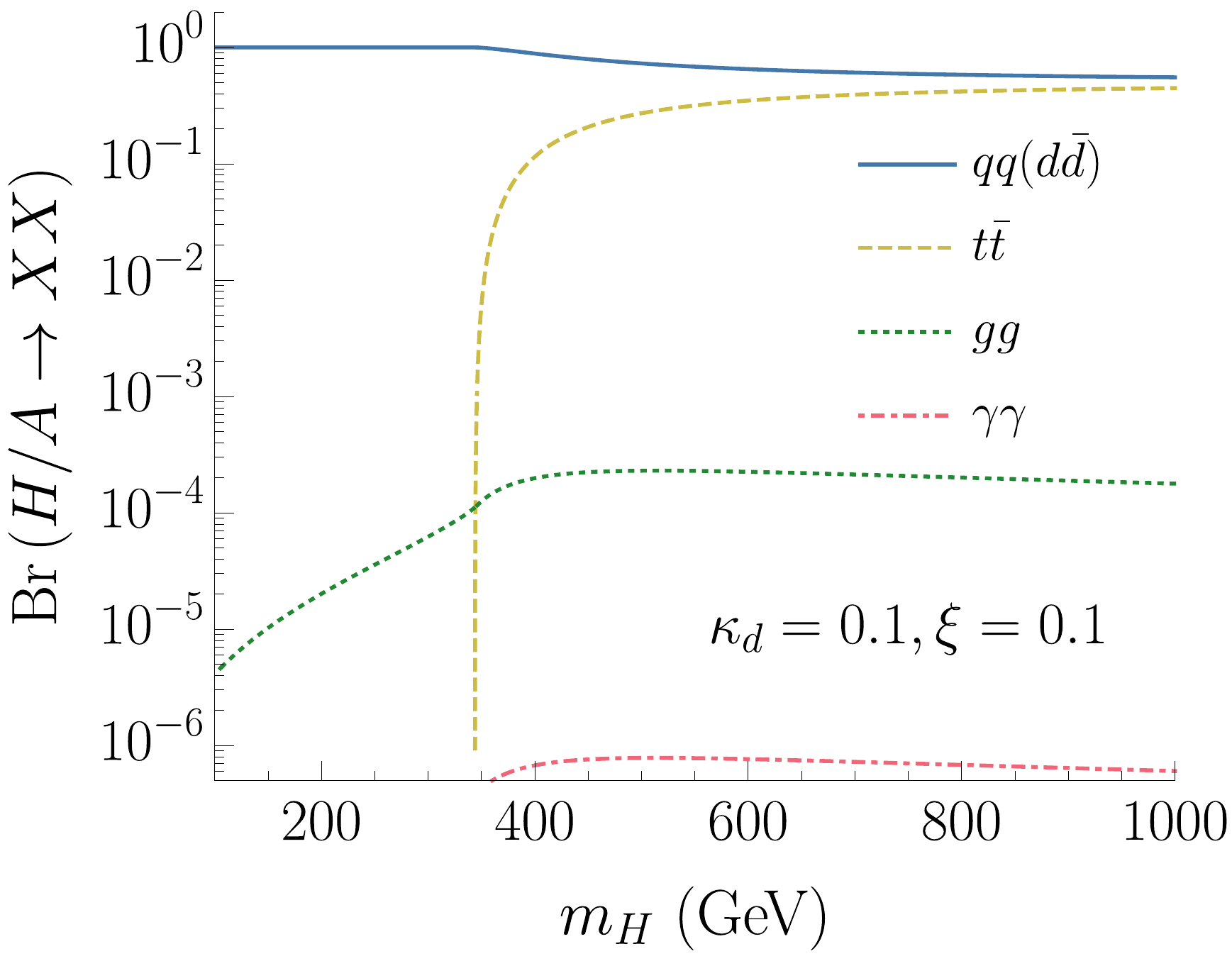} }}
~~
\subfloat{{\includegraphics[width=0.48\linewidth]{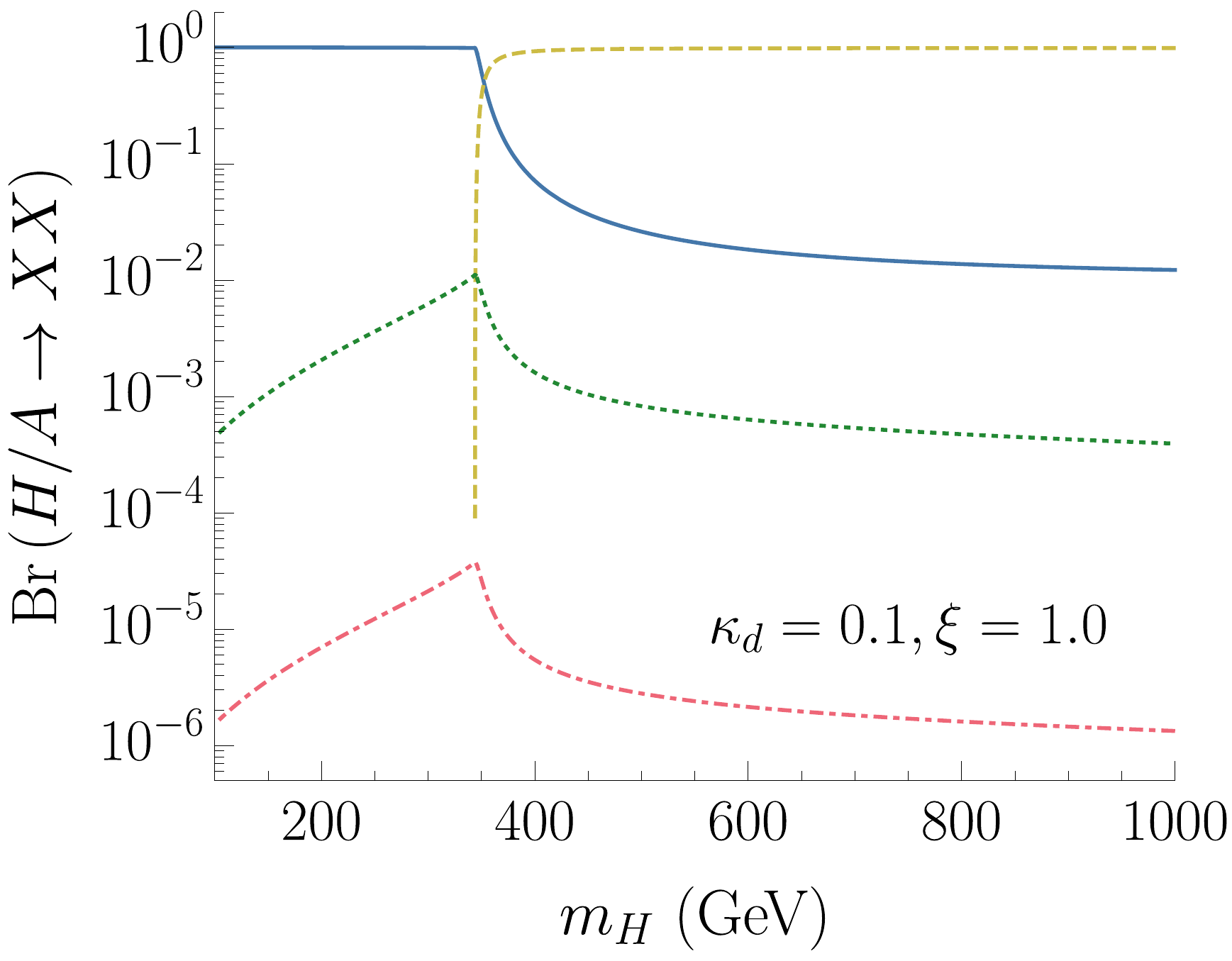} }}
\caption{
Plot of the branching fraction of the extra neutral Higgs bosons $H/A$ to $d\bar{d}$ (solid blue), $t\bar{t}$ (dashed yellow), $gg$ (dotted green) and $\gamma\gamma$ (dot-dashed red), as a function of $m_H$ with $\kappa_d = 0.1$, 
in the up-type SFV 2HDM.
In the left panel we show the branching fractions for $\xi = 0.1 $ and in the right panel for $\xi = 1$.
In both plots we have taken $\kappa_s = \kappa_b = 0$.
The behavior when replacing $\kappa_d$ with either $\kappa_s$ or $\kappa_b$ is similar, 
with the decays to $d\bar{d}$ replaced by $s\bar{s}$ or $b\bar{b}$ correspondingly. 
The couplings of the Higgs bosons to fermions needed to obtain the branching fractions are given in \tref{yukawaup}.
}
\label{fig:br_plots}
\end{figure*}

We start by discussing the decays of the neutral Higgs states.
The branching ratios for the neutral Higgses $H,A$ are shown in \figref{br_plots} for $\kappa_d=0.1, \kappa_s=\kappa_b=0$ and $\xi=0.1$ (left) or $\xi=1$ (right).
The neutral Higgses $H,A$ may decay at tree level to quarks
or at loop level to gluons and photons. 
From the figure, we see that for $\kappa_d=0.1$ and for both cases $\xi=0.1$ or $\xi=1$, 
the branching fractions to quarks are dominant.
In the case $\xi=1$ and for $m_H > 2m_t$, 
the neutral Higgs states decay mostly to a top pair, 
while for $\xi=0.1$ the dominant decays are to down quarks for all $m_H$.

\begin{figure}[h]
\centering
\includegraphics[width=0.8\linewidth]{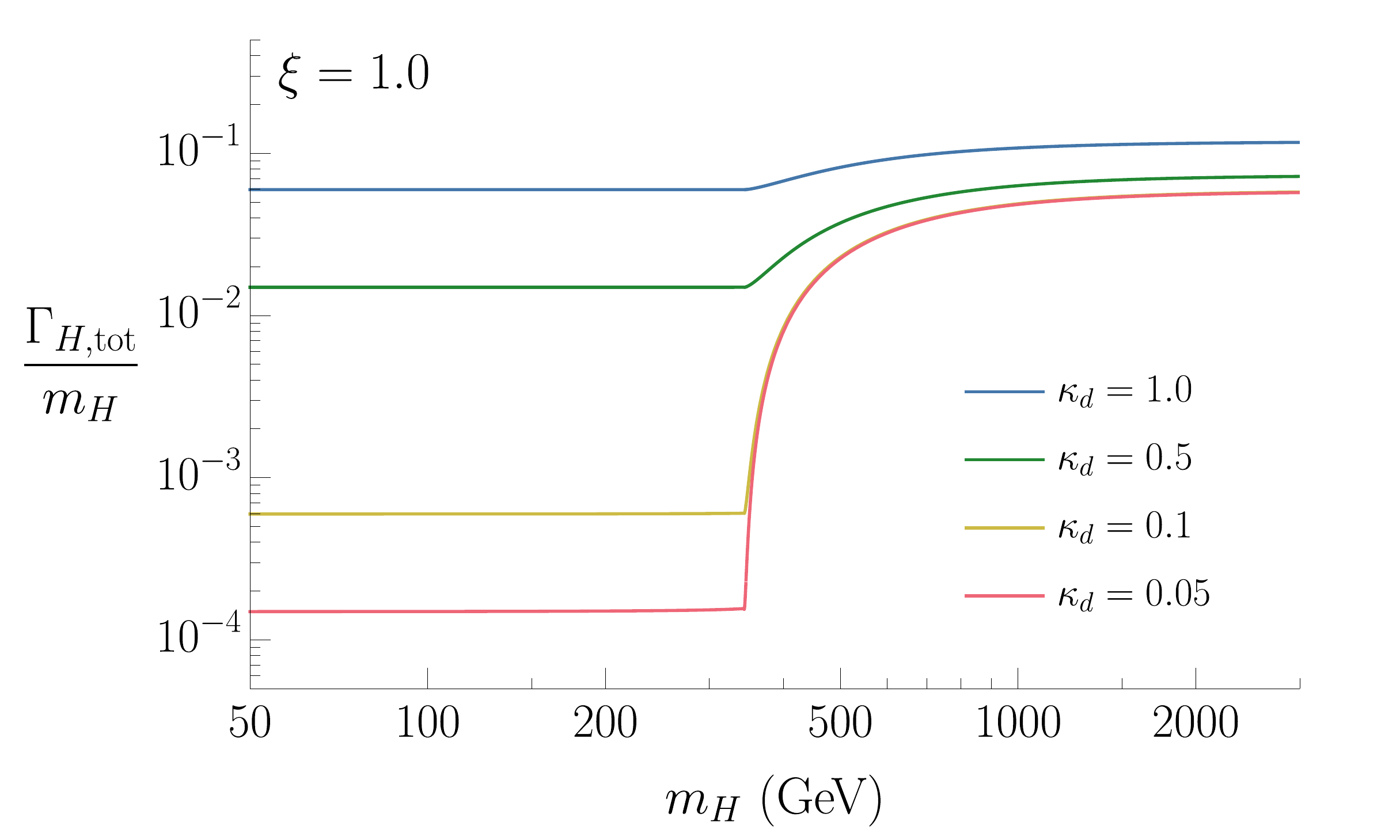}
\caption{Plot of the width-to-mass ratio, $\Gamma_{\text{tot},H} / m_H$ for the heavy neutral Higgs, $H$ for a variety of $\kappa_d$ values and $\xi = 1$. 
For values of $\Gamma/m \gtrsim 0.15$, resonance searches discussed in the text become less sensitive and the results should be interpreted with some care.
The couplings of the Higgs bosons to fermions needed to obtain the decay width are given in \tref{yukawaup}.
}
\label{fig:width_mass}
\end{figure}

An interesting aspect of having large couplings to the down type quarks is that the intrinsic width of the heavy Higgs can be quite large, in stark contrast to the case of the Standard Model Higgs.
In \figref{width_mass} we show the width-to-mass ratio, $\Gamma_{\text{tot},H/A}/m_H$, for the neutral Higgses for a variety of values of $\kappa_d$ with $\xi$ fixed to unity and $\kappa_s = \kappa_b = 0$.
We see that for values of $\kappa_d$ approaching $1$, the width of the resonance grows to $\gtrsim 10\%$ of the mass. 

The overall features for the charged Higgs decays are similar to the neutral Higgs case,
with the main difference being that the charged Higgs decays \textit{exclusively} to two quarks in our scenario. 
The charged Higgs width may also be sizable for large values of down-type Yukawa couplings $\kappa_j$, $j=d,s,b$.

%This behavior is similar to MFV theories for $m_H > 2m_t$ if $\xi = 1.0$.

%%%%%%%%%%%%%%%%%%%%%%%%%
%%%%%%%%%%%%%%%%%%%%%%%%%

%%%%%%%%%%%%%%%%%%%%%%%%%
%%%%%%%%%%%%%%%%%%%%%%%%%

%%%%%%%%%%%%%%%%%%%%%%%%%%%%%%%%%%%%%%%%%%%%%%%%%%%%%%%%
\subsection{Dijet searches}
\label{s.dijets}
%%%%%%%%%%%%%%%%%%%%%%%%%%%%%%%%%%%%%%%%%%%%%%%%%%%%%%%%
With sizable couplings to quarks, 
the SFV 2HDM is most efficiently probed at colliders via searches for dijet resonances. 
While the dijet backgrounds in hadron colliders are large,
sizable couplings to light quarks allow for abundant production of neutral and charged SFV Higgses. 
 
Searches for resonances in the dijet invariant mass spectrum have been carried out by 
the UA1 and UA2 experiments at the CERN $Sp\bar{p}S$ \cite{Albajar:1988rs, Alitti:1990kw, Alitti:1993pn}, 
the CDF and D0 experiments at the Tevatron \cite{Abe:1989gz, Abe:1993it, Abe:1995jz, Abe:1997hm, Abazov:2003tj, Aaltonen:2008dn}, 
and at the ATLAS and CMS experiments at the LHC \cite{Aad:2010ae, Aad:2011aj, Aad:2011fq, ATLAS:2012pu, Khachatryan:2010jd, Chatrchyan:2011ns, CMS:2012yf, Aad:2014aqa, Chatrchyan:2013qha, Khachatryan:2015sja, Khachatryan:2016ecr, ATLAS:2015nsi, Aaboud:2017yvp, Aaboud:2018zba, Aaboud:2018fzt, Khachatryan:2015dcf, Sirunyan:2016iap, Sirunyan:2017dnz, Sirunyan:2017nvi, Sirunyan:2018xlo, CMS:2018wxx}. 
To keep the phenomenology tractable, we again set limits in our model by treating the cases where each down-type quark Yukawa $\kappa_j$ ($j=d,s,b$) is separately dominant.
Regarding the up-type Yukawas, we consider two scenarios for the proportionality coefficient between the first and second-Higgs doublet Yukawa matrices, $\xi = 0.1$ and $\xi=1.0$.

In all the references that we consider, summarized in \tref{dijet_searches}, 
limits have been presented in terms of constraints on a parton-level cross section times branching fraction into dijets $\sigma \cdot B$, 
or times an additional parton-level kinematic acceptance factor, $\sigma \cdot B \cdot A$. 
In order to recast these limits in our model, 
we use \MadGraph~\cite{Alwall:2014hca} to obtain the parton-level $\sigma \cdot B \cdot A$ for the production of all our extra Higgs states $H$, $A$ and $H^{\pm}$  in $pp$ or $p\bar{p}$ collisions. 
We exclude events with a top in the final state (as in e.g. \cite{Aaboud:2017yvp,Sirunyan:2018xlo}), since top decays will generically not reconstruct into a single jet. 
We do include tops in the final state when calculating the Higgs widths and branching fractions. 

The results of the UA2,  CDF and D0 searches are presented in \cite{Alitti:1993pn,Abe:1997hm,Abazov:2003tj,Aaltonen:2008dn} as limits on $\sigma \cdot B$ or $\sigma \cdot B \cdot A$ for different resonance masses, 
and are thus straightforward to recast as limits in our model parameters. 
CMS constraints in \cite{Khachatryan:2016ecr,Sirunyan:2018xlo,CMS:2018wxx} are presented for a variety of final states -- $qq$, $qg$, and $gg$ at parton-level with kinematic cuts. 
Since we are interested in setting limits on an SFV Higgs with significant tree level couplings to quarks and in our model couplings to gluons arise only at loop level, 
we always use the $qq$ limits, with the corresponding kinematic cuts to obtain the acceptance.
The UA2, CDF, D0 and CMS limits apply when the dijet resonance is narrower than the dijet detector resolution, 
and are expected to deteriorate when the resonance is wide. 
To obtain a conservative limit, 
whenever the decay width of our extra Higgses exceeds the detector resolution for each experiment, 
we drop the corresponding dijet limits. 
For this purposes, we take the dijet detector resolution at UA2, CDF and D0 to be $10 \%$ of the dijet invariant mass, and at CMS to be $5\%$, 
as suggested by each one of the corresponding references \cite{Alitti:1993pn,Abe:1997hm,Abazov:2003tj,Aaltonen:2008dn,Khachatryan:2016ecr,Sirunyan:2018xlo,CMS:2018wxx}.

The constraints from ATLAS \cite{Aad:2014aqa,Aaboud:2017yvp}, on the other hand, are presented as limits on a Gaussian resonance in the $m_{jj}$ spectrum after all detector effects and cuts are applied. 
To apply these limits to our model, 
we follow the procedure detailed in Ref. \cite{Aad:2014aqa}, 
which involves truncating events outside $0.8 m_H < m_{jj} < 1.2 m_H$ for a given mass $m_H$ when computing the acceptance to avoid the effects of long tails.
The ATLAS results are presented for various values of the ratio between the decay width and mass of the resonance $\Gamma / m$, 
so in order to correctly apply these bounds we compute the width of the extra Higgs states across our parameter space, 
and take the limits for the next-largest value of $\Gamma/m$ to be conservative.

We present our dijet limits on the SFV 2HDM in Figs.~\ref{fig:dijetplot1}, \ref{fig:dijetplot2} and \ref{fig:dijetplot3} 
in the $\kappa_d-m_H$, $\kappa_s-m_H$ and $\kappa_b-m_H$ planes correspondingly. 
In each figure, 
the left panel corresponds to a proportionality constant in the up-type first- and second-doublet Yukawa matrices $\xi=0.1$,
while the right panel corresponds to $\xi=1$.
We now proceed to set limits from dedicated searches for $b$ quarks in the final state.

%%%%%%%%%%%%%%%%%%%%%%%%%
\begin{table*}%[h]
\centering
\begin{tabular}{c | c | c | c | c | l}
\hline
\begin{tabular}[c]{@{}c@{}}Collisions, \\ $\sqrt{s} (\tev)$\end{tabular} & Experiment & Ref. & \begin{tabular}[c]{@{}c@{}}Luminosity\\ ($\fb^{-1}$)\end{tabular} & \begin{tabular}[c]{@{}c@{}}Mass range \\ ($\gev$)\end{tabular} & Notes \\
\hline\hline
\multirow{1}{*}{$p\bar{p}, 0.63$} & \multirow{1}{*}{UA2} 	& \cite{Alitti:1993pn} 					& $1.09\times 10^{-2}$	& 140 -- 300 &  \\ 
\hline
\multirow{2}{*}{$p\bar{p}, 1.8$} & \multirow{1}{*}{CDF} 	& \cite{Abe:1997hm} 				& 0.106 						& 200 -- 1150 & \\ \cline{2-6}
 & D0 																			& \cite{Abazov:2003tj} 				& 0.109							& 200 -- 980 &  \\ 
 \hline
$p\bar{p}, 1.96$ & CDF 												& \cite{Aaltonen:2008dn} 			& 1.13 							& 260 -- 1400 &  \\ 
\hline
\multirow{2}{*}{$pp, 8$} & ATLAS 									& \cite{Aad:2014aqa} 				& 20.3 							& 400 -- 3400 &  \\ \cline{2-6}
 & \multirow{1}{*}{CMS} 												& \cite{Khachatryan:2016ecr} 	& 18.8 							& 500 -- 1600 & 	``Data scouting" \\
\hline
\multirow{4}{*}{$pp, 13$} 	& \multirow{2}{*}{ATLAS} 		& \cite{Aaboud:2017yvp} 			& 37 								& 1200 -- 6400 &  \\
 										&											& \cite{Aaboud:2018fzt} 			& 29.3 							& 450 -- 1800 & Trigger-level jets \\ \cline{2-6}
 										& \multirow{2}{*}{CMS}			& \cite{Sirunyan:2018xlo}			& 36								& 600 -- 8000 & \\
 										&											& \cite{CMS:2018wxx}				& 77.8							& 1800 -- 8000 & \\
 \hline
\end{tabular}
\caption{Table of inclusive dijet searches used to set limits in the $\kappa_j-m_H$ plane, $j=d,s,b$.}
\label{t:dijet_searches}
\end{table*}
%%%%%%%%%%%%%%%%%%%%%%%%%

%%%%%%%%%%%%%%%%%%%%%%%%%%%%%%%%%%%%%%%%%%%%%%%%%%%%%%%%
\subsubsection{Searches for $b$-tagged jets}\label{ss.bjet_searches}

In the case where the second Higgs doublet couples to $b$ quarks only, $\kappa_b \neq 0$, $\kappa_d, \kappa_s = 0$, quark fusion is suppressed due to the small $b$-quark parton luminosities. 
In this case, however, searches for resonances including a $b$-tagged jets help mitigate the background, and can be competitive with ordinary dijet searches.

Dedicated searches $b\bar{b}$ resonances have been carried out by the CDF experiment \cite{Abe:1998uz} at $1.8\tev$, CMS \cite{CMS:2012yf,Khachatryan:2015sja,Sirunyan:2018pas} and ATLAS \cite{Aaboud:2016nbq, Aaboud:2018tqo}. 
These searches are summarized in \tref{bjet_searches}, 
and are recast as follows. 
Tevatron sets constraints on the process $gb\rightarrow \phi b$, where $\phi$ decays to $b\bar{b}$ \cite{Aaltonen:2012zh}. 
Here, the kinematic acceptance and $b$-tagging efficiencies have been unfolded, 
so we can compute the parton-level cross section and branching ratio in our model and compare the results directly to the limits on $\sigma \cdot B$ to set constraints on $\kappa_b$. 
We find that Tevatron searches for $b\bar{b}$ resonances do not give any significant limits for our model. 
The CMS  constraints, on the other hand, are presented as limits on $\sigma \cdot B$ on a scalar, vector, and fermion resonance,
so it is straightforward to recast the limits for our scalar resonance.
ATLAS sets limits on resonances with $\geq 1$ $b$ tag, allowing an inclusive search for both $H, A \rightarrow b\bar{b}$ as well as $H^{\pm} \rightarrow bj$.
ATLAS reports the efficiency to tag the one and two-$b$ final state for a $Z'$ and also provides kinematic cuts to obtain the acceptance, 
so we simply calculate the total acceptance for our SFV 2HDM using the provided efficiency and implementing the kinematic cuts in \MadGraph. 
As in the ordinary dijet searches, 
ATLAS set constraints on a Gaussian resonance, 
so we use the same methodology as described in the previous section to set the limits.
The summary of constraints from $b$-tagged jets on the SFV 2HDM are presented in dashed lines in \figref{dijetplot3}.

%%%%%%%%%%%%%%%%%%%%%%%%%
\begin{table*}%[h]
\centering
\begin{tabular}{c|c|c|c|c|c}
\hline
Collisions, & \multirow{2}{*}{Experiment} & \multirow{2}{*}{Ref.} & Luminosity & Mass range & \multirow{2}{*}{Notes} \\
$\sqrt{s} (\tev)$ &  &  & ($\fb^{-1}$) & ($\gev$) &  \\ \hline \hline
$p\bar{p}, 1.8$ 					& CDF 								& \cite{Abe:1998uz} 					& $8.7\times 10^{-2}$ 	& 200 - 750 		& 2 $b$-jets \\ \hline
$p\bar{p}, 1.96$ 				& CDF+D0 						& \cite{Aaltonen:2012zh} 			& $2.6 + 5.2$ 				& 90 - 300 		& $gb \to \phi(b\bar{b}) b$ \\ \hline
$pp, 7$ 							& CMS 								& \cite{CMS:2012yf} 					& 5.0 							& 1000 - 4200 	& 2 $b$-jets \\ \hline
\multirow{2}{*}{$pp, 8$} 	& \multirow{2}{*}{CMS} 		& \cite{Khachatryan:2015sja} 	& 19.7 							& 1200 - 5500 	& 2 $b$-jets \\
					 					&  									& \cite{Sirunyan:2018pas} 		& 19.7 							& 325 - 1200 	& 2 $b$-jets \\ \hline
\multirow{2}{*}{$pp, 13$} 	& \multirow{2}{*}{ATLAS} 	& \cite{Aaboud:2016nbq} 			& 3.2 							& 1500 - 3550 	& 2 $+$ $\geq 1$ $b$-jet \\
 										&  									& \cite{Aaboud:2018tqo} 			& 36.1 							& 750 - 4800 	& 2 $+$ $\geq 1$ $b$-jet \\ \hline
\end{tabular}
\caption{The same as \tref{dijet_searches}, but for searches for $b$-tagged jets. We also indicate the number of $b$-tags required in each search.}
\label{t:bjet_searches}
\end{table*}
%%%%%%%%%%%%%%%%%%%%%%%%%

%%%%%%%%%%%%%%%%%%%%%%%%%%%%%%%%%%%%%%%%%%%%%%%%%%%%%%%%
\subsubsection{Boosted dijet searches}
 
\begin{figure*}[htbp!]
\centering
\subfloat{{\includegraphics[width=0.49\linewidth]{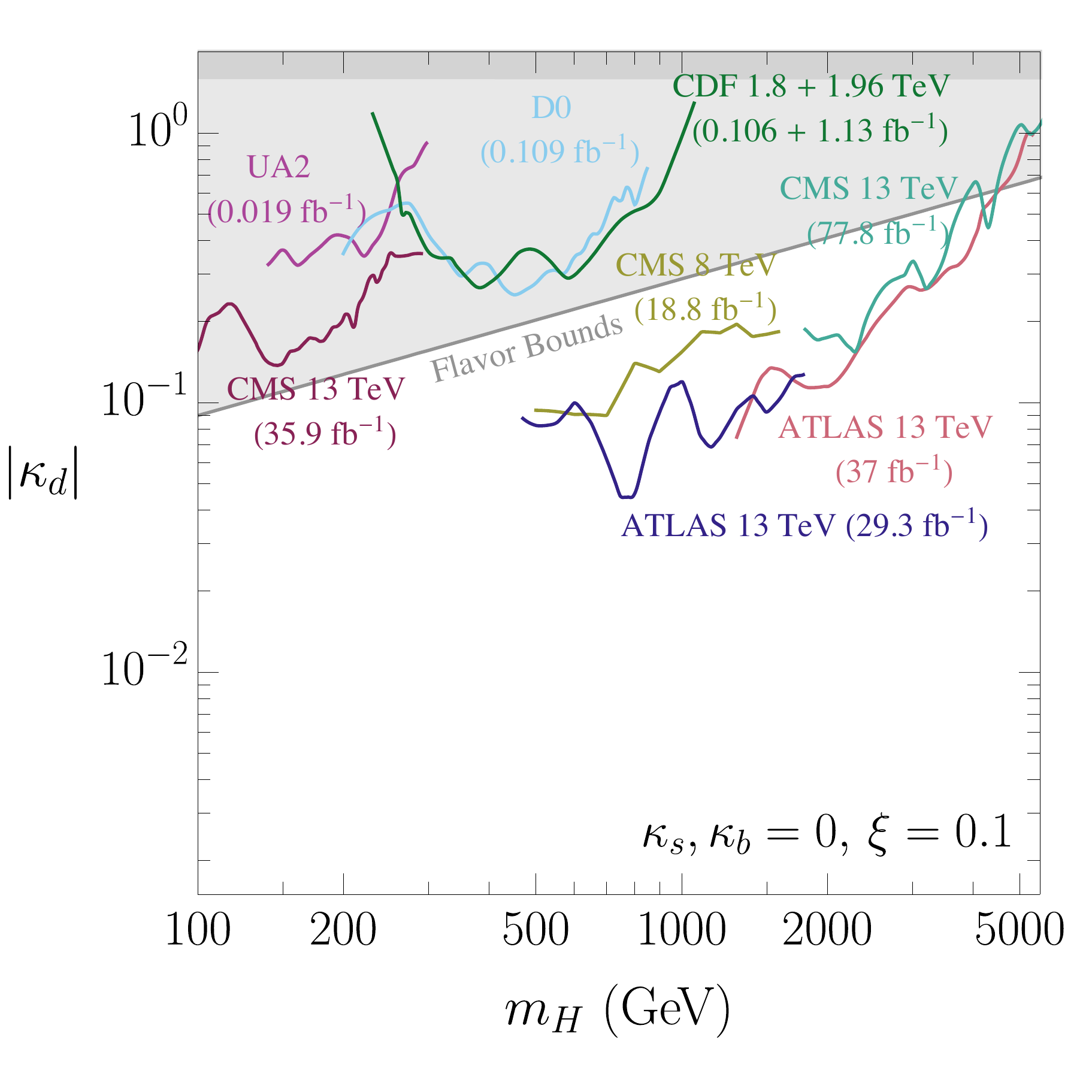} }}
~
\subfloat{{\includegraphics[width=0.49\linewidth]{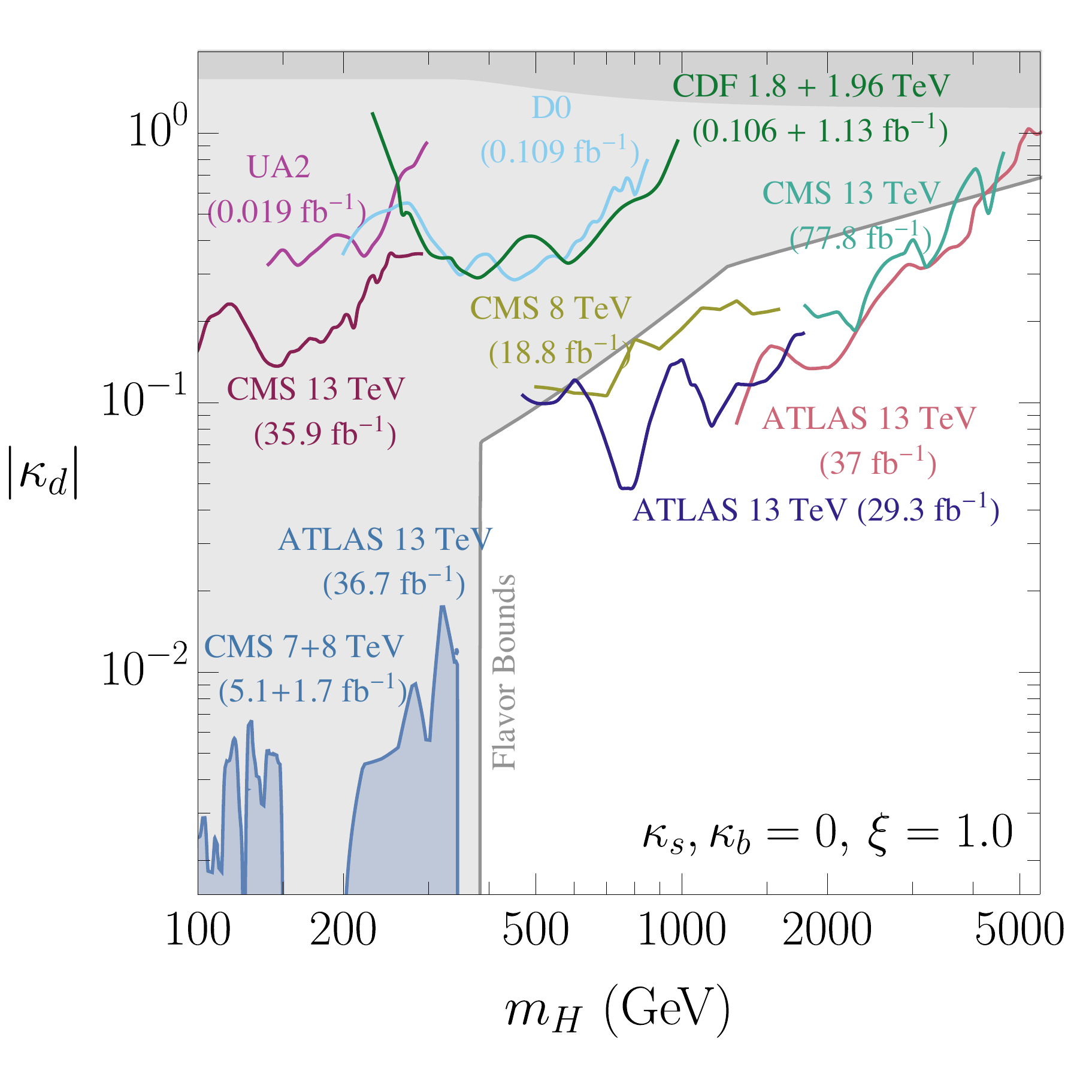} }}
\caption{
Constraints on the up-type SFV 2HDM from dijet and diphoton searches in the plane of the second-Higgs doublet mass scale $m_H$ vs. its Yukawa coupling to down quarks $\kappa_d$, 
assuming $\kappa_s = \kappa_b = 0$. 
The couplings of the second Higgs doublet to up-type quarks in SFV are universally proportional to the Standard Model ones, 
with proportionality constant $\xi = 0.1$ (left) and $\xi = 1.0$ (right).
Couplings of the second doublet to leptons have been set to zero.
All the Higgses in the second doublet, $H,A,H^\pm$ are taken to be mass degenerate.  
Constraints from flavor observables, detailed in Fig.~\ref{fig:flavor_kappad} are shown as the gray shaded region. 
The dark gray region above $\kappa_d \sim 1.0$ indicates values of $\kappa_d$ for which $\Gamma / m_H \gtrsim 0.15$ for the heavy neutral Higgs, at which point dijet searches become less reliable and the results should be interpreted with care.
}
\label{fig:dijetplot1}
\end{figure*}

\begin{figure*}[h]
\centering
\subfloat{{\includegraphics[width=0.49\linewidth]{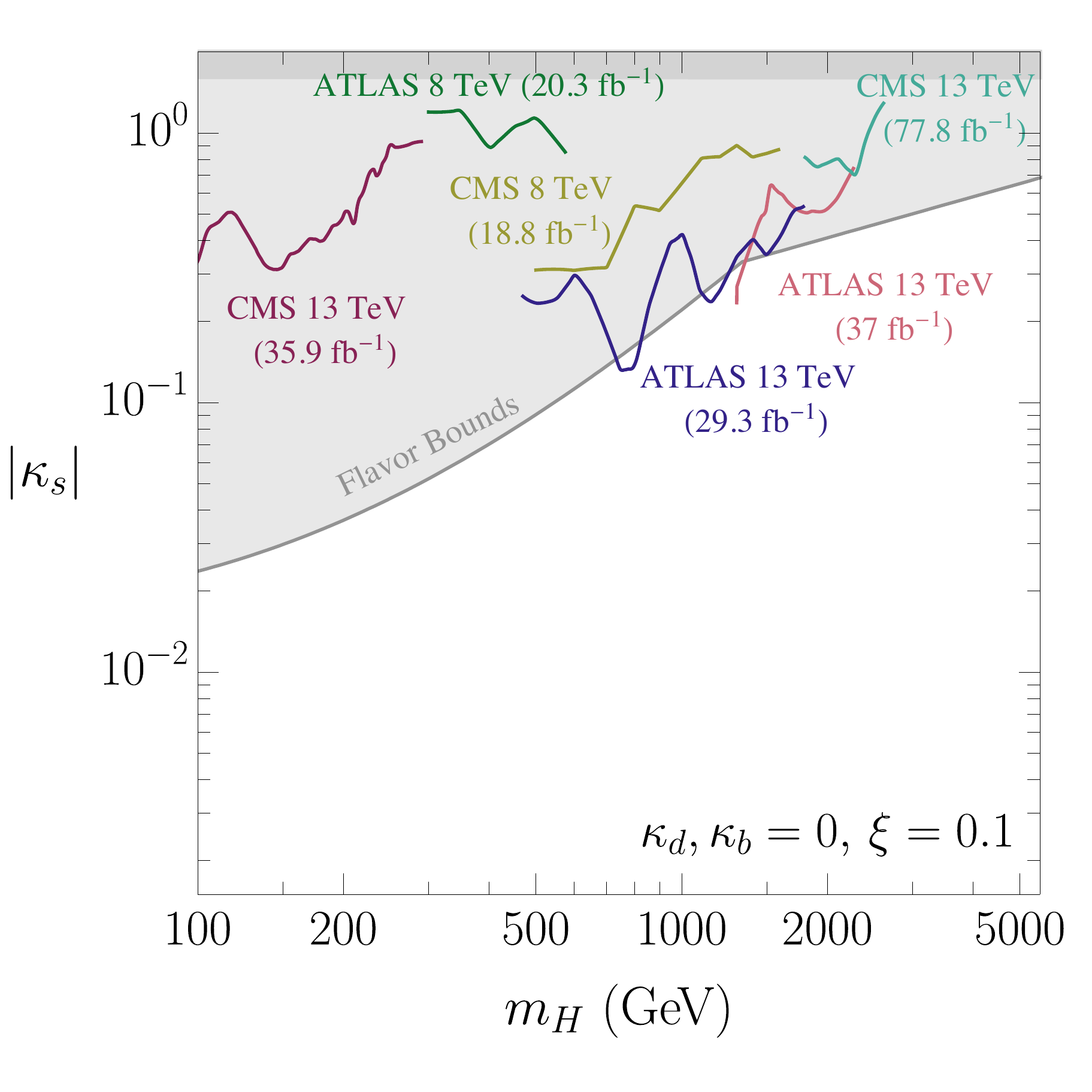} }}
~
\subfloat{{\includegraphics[width=0.49\linewidth]{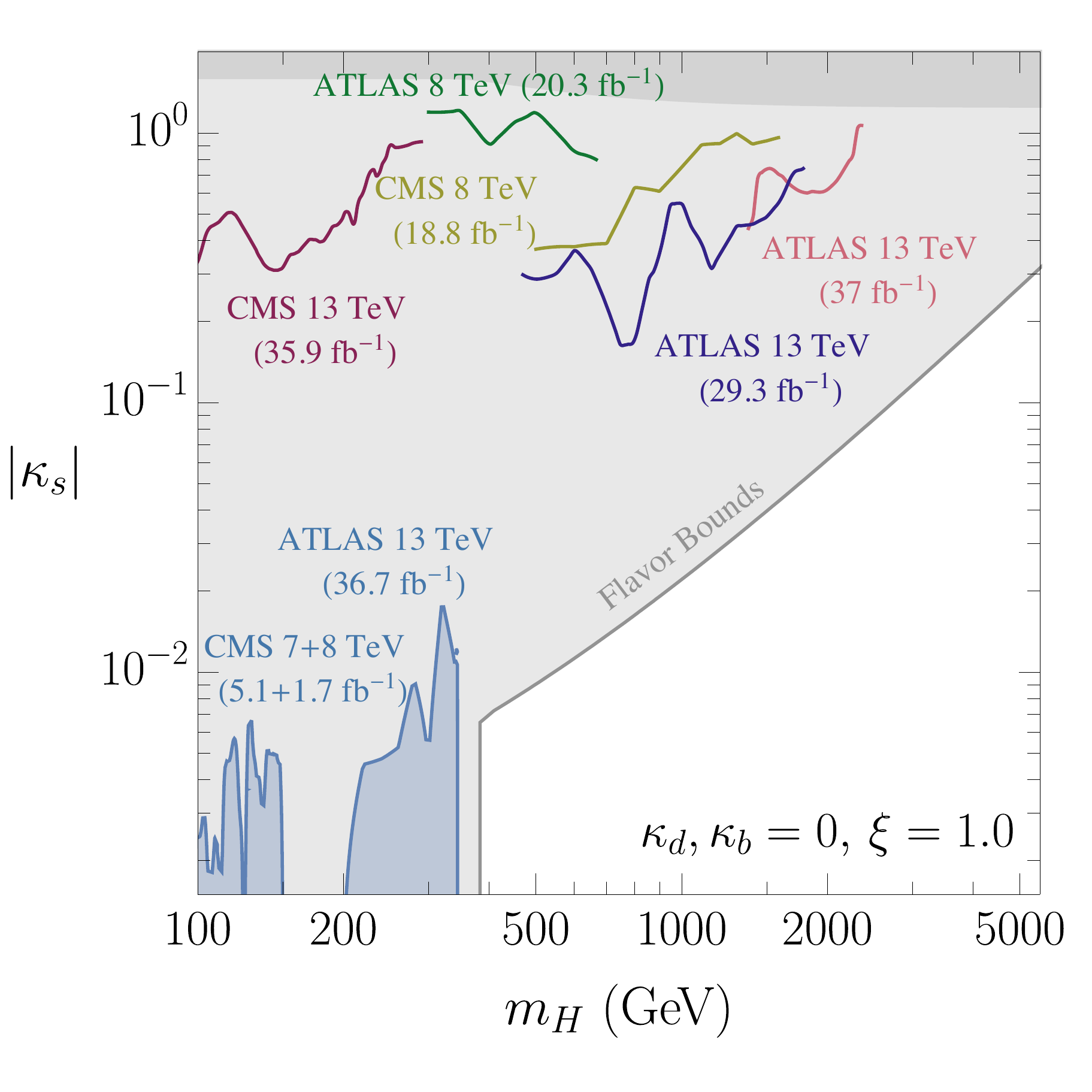} }}
\caption{The same as \figref{dijetplot1}, but for $\kappa_s$, with $\kappa_d = \kappa_b = 0$.}
\label{fig:dijetplot2}
\end{figure*}

\begin{figure*}[h]
\centering
\subfloat{{\includegraphics[width=0.49\linewidth]{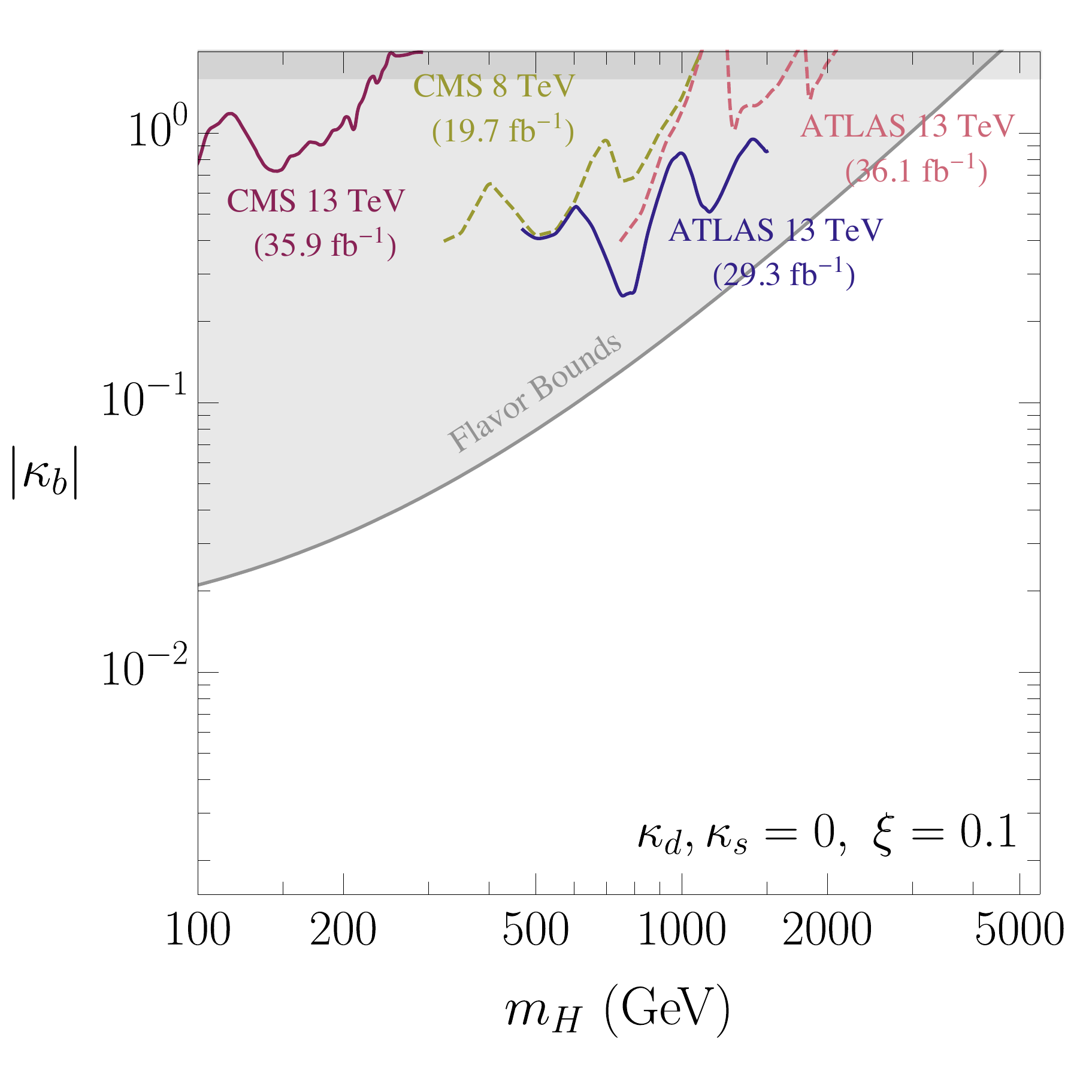} }}
~
\subfloat{{\includegraphics[width=0.49\linewidth]{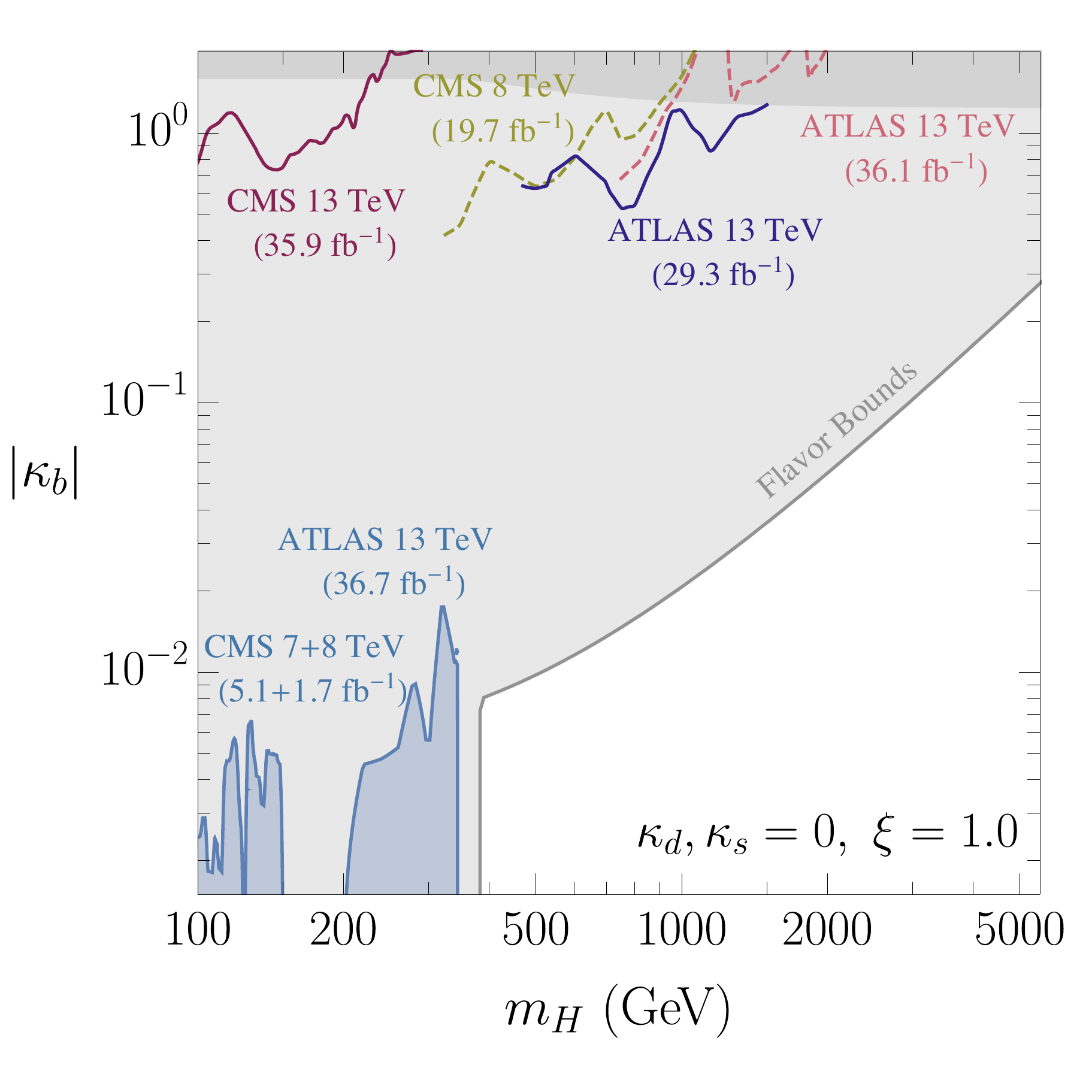} }}
\caption{The same as \figref{dijetplot1} but for $\kappa_b$, with $\kappa_d = \kappa_s = 0$. Solid lines indicate limits from ordinary dijet searches while the dashed lines indicate searches using $b$-tagging information (see text for details.)}
\label{fig:dijetplot3}
\end{figure*}

For second-Higgs doublet masses, $m_H \lesssim 300\gev$, the best collider constraints come from searches for the boosted topologies at the LHC~\cite{Aaboud:2018zba, Sirunyan:2017dnz, Sirunyan:2017nvi}. 
In this work we recast the CMS limits presented in ref.~\cite{Sirunyan:2017nvi}.
CMS presents constraints as limits on a $Z'$ resonance coupling universally to quarks. 
To recast the limits in the context of the SFV 2HDM, 
we simply rescale their limits by accounting for the difference in cross section times branching fraction to jets between the $Z'$ and SFV Higgs resonances. 
We show the results in Figs. \ref{fig:dijetplot1}-\ref{fig:dijetplot3}. 

 %%%%%%%%%%%%%%%%%%%%%%%%%%%%%%%%%%%%%%%%%%%%%%%%%%%%%%%%
 %%%% Figures
%% three two-panel figures showing flavor constraints in the \kappa_i vs. m_H plane for \xi = 1.0 and 0.1 

%%%%%%%%%%%%%%%%%%%%%%%%%%%%%%%%%%%%%%%%%%%%%%%%%%%%%%%%

%%%%%%%%%%%%%%%%%%%%%%%%%%%%%%%%%%%%%%%%%%%%%%%%%%%%%%%%
\subsection{Other search channels}
\label{s.othersearches}
%%%%%%%%%%%%%%%%%%%%%%%%%%%%%%%%%%%%%%%%%%%%%%%%%%%%%%%%
Aside from dijets, there are several other decay channels for the heavy Higgs that may be interesting at the LHC or future colliders.
In particular, as noted in the previous section, for small $\kappa_j$ and $m_H$ and $\xi = 1.0$, the neutral Higgs production becomes $\textrm{SM}$-like, with the production dominated by gluon fusion and a large decay branching fraction to diphotons.

The leading constraints on the SFV 2HDM with decays into diphoton resonances come from ATLAS ~\cite{Aaboud:2017yyg} and CMS \cite{Khachatryan:2014ira}. 
ATLAS reports limits on a fiducialized Higgs production cross section times branching fraction to diphotons  \cite{Aaboud:2017yyg}, 
which is straightforward to recast in our model by implementing the fiducial cuts in \MadGraph.
In the mass range $110 - 150\gev$, the best constraint arises from a CMS search for additional Higgs-like states~\cite{Khachatryan:2014ira}, which sets limits on the ratio $\sigma / \sigma_{SM}$ for a new Higgs state produced in gluon fusion. 
We recast these limits for our model by weighting the numerator and denominator by the branching ratio for $H, A \to \gamma\gamma$ in our 2HDM and $h \to \gamma\gamma$ in the SM respectively.
While there are other searches for diphoton resonances between $150$ and $200\gev$ (e.g.,~\cite{Aad:2014ioa, Khachatryan:2015qba}), they are not yet sensitive enough to provide constraints here. 
\footnote{Below $m_H=65\gev$, a region which we do not explore in this work, the best bounds on diphoton resonances come from reinterpreting the fiducial cross section measurement of inclusive $\gamma\gamma$ production at the LHC~\cite{Mariotti:2017vtv}. }
The constraints from diphoton searches are illustrated alongside the dijet searches in Figs. \ref{fig:dijetplot1}-\ref{fig:dijetplot3}. 

For $\xi \sim 1.0$, searches for $t\bar{t}$ or $t + j$ resonances may also be interesting. 
We have checked constraints from searches at 8 and $13\tev$~\cite{Aad:2015fna, Aaboud:2017hnm, Aaboud:2018mjh} for the heavy neutral Higgs in the SFV 2HDM, but they are not yet sensitive enough to put limits on the SFV parameter space. 
This direction may be particularly interesting in the context of $t + j$ resonance searches from the charged Higgs, where the large production cross section specific to the SFV model can be exploited.
While thus far we have considered only resonant searches for the heavy Higgs, we should note that $t$-channel exchange of the neutral or charged Higgs might also lead to interesting deviations in the angular distributions of dijets at high mass. We leave this interesting avenue to a future analysis.

%%%%%%%%%%%%%%%%%%%%%%%%%%%%%%%%%%%%%%%%%%%%%%%%%%%%%%%%
\subsection{Summary and discussion of collider bounds}
\label{s.discussioncollider}
%%%%%%%%%%%%%%%%%%%%%%%%%%%%%%%%%%%%%%%%%%%%%%%%%%%%%%%%
We summarize the collider constraints on the SFV 2HDM in Figs. \ref{fig:dijetplot1}-\ref{fig:dijetplot3}, presented in the same parameter space as Figs. \ref{fig:flavor_kappad} -- \ref{fig:flavor_kappab}. 
As before, we consider only the case where one of $\kappa_{j}$ ($j = d, s, b$) is nonzero at a time, and present limits both for $\xi = 0.1$ and $1.0$. 
The flavor constraints described in \sref{flavor} are depicted together as a grey shaded region in each case. 

In Figs.~\ref{fig:dijetplot1} and \ref{fig:dijetplot2} we start by presenting the results for the couplings to light quarks, $\kappa_d$ or $\kappa_s$, being nonzero. 
From these figures, 
we see that extra Higgs bosons as light as $~100 \,$ GeV with $\sim 0.1$ couplings to down quarks and  $\sim 2\times 10^{-2}$ to strange quarks remain consistent with both dijet and flavor searches. 
Limits from dijet searches improve at higher masses where QCD backgrounds are smaller.
For $\kappa_d$ is nonzero, 
dijet searches set the most stringent bounds for $m_H > 500\gev$ both for $\xi = 0.1$ and $\xi = 1.0$. 
For $\kappa_s$ nonzero, 
dijet searches are comparatively weaker due to the smaller strange-quark PDF,
which leads to a smaller production cross section.
As a consequence, for $\kappa_s$ non zero, we see that the best limits come mostly from flavor, especially in the case $\xi = 1.0$. 
For $\xi = 0.1$ the dijet constraints are already nearly as strong for $m_H \gtrsim 1 \tev$, and may set the most stringent bounds with data from the HL-LHC.
The results in \figref{dijetplot1} and \ref{fig:dijetplot2} illustrate the inherent complementarity of flavor and collider observables present in models with spontaneous flavor violation.
Regarding projected limits on light-quark dijet resonances at future hadron colliders, such constraints have been extensively studied~\cite{Yu:2013wta, Chekanov:2017pnx, CidVidal:2018eel}. 
In particular, with $3\,\mathrm{ab}^{-1}$ integrated luminosity at the HL-LHC, the limits on $\sigma \cdot B$ are expected to improve by a factor of $10$ for $\gtrsim \mathrm{TeV}$ resonances, pushing the limit on $\kappa_d$ to $\sim 2\times 10^{-2}$ for a $1\tev$ heavy Higgs (assuming $\kappa_s = \kappa_b = 0, \xi=0.1$). 
For a heavy Higgs coupled predominantly to strange quarks, the dijet bounds would surpass current flavor constraints above $\sim 1\tev$, assuming $\xi = 0.1$.

Current LHC constraints in the $\kappa_b$ vs. $m_H$ plane, assuming $\kappa_d, \kappa_s = 0$ are shown in \figref{dijetplot3} for the various searches in \tref{bjet_searches}. 
In addition to the $b$-jet searches described in \ssref{bjet_searches}, 
indicated by dashed lines, 
we also include the bounds from the inclusive dijet searches without using any additional $b$-tagging information (solid lines). 
For high masses, where the backgrounds from QCD dijets are already somewhat reduced, 
the inclusive searches are competitive with the dedicated $b$-jet searches, especially since they include the additional production of $H^{\pm}$. 
The collider constraints shown in \figref{dijetplot3} are weak in comparison with the stringent flavor bounds, 
in large part because of the small bottom-quark PDF. 
In principle, a small but nonzero value of $\kappa_d$ or $\kappa_s$ could lead to a significant cross-section enhancement, while preserving a significant branching fraction to $b$-quarks, potentially leading to more sensitivity from $b$-tagged searches. 
We leave such a consideration of the full five-dimensional parameter space to future work.
The sensitivity of the HL- and HE-LHC to searches for $b$-jet resonances was also studied in Ref.~\cite{Chekanov:2017pnx}. 
In addition to the improved limit on $\sigma \cdot B$, the contribution of the $b$-quark PDF is enhanced at low $x$ for higher energies, making collider searches potentially competitive with flavor constraints even in this scenario. 

Searches for diphoton resonances provide constraints in the $\kappa_j$ vs. $m_H$ plane for masses between $65$ and $350\gev$. 
These constraints depend largely on the branching ratio to two photons driven by $\xi$, which is largely independent of any hierarchies in the $\kappa_j$, so long as all of the $\kappa_j \lesssim 10^{-2}$.
However, because they depend explicitly on the loop-induced production and decay mediated by the top quark, they apply only for $\xi = 1.0$, and vanish for $\xi = 0.1$. 
For $\xi \sim 1.0$, however, $B_d$-mixing constraints already forbid $m_H \lesssim 400\gev$, independent of the $\kappa_j$, making diphoton constraints largely redundant.

Finally, we note again that the decay channels change if the alignment parameter, $\cos(\beta - \alpha)$ is allowed to be nonzero, 
or if decays between different Higgs states are allowed. 
For brevity, we will not consider such scenarios here. 
Instead, we now move on to a perhaps more interesting consequence of having nonzero alignment parameter,
namely the possibility of a substantial enhancement of the Yukawas of the $125$ GeV Higgs to light quarks.

%%%%%%%%%%%%%%%%%%%%%%%%%%%%%%%%%%%%%%%%%%%%%%%%%%%%%%%%
%%%%%%%%%%%%%%%%%%%%%%%%%%%%%%%%%%%%%%%%%%%%%%%%%%%%%%%%
%% SECTION BEGINS
%%%%%%%%%%%%%%%%%%%%%%%%%%%%%%%%%%%%%%%%%%%%%%%%%%%%%%%%
%%%%%%%%%%%%%%%%%%%%%%%%%%%%%%%%%%%%%%%%%%%%%%%%%%%%%%%%
\section{Light Higgs Yukawa enhancement in the SFV 2HDM}
\label{s.light_higgs}

Thus far we have focused only on the collider phenomenology of the new Higgs states, 
assuming no mixing between the SM-like Higgs and the extra neutral scalars. 
We now relax this assumption by allowing nonzero alignment parameter $\cos(\beta-\alpha)$.
This leads to interesting consequences for the phenomenology of the light Higgs, 
particularly via enhanced Yukawa couplings to light quarks inherited from the mixing.

%%%%%%%%%%%%%%%%%%%%%%%%%%%%%%%%%%%%%%%%%%%%%%%%%%%%%%%%
\subsection{Enhancements to down-type quark Yukawas}
\begin{figure*}%[htbp]
\begin{center}
\includegraphics[width=\textwidth]{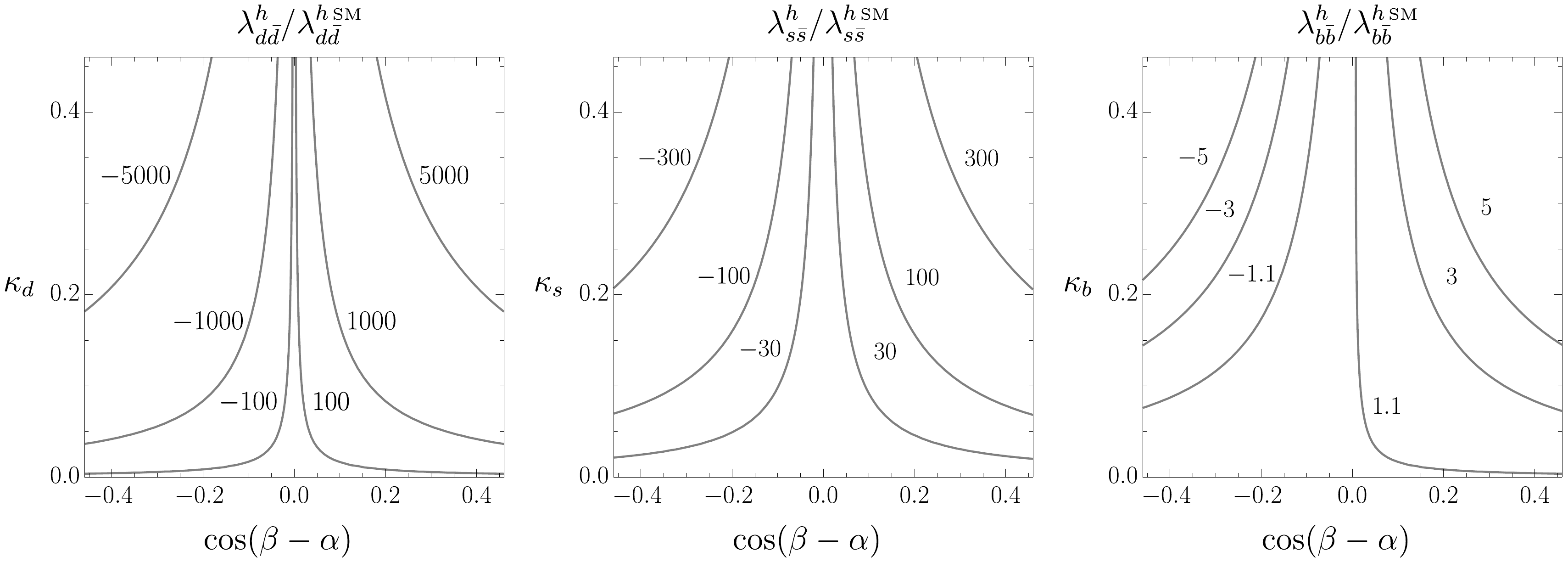}
\caption{
Enhancement contours of the $125\gev$ Higgs boson Yukawa couplings to down (left panel), strange (middle) and bottom quarks (right) in the up-type SFV 2HDM,
as a function of the alignment parameter $\cos(\beta - \alpha)$ and the Yukawa couplings of the second Higgs doublet to these quarks $\kappa_j$, $j=d,s,b$. 
We note that only the relative sign between $\kappa_j$ and $\cos(\beta - \alpha)$ is physical. }
\label{fig:yukawa_enhancement_downtype}
\end{center}
\end{figure*}

%%%%%%%%%%%%%%%%%%%%%%%%%%%%%%%%%%%%%%%%%%%%%%%%%%%%%%%%
In the up-type SFV 2HDM discussed so far, 
the Yukawa couplings of the second Higgs doublet to down-type quarks can be large,
so mixing can lead to enhancements of the down-type quark Yukawas of the $125$ GeV Higgs.
This can be seen explicitly from the couplings in \tref{yukawaup}.
We illustrate this feature in \figref{yukawa_enhancement_downtype} by plotting 
contours of the Yukawa enhancements for the down-type quarks 
as a function of the second-doublet Yukawas  $\kappa_j$ ($j=d,s,b$) 
and the alignment parameter $\cos(\beta-\alpha)$. 
We see that in the up-type SFV 2HDM 
the $125\gev$ Higgs Yukawas may in principle be enhanced by several orders of magnitude with respect to the SM expectations. 
The effect is particularly dramatic in the couplings to down and strange quarks
for large values of  $\kappa_{d,s}$ and $\cos(\beta-\alpha)$.

%%%%%%%%%%%%%%%%%%%%%%%%%
%%%%%%%%%%%%%%%%%%%%%%%%%

The $\kappa_j-\cos(\beta - \alpha)$ parameter space, however, is constrained both by limits on the new extra Higgs states, 
described in sections \ref{s.flavor} and \ref{s.collider}, 
and by 
measurements of the $125\gev$ Higgs properties.
We now study if such constraints are compatible with large enhancements to the down and strange couplings.
For this purpose, we set only $\kappa_d$ or $\kappa_s$ to be nonzero and fix $\kappa_b=0$.
We also set the couplings of the second doublet to up-type quarks to zero by taking $\xi=0$, 
in order to relax bounds from flavor discussed in the previous sections (in particular, from radiative B-meson decays, see \tref{2hdm_operators}).
Regarding the Higgs potential parameters we make the following assumptions.
First, 
we fix the heavy Higgs mass to $m_H = 500\gev$.
With $m_H > v = 246 \, \textrm{GeV}$ and Higgs potential parameters not much larger than one, the alignment parameter may be approximately expressed as \cite{Egana-Ugrinovic:2015vgy}
%boe%
\begin{equation}
\cos(\beta-\alpha)
=
-\lambda_6
\frac{v^2}{m_H^2} 
\bigg[
1+ 
\mathcal{O}\bigg(\frac{ v^4}{m_H^4}
\bigg)
\bigg] 
\quad ,
\end{equation} 
%eoe%
where $\lambda_6$ is defined in the Higgs potential \eqref{2HDMpotential}.
Note that large mixing can only be obtained with large $\lambda_6$. 
We have checked that for $m_H = 500\gev$, values of $|\cos(\beta - \alpha)|$ up to 0.45 are attainable 
with $\lambda_6$ remaining perturbative at scales $\lesssim 5\tev$.
Finally, 
for simplicity and as in the previous sections, 
we take the Higgs potential parameters $\lambda_4 = \lambda_5=0$. 
This makes the heavy pseudoscalar and charged Higgs bosons degenerate $m_A=m_H^\pm$.
The nonzero value of $\lambda_6$, 
on the other hand, 
generates a splitting between the mass of the neutral heavy Higgs state $m_H$ and $m_A=m_H^\pm$,
but which is only of order $\lambda_6^2 v^4/m_H^2 \sim \cos(\beta-\alpha)^2 v^2$ \cite{Egana-Ugrinovic:2015vgy} and is neglected in what follows.

\begin{figure*}%[htbp]
\begin{center}
\includegraphics[width=0.45\textwidth]{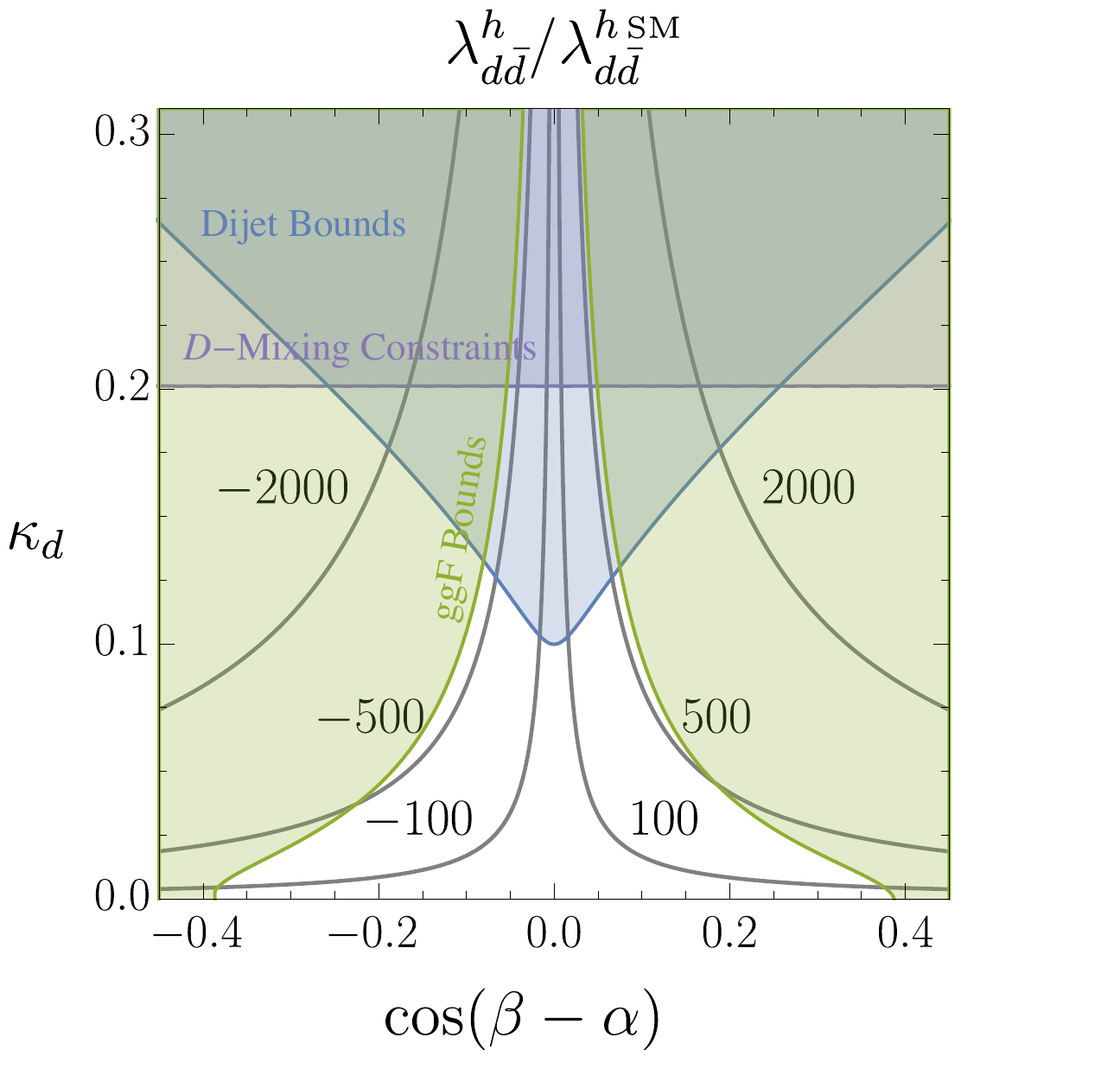}
\includegraphics[width=0.45\textwidth]{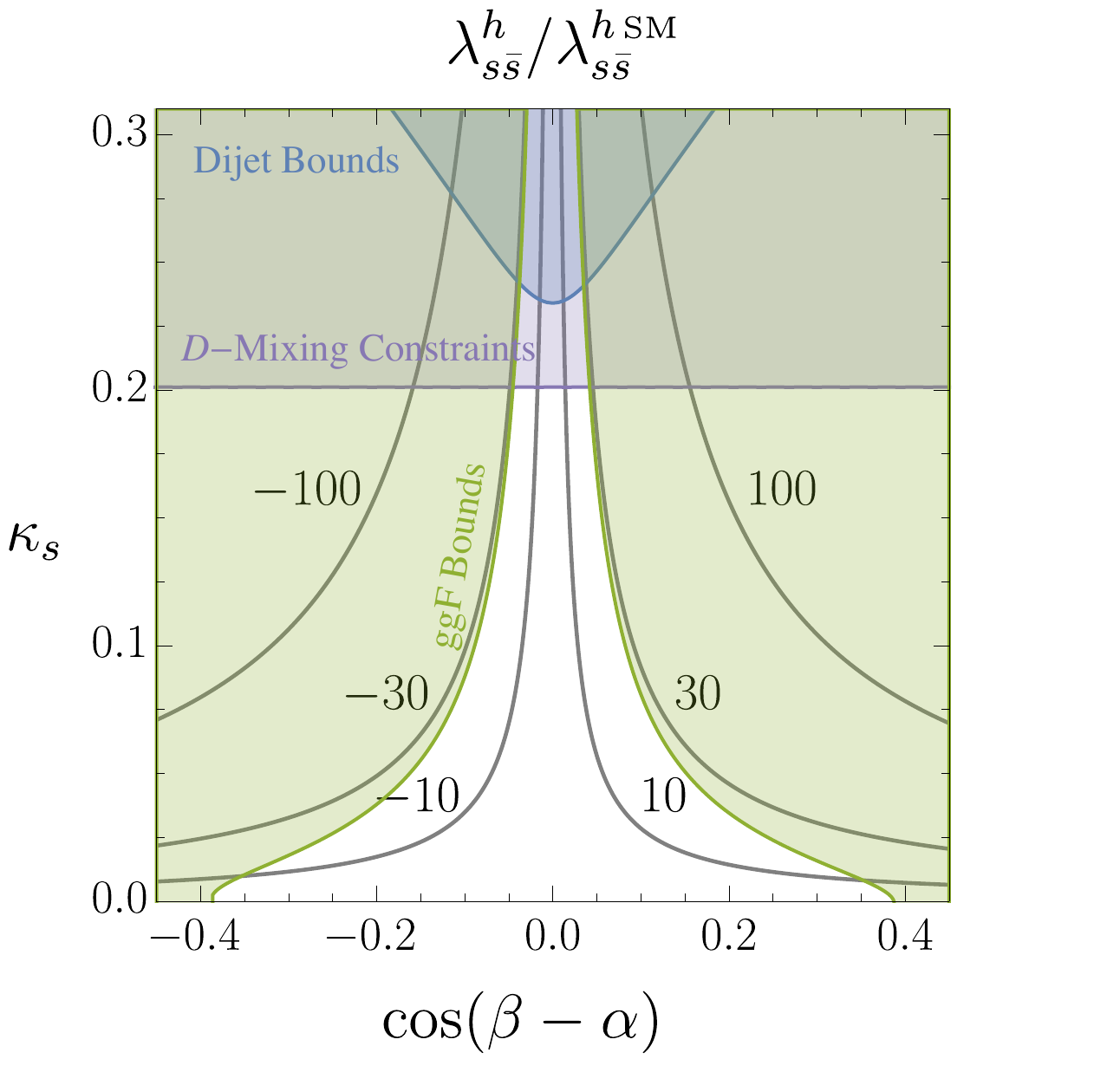}
\caption{\textit{Black contours:} enhancement of the down (left) and strange-quark (right) $125\gev$ Higgs Yukawa couplings in the up-type SFV 2HDM,
as a function of the alignment parameter $\cos(\beta-\alpha)$ and the Yukawa coupling of the second Higgs doublet to down quarks $\kappa_d$ (left) or strange quarks $\kappa_s$ (right).
\textit{Purple:} $D-\bar{D}$ mixing bounds on the  extra Higgs bosons providing the enhancements via mixing. 
\textit{Blue:} dijet bounds from production of the extra Higgs bosons at the LHC.
\textit{Green:} constraints from the measured inclusive gluon-fusion signal strength for the $125\gev$ Higgs \cite{ATLAS-CONF-2019-005}.
In both panels we have set the couplings of the second Higgs doublet to leptons, up-type quarks and to the bottom quark to zero,
and we have fixed the heavy Higgs mass scale to $m_H=500 \, \textrm{GeV}$.
In addition, 
in the left panel we set the Yukawa coupling of the second Higgs doublet to the strange quark to zero, $\kappa_s=0$, 
while on the right panel we have instead set the coupling to the down quark to zero, $\kappa_d=0$.
}
\label{fig:yukawa_enhancement_benchmark}
\end{center}
\end{figure*}

With these simplifications and keeping $m_H=500 \, \textrm{GeV}$ fixed,  we start by recasting the leading flavor and collider constraints  of the previous sections on the extra Higgs bosons to the $\kappa_j$, $\cos(\beta - \alpha)$ plane.
We present the limits in \figref{yukawa_enhancement_benchmark} in blue and purple,
for enhanced down-quark Yukawas on the left panel, 
and strange-quark Yukawas on the right panel.
While in the previous sections we worked in the limit $\cos(\beta-\alpha)=0$,
the $D-\bar{D}$ mixing bounds presented in the figure are not affected by nonzero mixing amongst neutral Higgs states, 
since meson oscillations at one loop  arise entirely from box diagrams involving charged Higgses only. 
The dijet constraints, on the other hand, become weaker in the presence of mixing, 
as production of the heavy neutral Higgses to dijets is depleted by a factor $\sin^2(\beta-\alpha)$.

We now turn to the analysis of constraints from the measured $125\gev$ Higgs properties.
Four important types of modifications to the Higgs properties arise in the presence of mixing.
First,
large alignment parameter dilutes the Higgs-gauge boson couplings
and the couplings of the Higgs to third-generation fermions (in our scenario with $\kappa_b=0,\xi=0$) by a factor $\sin(\beta-\alpha)$ \cite{Egana-Ugrinovic:2015vgy,Gunion:2002zf}.
Second, 
enhancements on the $125\gev$ Higgs-light quark Yukawas increase the width of unmeasured Higgs decays to dijets,
universally diluting the branching ratios to all the measured final states.
The overall dilution of the measured branching fractions due to the enhanced Higgs Yukawas is
%boe%
\begin{equation}
\bigg[1 
+ \frac{\Gamma^{\mathrm{SM}}_{q\bar{q}}}{\, \Gamma^{\mathrm{SM}}_{\mathrm{tot}}}
\bigg(
\bigg[
\frac{\lambda^{h}_{q\bar{q}}}{\sin(\beta-\alpha)  \lambda^h_{q\bar{q},\mathrm{SM}}} \bigg]^2
-1
\bigg)
\bigg]^{-1},
\label{eq:dilution}
\end{equation}
%eoe%
where $q$ represents the light quark, $s,d$. 
Third, 
in the presence of the heavy Higgs bosons, 
several of the $125$ GeV Higgs production modes measured at the LHC receive new contributions,
both from the enhanced Yukawas of the $125$ GeV Higgs itself,
and from diagrams mediated by the extra Higgs bosons. 
For instance, 
the gluon fusion channel receives new contributions from quark fusion due to the enhanced $125$ GeV Higgs Yukawa to down or strange quarks.
Finally, for large $\kappa_d,\kappa_s$ and nonzero mixing,
there is a significant contribution to SM Higgs pair production via quark fusion production of heavy Higgs states decaying to two SM Higgses,
which may be already constrained by current di-Higgs measurements \cite{Aad:2019uzh, Sirunyan:2018two}. 
This important effect has not been pointed out previously in the literature, 
so we leave the corresponding study for a dedicated forthcoming paper~\cite{Egana-Ugrinovic2019xxx}.

A full analysis of the above effects on all the measured Higgs properties is beyond the scope of this work.
For brevity,
here we only obtain a conservative limit on the $\kappa_j$, $\cos(\beta - \alpha)$ plane by simply requiring that the inclusive gluon-fusion signal strength $\mu_{\mathrm{ggF}} \equiv \sigma_{\textrm{ggF}} / \sigma^{\mathrm{SM}}_{\mathrm{ggF}}$,
which is the most precisely measured at ATLAS and CMS, 
lies within the bound in ~\cite{ATLAS-CONF-2019-005}.
We compute the signal strength using a \MadGraph{} implementation of our model, 
taking into account the dilution of the one loop ggF triangle diagrams due to Higgs mixing,
the dilution of the measured branching fractions in \eqref{dilution}
and the extra contribution to ggF from indistinguishable quark fusion production due to the enhanced Higgs Yukawas.
The resulting  constraints are presented in green in \figref{yukawa_enhancement_benchmark}. 

With all the constraints in place, from \figref{yukawa_enhancement_benchmark}
we see that flavor and collider limits on the extra Higgs states together with limits on the measured Higgs properties,
already restrict possible enhancements of the Higgs down- and strange-quark Yukawas.
Within our 2HDM,
enhancements on the down-quark Yukawa larger than $\sim 500$ or on the strange-quark Yukawa larger than $\sim 30$ are not possible.
While in reaching this conclusion we have fixed the heavy Higgs mass to $m_H=500 \, \textrm{GeV}$, 
we do not expect that significantly larger enhancements would be allowed by relaxing this assumption,
as limits on the heavy Higgses are similar over a wide range of $m_H$ (see Figs. \ref{fig:dijetplot1} and \ref{fig:dijetplot2}),
and limits from the ggF signal strength are largely independent of $m_H$ for fixed Yukawa enhancements.

We conclude by commenting on the prospect for measuring enhanced Higgs Yukawas.
Only the Yukawa couplings to third-generation quarks have been measured at the LHC thus far~\cite{Aaboud:2018urx, Aaboud:2018zhk, Aaboud:2018pen, Sirunyan:2018koj}.
However, 
limits on Higgs Yukawas to light quarks at current and future colliders have been discussed in 
refs.~\cite{Bodwin:2013gca,Kagan:2014ila,Perez:2015lra,Zhou:2015wra,Brivio:2015fxa,Delaunay:2016brc,Bishara:2016jga,Soreq:2016rae,Yu:2016rvv,Aaboud:2017xnb,Duarte-Campderros:2018ouv}.
Limits on the Higgs Yukawa couplings to down and strange quarks may be obtained using limits on the total Higgs width \cite{Sirunyan:2017exp}.
The bounds are at the level 
$\lambda^h_{d\bar{d}}  \lesssim 10^4  y_{d}^{\textrm{SM}}$ 
and $\lambda^h_{s\bar{s}}  \lesssim 10^3  y_{s}^{\textrm{SM}}$
\cite{Zhou:2015wra} correspondingly. 
A fit to Higgs data allowing for only $\lambda^h_{d\bar{d}}$ or $\lambda^h_{s\bar{s}}$ to vary with respect to their SM values gives a limit that is an order of magnitude better \cite{Kagan:2014ila}, 
but which cannot yet probe our benchmark scenario. 
Regarding direct probes, 
searches for $h \to \phi\gamma$ decays provide the only direct bound on $\lambda^h_{s\bar{s}}$~\cite{Aaboud:2017xnb}.
These searches are not sensitive to an $\sim 30$ enhancement factor,
both with current data or at the HL-LHC \cite{Kagan:2014ila,Bishara:2016jga}.
Other studies indicate that such an enhancement would be discoverable at the HL-LHC using Higgs kinematic distributions  \cite{Bishara:2016jga}  or at an $e^+ e^-$ machine with strange tagging ~\cite{Duarte-Campderros:2018ouv}, making the SFV 2HDM an ideal target for such searches.\\

%%%%%%%%%%%%%%%%%%%%%%%%%
%%%%%%%%%%%%%%%%%%%%%%%%%

%%%%%%%%%%%%%%%%%%%%%%%%%%%%%%%%%%%%%%%%%%%%%%%%%%%%%%%%
\subsection{Enhancements to up-type quark Yukawas}
\begin{figure}%[htbp]
\begin{center}
\includegraphics[width=\textwidth]{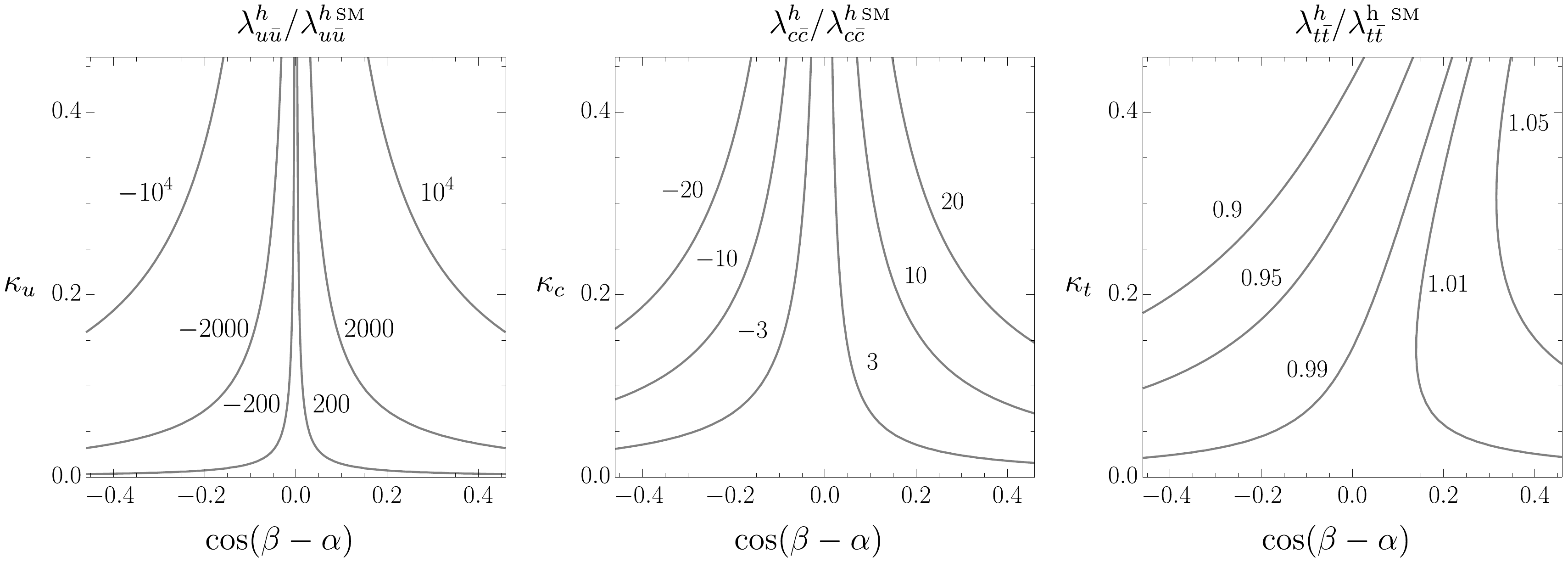}
\caption{
The same as \figref{yukawa_enhancement_downtype}, but for the up-type quark Yukawa enhancements, which are possible in down-type SFV.
}
\label{fig:yukawa_enhancement_uptype}
\end{center}
\end{figure}

%%%%%%%%%%%%%%%%%%%%%%%%%%%%%%%%%%%%%%%%%%%%%%%%%%%%%%%%
In this and the previous two sections, we have focused entirely on up-type SFV, 
which leads to enhancements of the down-type quark Yukawas. 
To understand potential enhancements to the up-type Yukawas instead, 
we now briefly consider the down-type SFV 2HDM, defined by \eqref{downtypeSFV}. 
In this theory there are three new up-type Yukawas, $\kappa_u$, $\kappa_c$ and $\kappa_t$, coupling the second Higgs doublet and up-type quarks, while the down-type quark couplings are universally proportional to the SM ones. 
The Yukawa couplings for the $CP$-even neutral Higgs in down-type SFV are given in \tref{yukawadown}.
The resulting enhancements in the $125$ GeV Higgs up-quark Yukawas are plotted in \figref{yukawa_enhancement_uptype}. 
Note that because $y_t^{\textrm{SM}} \sim 1$ is already large in the SM, 
mixing among Higgses can result in a suppression rather than an enhancement of this Yukawa,
due to cancellations. 

\begin{figure*}%[htbp]
\begin{center}
\includegraphics[width=0.45\textwidth]{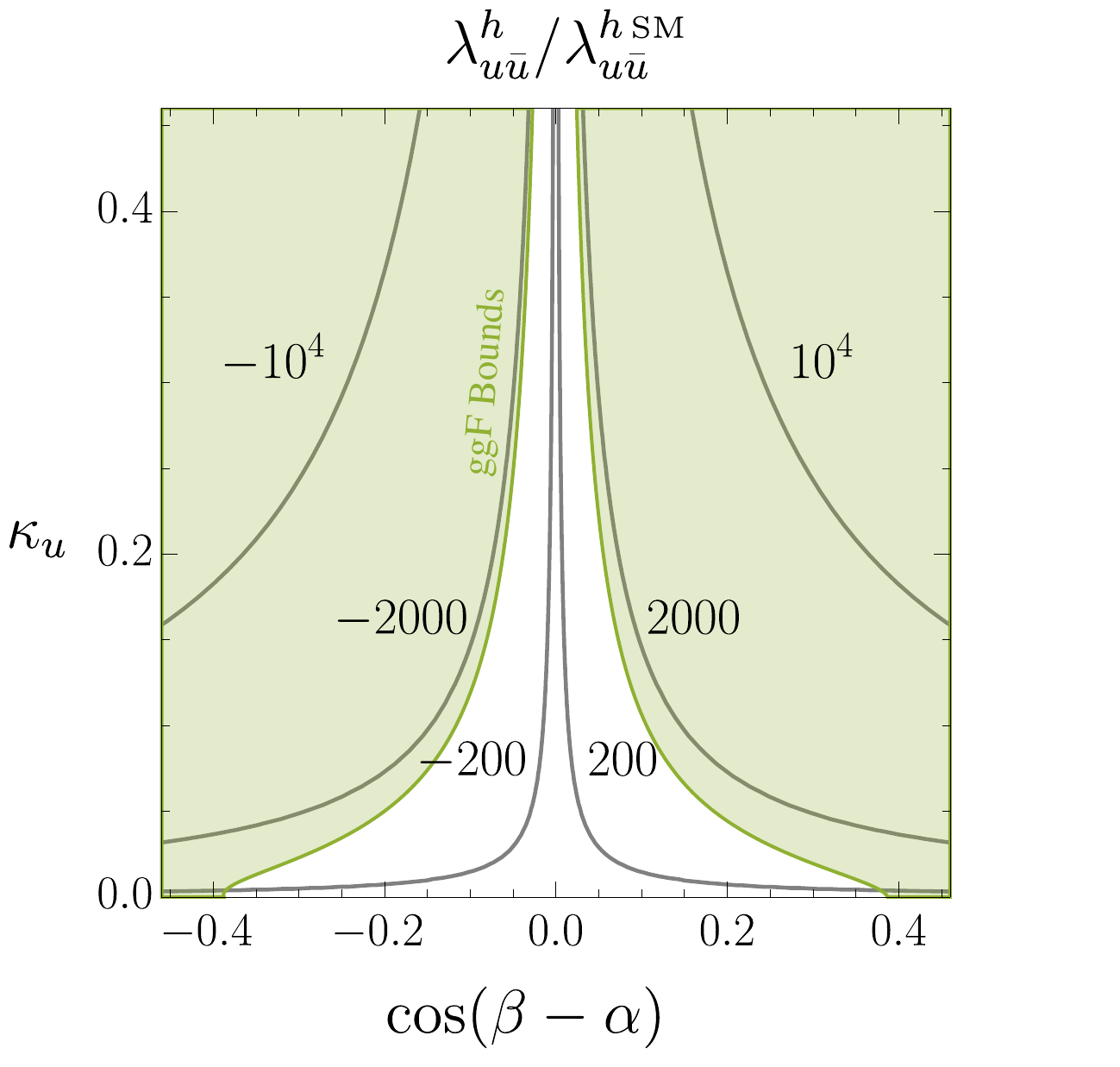}
\includegraphics[width=0.45\textwidth]{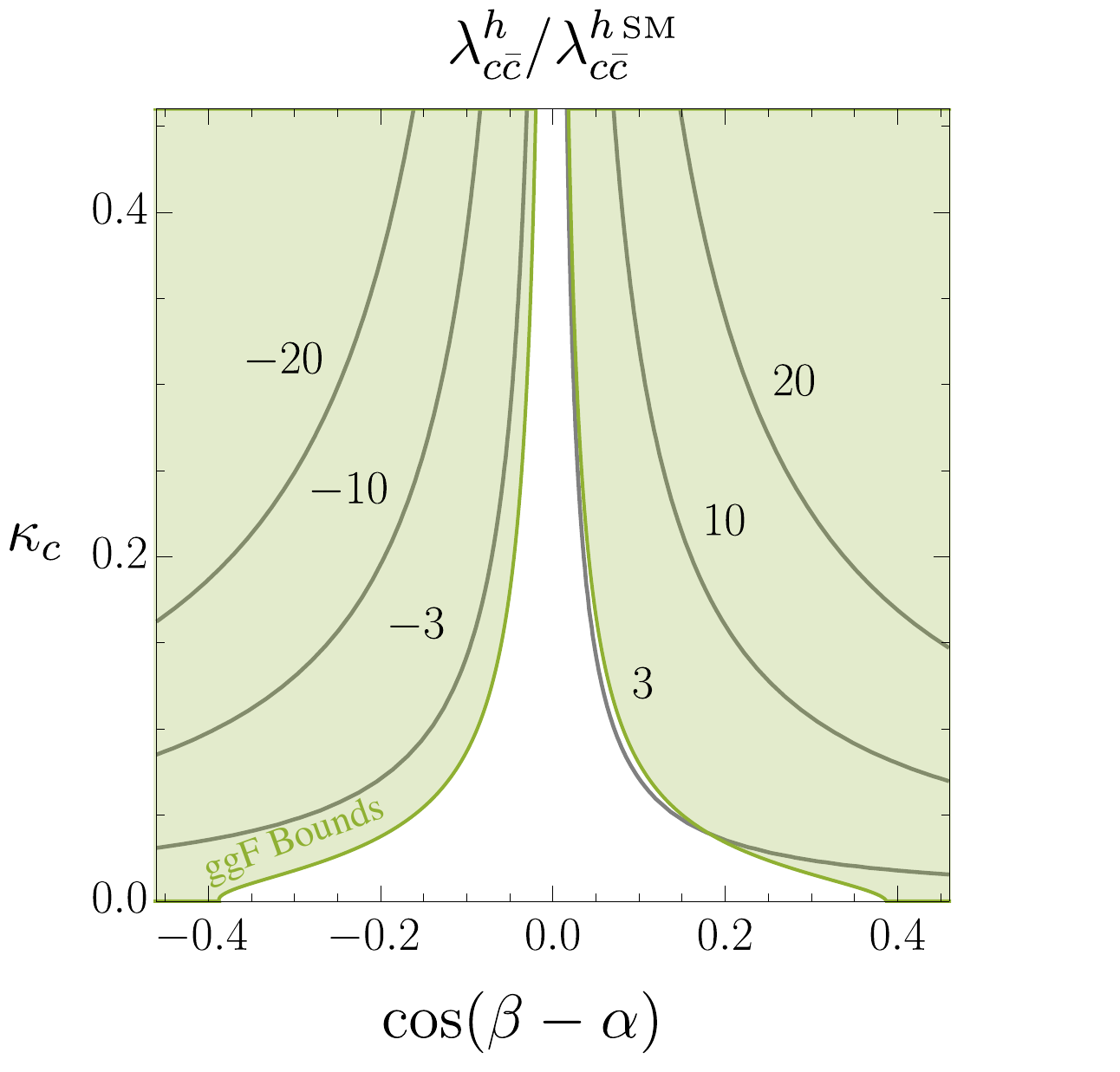}
\caption{
Same as \ref{fig:yukawa_enhancement_benchmark}, but for the up-type quark Yukawa enhancements, which are possible in down-type SFV.
Constraints from the measured inclusive gluon-fusion signal strength for the $125\gev$ Higgs \cite{ATLAS-CONF-2019-005} are shown in green.
Differently from the case of up-type SFV in \figref{yukawa_enhancement_benchmark}, 
in these figures we do not present constraints due to the extra Higgs states from flavor and dijets, 
which we have not recast here and are left for future work. 
In the left panel we set the Yukawa coupling of the second Higgs doublet to the charm quark to zero, $\kappa_c=0$, 
while on the right panel we have instead set the coupling to the up quark to zero, $\kappa_u=0$.
}
\label{fig:yukawa_enhancement_benchmark_downtype}
\end{center}
\end{figure*}

The most interesting effect in the down-type SFV 2HDM is again the possibility of large enhancements of the Yukawas to the light quarks,
in this case to the up- and charm-quarks, 
when the alignment parameter and second-Higgs doublet Yukawas $\kappa_u, \kappa_c$ are large. 
In this work we have not studied flavor or collider limits on the extra Higgs states. 
These limits set constraints on $\kappa_u$ and $\kappa_c$, 
but for brevity their analysis is left for future work. 
In the absence of a rigorous analysis on the extra Higgs states,
we only present limits from the measured gluon-fusion $125\gev$ Higgs signal strength, as described in the previous section.
The resulting constraints are presented in \figref{yukawa_enhancement_benchmark_downtype} in green, 
where we have set the Higgs potential parameters as in the previous section, $m_H=500\gev$, couplings of the second doublet to down-quarks, leptons, and the top-quark to zero. From the figure, we see that enhancements factors of order $\sim 2000$ and $\sim 3$ are possible on the up and charm-quark Yukawas correspondingly. Larger enhancements are ruled out by the measured gluon-fusion signal strengths. 

Current limits on the up-quark Yukawa from the total Higgs width are at the level of $\lambda^h_{u\bar{u}}  \lesssim 10^4\,  y_{u}^{\textrm{SM}}$,
while a global fit to data sets a limit that is an order of magnitude better \cite{Kagan:2014ila}.
Regarding the charm-quark Yukawa, 
direct limits using charm taggers currently set a constraint of the order $\lambda^h_{c\bar{c}}  \lesssim 10^2\,  y_{c}^{\textrm{SM}}$ \cite{Aaboud:2018fhh}.
A global analysis of LHC data sets a bound~\cite{Perez:2015aoa} $\lambda^h_{c\bar{c}}  \lesssim 6.2\,  y_{c}^{\textrm{SM}}$.
These limits are expected to drastically improve at the HL-LHC~\cite{Bodwin:2013gca, Perez:2015lra, Brivio:2015fxa}.
The down-type SFV 2HDM provides a well-motivated target for such searches.

%%%%%%%%%%%%%%%%%%%%%%%%%
%%%%%%%%%%%%%%%%%%%%%%%%%

%%%%%%%%%%%%%%%%%%%%%%%%%
%%%%%%%%%%%%%%%%%%%%%%%%%

%%%%%%%%%%%%%%%%%%%%%%%%%%%%%%%%%%%%%%%%%%%%%%%%%%%%%%%%
%%%%%%%%%%%%%%%%%%%%%%%%%%%%%%%%%%%%%%%%%%%%%%%%%%%%%%%%
%% SECTION BEGINS
%%%%%%%%%%%%%%%%%%%%%%%%%%%%%%%%%%%%%%%%%%%%%%%%%%%%%%%%
%%%%%%%%%%%%%%%%%%%%%%%%%%%%%%%%%%%%%%%%%%%%%%%%%%%%%%%%
\section{Conclusions}

In this work we presented two theories of extended Higgs sectors, 
the up- and down-type spontaneous flavor violating (SFV) two-Higgs-doublet models.
In these theories, 
extra Higgs states can have significant couplings to any quark generation.
They are free from tree level FCNCs due to flavor alignment,
while at one loop such currents are further suppressed by CKM matrix elements. 
While flavor alignment usually is accompanied by significant tuning, 
the SFV structure can be ensured by a UV completion in a technically natural way~\cite{Egana-Ugrinovic:2018znw}, and demonstrated explicitly for the 2HDM in Appendix~\ref{s.uv_completion}.

We performed a comprehensive study of bounds from FCNCs on the extra Higgs states in the up-type SFV 2HDM, 
where such Higgses can have generation specific couplings to down quarks. 
At the LHC, such Higgses are produced via quark fusion and decay mostly to dijet, 
but also to diphoton final states.
We then performed a detailed study of dijet and diphoton bounds,
and, whenever necessary, 
also of bounds on resonances of two bottom quarks. 
We concluded that a neutral, a pseudoscalar and a charged Higgs can all together have a mass of $100$ GeV, 
and Yukawa couplings to down quarks as large as $10^{-1}$,
and to strange quarks as large as $10^{-2}$, without being ruled out by flavor or collider bounds.
These couplings are orders of magnitude larger than the corresponding SM Higgs Yukawas. 
LHC collider bounds are the most stringent on Higgses coupling to down-quarks,
due to their large quark-fusion production cross section.

If the $125$ GeV Higgs is partially composed of such extra Higgses with large couplings to light quarks,
we find that there can be dramatic enhancements to its Yukawas,
with respect to the SM expectations. 
We find that while keeping consistency with all collider and flavor bounds on our Higgs sector,
and with currently measured Higgs signal strengths,
enhancements of the down-quark and strange-quark Yukawa couplings up to $\sim 500$ and $\sim 30$ correspondingly can be obtained in up-type SFV.
Larger enhancements seem unlikely within a realistic construction,
due to collider bounds on the extra Higgses providing the Yukawa enhancements, 
and due to bounds on the measured Higgs signal strengths.

In down-type SFV on the other hand, enhancements in the up-type quark Yukawas may instead be obtained.
While we do not perform a comprehensive analysis of flavor and collider constraints in down-type SFV,
we find that at least while keeping consistency with the currently measured ggF Higgs signal strength, 
enhancements of order $\sim 2000$ and $\sim 3$ to the up- and charm-quark Yukawas correspondingly are possible.

We see several avenues of progress for the future. 
First and foremost,
our work motivates studying new physics with preferential couplings to light quarks. 
Such scenarios have been widely overlooked, 
mostly due to theoretical expectations on how the flavor symmetries are broken,
which may be misleading. 
Theories of axions, $Z'$ gauge bosons, leptoquarks, vectorlike fermions and others, 
with specific couplings to light quarks, 
could lead to interesting new phenomenology and give insight for new experimental probes. 
Secondly, 
while here we mostly studied an extended Higgs sector with large couplings to down-type quarks,
it is rather obvious that an analysis of large couplings to up-type quarks is also needed.
Finally, 
given the difficulties of finding extra Higgses in dijet final states and extracting their flavor content, 
efforts in the direction of light quark jet taggers at current and future colliders are valuable. 
As the LHC moves into its third run and discussion on future colliders continues, 
it is important to remain open to well-motivated new physics that could lead to unexpected signatures,
that have remained hidden under the vast amounts of data. 

%%%%%%%%%%%%%%%%%%%%%%%%%%%%%%%%%%%%%%%%%%%%%%%%%%%%%%%%
%%%%%%%%%%%%%%%%%%%%%%%%%%%%%%%%%%%%%%%%%%%%%%%%%%%%%%%%
%% SECTION BEGINS
%%%%%%%%%%%%%%%%%%%%%%%%%%%%%%%%%%%%%%%%%%%%%%%%%%%%%%%%
%%%%%%%%%%%%%%%%%%%%%%%%%%%%%%%%%%%%%%%%%%%%%%%%%%%%%%%%
\section*{Acknowledgments}

We would like to thank Sally Dawson, Kohsaku Tobioka and Jared Evans for useful discussions.  The work of DE, SH and PM was supported in part by the National Science Foundation grant PHY-1620628 and in part by PHY-1915093.
DE and PM thank the Galileo Galilei Institute for Theoretical Physics for their hospitality and the INFN for partial support during the completion of this work, as well as support by a grant from the Simons Foundation (341344, LA).
PM would like to thank the Center for Theoretical Physics at Columbia University for its hospitality during completion of part of this work.
The work of SH was also supported by the U.S. Department of Energy, Office of Science, Office of Workforce Development for Teachers and Scientists, Office of Science Graduate Student Research (SCGSR) program. The SCGSR program is administered by the Oak Ridge Institute for Science and Education (ORISE) for the DOE. ORISE is managed by ORAU under contract number DE-SC0014664.

%%%%%%%%%%%%%%%%%%%%%%%%%%%%%%%%%%%%%%%%%%%%%%%%%%%%%%%%
%%%%%%%%%%%%%%%%%%%%%%%%%%%%%%%%%%%%%%%%%%%%%%%%%%%%%%%%

\appendix

%%%%%%%%%%%%%%%%%%%%%%%%%%%%%%%%%%%%%%%%%%%%%%%%%%%%%%%%
%%%%%%%%%%%%%%%%%%%%%%%%%%%%%%%%%%%%%%%%%%%%%%%%%%%%%%%%

%%%%%%%%%%%%%%%%%%%%%%%%%%%%%%%%%%%%%%%%%%%%%%%%%%%%%%%%
%%%%%%%%%%%%%%%%%%%%%%%%%%%%%%%%%%%%%%%%%%%%%%%%%%%%%%%%
%% SECTION BEGINS
%%%%%%%%%%%%%%%%%%%%%%%%%%%%%%%%%%%%%%%%%%%%%%%%%%%%%%%%
%%%%%%%%%%%%%%%%%%%%%%%%%%%%%%%%%%%%%%%%%%%%%%%%%%%%%%%%
\section{UV completion}
\label{s.uv_completion}

A UV completion for a general theory satisfying the SFV Ansatz was presented in ref.~\cite{Egana-Ugrinovic:2018znw}. For completeness, we present here an adaptation of this UV completion for the up-type SFV 2HDM.
The goal is to build a UV completion in which flavor alignment arises in a technically natural way,
to avoid large tunings in the flavor structure of our extended Higgs sector.

%%%%%%%%%%%%%%%%%%%%%%%%%
\begin{table} [ht!]
\begin{center}
$
\begin{array}{c|ccccc}
			&	U(3)_U &	U(3)_{\bar{U}} 	& U(3)_{\bar{u}}	&	U(1)_B	&	\mathbb{Z}_2 \\ \hline
U			&	3			&							&							&	1/3		&	-1	\\
\bar{U}	&				&	3						& 							&	-1/3		&	-1	\\
S			&	\bar{3}	&							&	\bar{3}				&				&	-1
\end{array}
$
\end{center}
\caption{Charge assignments for the vectorlike quarks and gauge singlet. SM fields are neutral under the $\mathbb{Z}_2$ symmetry.}
\label{t:uv_completion_charges}
\end{table}
%%%%%%%%%%%%%%%%%%%%%%%%%

To do so, we extend the 2HDM with a pair of vectorlike right-handed up-type quarks, $U_A, \bar{U}_A$, where $A = 1,\dots 3$ and $\bar{U}_A$ has the same gauge quantum numbers as the right-handed SM up-type quark, $\bar{u}_i$. The vectorlike quarks transform under their own flavor group, $U(3)_{U} \times U(3)_{\bar{U}}$ that is distinct from the SM flavor group.
We also include new gauge singlets, $S_{iA}$, which transform as triplets of the $U(3)_{\bar{u}}$ and $U(3)_{U}$ flavor groups.
We consider a Lagrangian with canonically normalized kinetic terms and the following renormalizable quark interactions
%boe%
\begin{equation}\label{eq:uv_completion_lagrangian}
\mathcal{L} \supset 
M_{AB} U_A \bar{U}_B 
+ \zeta  S_{iA} U_{A} \bar{u}_i 
+ \eta^u_{aij} Q_i H_a \bar{u}_j 
- \eta^d_{aij} Q_i H_a^c \bar{d}_j 
+ \textrm{h.c.}
\end{equation}
%eoe%
The last two terms in the Lagrangian correspond simply to the Yukawa couplings of the 2HDM,
the first term is a vectorlike mass for the extra right-handed quarks, 
while the second term is the only renormalizable  coupling that we allow between such quarks and the SM right-handed quarks. 
Additional renormalizable reactions can be forbidden by imposing additional discrete or continuous symmetries, as in \tref{uv_completion_charges}. 
In particular, these symmetries forbid couplings between the new vectorlike quarks and the two Higgs doublets $H_a$, $a = 1, 2$ at the renormalizable level. 
Without loss of generality, 
we may diagonalize the matrix $M_{AB}$ via a vectorlike quark rotation, 
$M_{AB} = \delta_{AB} M_A$. 
As in \sref{2hdm_basics} we work in the Higgs basis, where the first doublet $H_1$ is the SM doublet breaking electroweak symmetry, cf. \eqref{higgs_basis}.

We now impose two important constraints. 
First, we impose that $CP$ and the quark family number group $U(1)^3_f$ are conserved symmetries of the theory \eqref{uv_completion_lagrangian}.
In this case, there is of course no flavor mixing amongst SM quarks nor $CP$ violation,
making the theory unrealistic. 
We will solve this issue below.
Second, we allow two down-type Yukawa matrix spurions in the theory,
but only one up-quark Yukawa matrix. 
In this case, the up-type Yukawa matrices of the two Higgs doublets are necessarily proportional, $\eta^u_1 \propto \eta^u_2$. 

Now, due to $U(1)^3_f$ and $CP$ conservation,
there exists a flavor basis in which the Yukawa matrices $\eta_a^{u,d}$, $a = 1,2$ are real and diagonal, 
and are thus trivially flavor-aligned. 
Moreover, 
these matrices remain real diagonal under RGE evolution or threshold corrections from the UV due to the  $U(1)^3_f$ and $CP$ symmetries.
In this real diagonal flavor basis, the Yukawa matrices for the two Higgs doublets are
%boe%
\begin{equation}
\label{eq:uv_completion_basis}
\eta^{d}_{aij} = \delta_{ij} \eta^d_{ai}
\qquad , \qquad 
\eta^{u}_{aij} = \delta_{ij} \xi_a \eta_i^u \qquad ,
\end{equation}
%eoe%
where $\xi_{1,2}$ are real proportionality constants. 
Again, note that we allow two down-type Yukawa spurions, $\eta^{d}_{1ij}$ and $\eta^{d}_{2ij}$,
but only one up-type Yukawa spurion $\eta_i^u$. 
We commit to the basis \eqref{uv_completion_basis} in what follows.

%%%%%%%%%%%%%%%%%%%%%%%%%
\begin{figure} [h!]
\begin{center}
\includegraphics[width=0.45\linewidth]{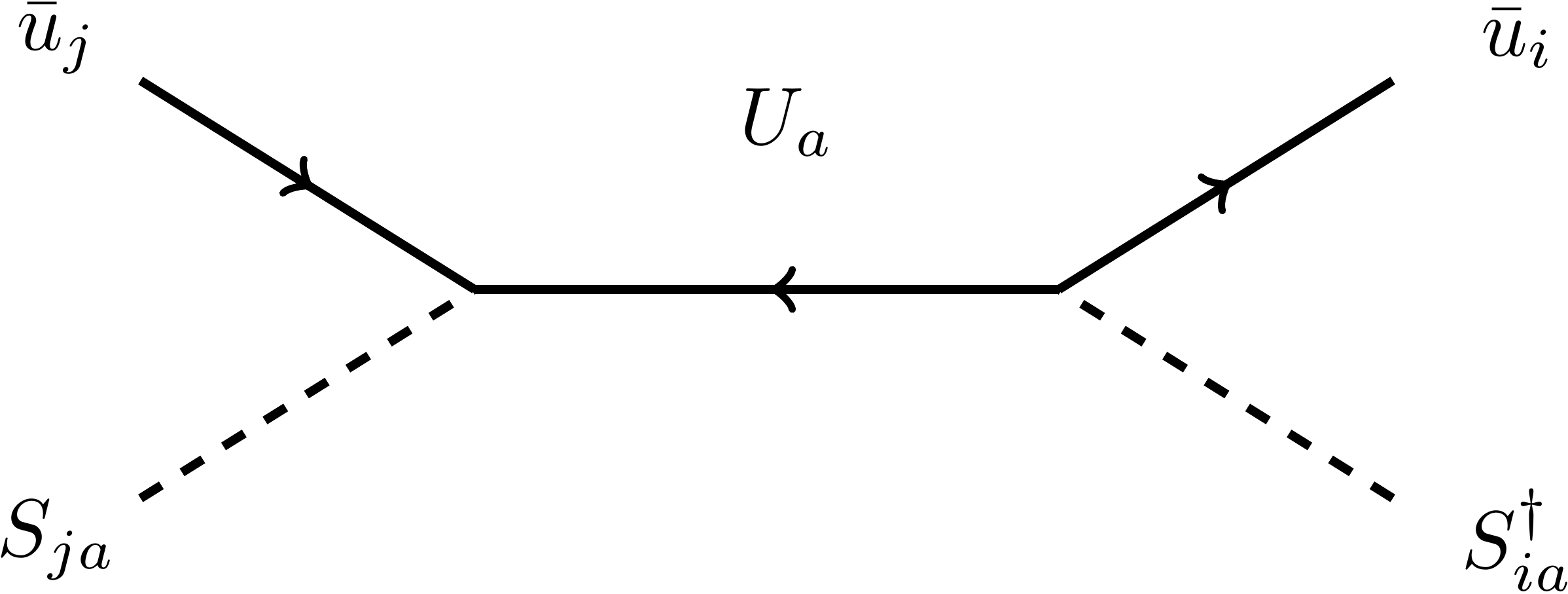}
\end{center}
\caption{Tree level diagram leading to the effective theory \eqref{uv_EFT}.
At leading order in the momentum expansion, this diagram is the only contribution to the dimension six EFT. 
Diagrams with gauge bosons are trivially related by gauge invariance.
}
\label{fig:uv_treediag}
\end{figure} 
%%%%%%%%%%%%%%%%%%%%%%%%%

We must now introduce $CP$ and family number breaking in the theory in order to allow for flavor mixing and a CKM phase. 
We do so by breaking the $\textrm{CP} \times U(1)^3_f$ symmetry only {\em spontaneously}, 
by condensates of the singlet field $S_{iA}$.
Note that since this amounts only to soft breaking, 
the 2HDM Yukawas are protected from flavor misalignment at scales above  $\sim S_{iA}$ by the $\textrm{CP} \times U(1)^3_f$ symmetry,
and it is only below this scales that we need to worry about possible misalignment effects. 

The effects of the condensates in the infrared are most easily understood by treating $S_{iA}$ as a flavor breaking spurion and integrating out the right-handed quarks $U_A$, $\bar{U}_A$. 
At tree level the only contributing diagram to the low-energy theory is given by \figref{uv_treediag} (plus diagrams related by gauge invariance). 
The effects of this diagram in the infrared are suppressed if the vectorlike quark masses are much larger than the singlet condensates. 
Since the singlet condensates are the origin of $\textrm{CP} \times U(1)^3_f$ breaking in our theory, 
in order to allow for sizable mixing angles and CKM phase, 
we must then take $S \sim M$.
This motivates organizing the low-energy theory as an expansion in terms of an effective operator dimension that counts powers of the singlet condensate $S$ and vectorlike quark masses $M$ in the operator coefficient \cite{Egana-Ugrinovic:2018znw}.
Our effective dimension is then defined as
\begin{equation}
n_{ED}=4+n_{M^2}-n_{S^2} \quad .
\end{equation}
The leading effects in the infrared are obtained by working up to effective-dimension four.
Higher effective-dimension operators lead to effects that are suppressed by infrared momenta over vectorlike quark masses, $\sim p / M$, and we drop them.
At effective-dimension four,
the diagram  \figref{uv_treediag} leads only to wave-function renormalization of the right-handed up quarks.
The low-energy effective theory is then, up to this dimension,
\footnote{One may worry that operators of higher effective-dimension may lead to FCNCs. 
However, it is easy to check that such operators, after using the equations of motion, lead only to $\Delta F~=~1$ four-fermion operators suppressed by up-type Yukawas.
This allows the vectorlike quark masses to lie far below the scales of order $\sim 10^5 \, \textrm{TeV}$ dictated by typical up-type FCNC bounds. }
%boe%
\begin{eqnarray}
\begin{split}
\mathcal{L}  ~\supset ~& 
D_{\mu}H_a^{\dagger} D^{\mu} H_a 
+ i\, Z^u_{ij} \bar{u}_i^{\dagger} \bar{\sigma}^{\mu}D_{\mu} \bar{u}_j
+ i\, Q_i^{\dagger} \bar{\sigma}^{\mu} D_{\mu} \bar{d}_j
+ i\, Q_i^{\dagger} \bar{\sigma}^{\mu} D_{\mu} Q_i \\ 
 & + \Big[ 
{\eta}^u_{aij}\, Q_i H_a \bar{u}_j
- {\eta}^{d\dagger}_{aij} Q_i H_a^c \bar{d}_j
+ \textrm{h.c.}
\Big]
\label{eq:uv_EFT}
\end{split}
\end{eqnarray}
%eoe%
where
%boe%
\begin{equation}
Z^u_{ij} = \delta_{ij} + \frac{\zeta^* \zeta}{M_A^* M_A} S_{iA}^{\dagger} S_{jA}.
\label{eq:uv_wf_renorm}
\end{equation}
%eoe%
The wave-function renormalization matrix $Z^u$ is not diagonal in quark flavor space and is the only source of individual quark number and $CP$ violation in the tree level effective theory at dimension four.
The expression \eqref{uv_wf_renorm} is actually tree level exact at all effective dimensions: 
higher effective-dimension terms in the EFT expansion generate other operators but do not lead to corrections to the wave-function matrix \eqref{uv_wf_renorm}.
Note that in this effective theory,
the strong-$CP$ problem is automatically solved via the Nelson-Barr mechanism \cite{Nelson:1983zb,Barr:1984qx,Barr:1984fh,Bento:1991ez}.

In order to check if our theory remains flavor-aligned in the low energy effective theory, we must go to the canonical kinetic basis for the right-handed SM up-quarks. 
We do so by defining the square root matrix
%boe%
\begin{equation} 
Z^u \equiv \sqrt{Z^u}^{\,\dagger}\,\sqrt{Z^u}
\label{eq:sqrt_matrix}
\end{equation}
%eoe%
and perform a field redefinition 
%boe%
\begin{equation}
\bar{u}'_i = \big( \sqrt{Z^u} \,\big)_{ij} \, \bar{u}_j.
\end{equation}
%eoe%
In terms of the redefined quark fields (dropping the primes), the low energy theory is
%boe%
\begin{eqnarray}
\begin{split}
& D_{\mu} H_a^{\dagger} D^{\mu} H_a 
+ i\, \bar{u}_i^{\dagger} \bar{\sigma}^{\mu} D_{\mu} \bar{u}_j
+ i\, \bar{d}_i^{\dagger} \bar{\sigma}^{\mu} D_{\mu} \bar{d}_i
+ i\, Q_i^{\dagger} \bar{\sigma}^{\mu} D_{\mu} Q_i & \\ &
+ \Big[
{\lambda}^u_{aij} ~ Q_i H_a \bar{u}_j
- {\lambda}^{d\dagger}_{aij} Q_i H_a^c \bar{d}_j 
+{\rm h.c.}
\Big] &
\label{eq:uv_eft_2}
\end{split}
\end{eqnarray}
%eoe%
where
%boe%
\begin{equation}
\lambda^d_a = \eta^d_a
\qquad , \qquad 
\lambda^u_a = \xi_a \eta^u \big(\sqrt{Z^u}\, \big)^{-1} \qquad ,
\label{eq:uv_yukawas_eft1}
\end{equation}
and the matrices $\eta^u$ and $\eta^d_{1,2}$ defined in \eqref{uv_completion_basis} are real and diagonal in our flavor basis.
Note that upon renormalization of the up-quarks, 
the first and second-doublet down-quark Yukawa matrices $\eta_{1,2}^d$ are unaffected, 
so they remain real diagonal and flavor aligned, 
but they are not necessarily proportional to each other. 
On the other hand, 
the first and second-doublet up-type Yukawa matrices in the effective theory are proportional to each other,
so they are also simultaneously diagonalizable, 
i.e., 
they are also flavor aligned. \footnote{Note that if we would have added a new Yukawa spurion in the up-sector for the second Higgs doublet $\eta^u_2$ not proportional to the one of the first Higgs doublet $\eta^u_1$, 
flavor alignment in the up sector (simultaneous diagonalizability of the Yukawas) would have been spoiled after applying the wave-function renormalization factor in both spurions.
This is the reason why up-type SFV requires the first- and second-doublet up-sector Yukawa spurions to be proportional.}
We conclude that in our theory, 
flavor alignment is preserved in the low-energy effective theory,
even after introducing $\textrm{CP} \times U(1)_f^3$ breaking spontaneously. 
Misalignment arises from RGE running below the vectorlike quark matching scale, 
but such corrections are suppressed by a loop factor, 
CKM matrix elements and SM Yukawas, 
and can be calculated explicitly within the 2HDM.
We dedicate \aref{radiative_corrections} to a detailed study of RGE misalignment corrections. 

The Yukawa structure \eqref{uv_yukawas_eft1} corresponds precisely to the up-type SFV 2HDM defined in \ssref{SFV2HDM}.
To provide direct contact with our notation in \ssref{SFV2HDM}, 
we first express the Yukawas for the SM Higgs doublet, $H_1$ as functions of the CKM matrix and quark Yukawa couplings by directly comparing \eqref{uv_yukawas_eft1} with \eqref{uptypeflavorbasis},
%boe%
\begin{eqnarray}\label{eq:Zsol} 
 {\lambda}^u_{1}
 & = \xi_1 {\eta}^u \big( \sqrt{Z^u}  \, \big)^{-1}  = V^T~Y^u &  \quad ,\\
& {\lambda}^d_{1} =  {\eta}^d_{1}  = Y^d & \quad ,
\label{eq:etadsol}
\end{eqnarray}
%eoe%
where the SM Yukawa couplings $Y^{u,d}$ are defined in \eqref{sm_yukawas_svd}.
From \eqref{Zsol}, we can extract the relationship between the wave-function renormalization matrix and the CKM matrix and up-type Yukawas:
%boe%
\begin{equation}
\big(\sqrt{Z^u}\,\big)^{-1} = \frac{1}{\xi_1}\, \big[ \eta_u^{-1} V^T\, Y^u\,\big] \quad .
\label{eq:Zsol2}
\end{equation}
%eoe%
Finally, using \eqref{Zsol2} in \eqref{uv_yukawas_eft1}, we obtain the Yukawas for the second doublet in terms of the CKM matrix and quark Yukawa couplings
%boe%
\begin{eqnarray}
\lambda^{u}_2  & = & \xi \lambda^{u}_1 = \xi \, V^T\,Y^u\, \label{eq:YukawasEFT2} \quad , \\
\lambda^{d\dagger}_2 &=&\eta^d_2= K^d \quad ,
\label{eq:YukawasEFT2p}
\end{eqnarray}
%eoe%
where we have defined the real coefficient $\xi = \xi_2\,/\,\xi_1$ and denoted by $K^d$ the real-diagonal matrix containing three new quark Yukawa couplings, 
as in \eqref{uptypeSFV}.

We conclude by pointing out that other UV completions leading to flavor and/or $CP$ breaking only in quark wave-function renormalization have been studied in e.g. 
\cite{Nelson:1983zb,Barr:1984qx,Barr:1984fh,Bento:1991ez,Hiller:2001qg,Hiller:2002um,Davidson:2007si}. 
Other constructions leading to flavor alignment in the context of supersymmetry  or extra-dimensional models can be found in \cite{Nir:1993mx,Leurer:1993gy} and \cite{Cacciapaglia:2007fw,Csaki:2008eh,Csaki:2009wc} correspondingly.

%%%%%%%%%%%%%%%%%%%%%%%%%%%%%%%%%%%%%%%%%%%%%%%%%%%%%%%%
%%%%%%%%%%%%%%%%%%%%%%%%%%%%%%%%%%%%%%%%%%%%%%%%%%%%%%%%
%% SECTION BEGINS
%%%%%%%%%%%%%%%%%%%%%%%%%%%%%%%%%%%%%%%%%%%%%%%%%%%%%%%%
%%%%%%%%%%%%%%%%%%%%%%%%%%%%%%%%%%%%%%%%%%%%%%%%%%%%%%%%
\section{Flavor misalignment and tuning from RGE}
\label{a.radiative_corrections}

Below the scale of spontaneous flavor violation given by the singlet condensates, 
RGE corrections for the 2HDM Yukawas spoil flavor alignment, 
as discussed in  \sref{uv_completion}.
In addition, 
these corrections may lead to significant contributions to the small first-generation quark Yukawas, 
which would lead to a source of tuning in the theory.
We dedicate this appendix to quantify the flavor misalignment and tuning from RGE running of the 2HDM Yukawas. 
The beta functions for the up- and down-type Yukawa couplings in the 2HDM are given by~\cite{Cvetic:1997zd}
%boe%
\begin{eqnarray}
\nonumber
16 \pi^2 \frac{d}{d \mu}\lambda^{d\dagger}_a &=&\,
\sum_{b=1}^2
\bigg[ 3\,\textrm{Tr}\big(\,
\lambda^{d\dagger}_a \lambda^{d}_b 
+ \lambda^u_b \lambda^{u\dagger}_a\, \big)\,\lambda^{d\dagger}_b\,
+\,\frac{1}{2}\big(\lambda^u_b \lambda^{u\dagger}_b
+ \lambda^{d\dagger}_b \lambda^{d}_b \big)\, 
\lambda^{d\dagger}_a \\
&& +
\lambda^{d\dagger}_a\, \lambda^{d}_b \lambda^{d\dagger}_b 
- 2 \lambda^{u}_b \lambda^{u\dagger}_a \,\lambda^{d\dagger}_b
\bigg]
~ - ~
A_D\, \lambda^{d\dagger}_a
\qquad ,
\label{eq:betadown}
\end{eqnarray}
%eoe%
%boe%
\begin{eqnarray}
\nonumber
16 \pi^2 \frac{d}{d \mu}\lambda^u_a &=& \,
\sum_{b=1}^2
\bigg[ 3\, \textrm{Tr} \big( \,
\lambda^u_a \lambda^{u\dagger}_b
+ \lambda^{d\dagger}_b \lambda^{d}_a\,\big) \,\lambda^u_b\,
+ \, \frac{1}{2} \big(\lambda^u_b \lambda^{u\dagger}_b
+ \lambda^{d\dagger}_b \lambda^{d}_b \big)\,
\lambda^u_a\\
&& +
\lambda^u_a \, \lambda^{u\dagger}_b \lambda^{u}_b
- 2 \lambda^{d\dagger}_b \lambda^{d}_a\, \lambda^u_b
\bigg]
~ - ~
A_U\, \lambda^u_a
\qquad ,
\label{eq:betaup}
\end{eqnarray}
%eoe%
where $\mu=\log \Lambda$ and
%boe%
\begin{equation}
A_U= 8 g_3^2 + \frac{9}{4}g_2^2 + \frac{17}{12}g_1^2, \qquad , \qquad  A_D=A_U-g_1^2 \qquad .
\end{equation}
%eoe%

%%%%%%%%%%%%%%%%%%%%%%%%%%%%%%%%%%%%%%%%%%%%%%%%%%%%%%%%
\subsection{Tuning due to radiative corrections to light-quark masses}
\label{ss.radiative_tuning}
%%%%%%%%%%%%%%%%%%%%%%%%%%%%%%%%%%%%%%%%%%%%%%%%%%%%%%%%
Large Yukawa couplings of a second doublet to first generation quarks lead to sizable RGE corrections to the corresponding SM Higgs Yukawas. 
This is an irreducible source of tuning in the theory. For instance, at zero-th order in off-diagonal CKM elements the corrections to the first generation down-quark Yukawas in up-type SFV are
%boe%
\begin{equation}
\delta y_{d,s,b} \sim \frac{1}{16\pi^2} y_t^2 \xi \kappa_{d,s,b} \log\left(\frac{\Lambda_{\textrm{UV}}^2}{\Lambda_{\textrm{IR}}^2}\right).
\end{equation}
%eoe%
The largest tuning comes from fine cancellations needed to obtain the SM down quark mass. 
We then define tuning as $\textrm{max}\, \big[\frac{d\log y_d}{d\log \alpha_i}\big]$ \cite{Barbieri:1987fn}, with $\alpha_i=(\kappa_d,\xi_i)$.
With this measure, we find that for $\kappa_d = 1$, $\xi = 1$, the tuning is $\mathcal{O}(10^{-3})$, but the tuning decreases linearly with both $\kappa_d$ and $\xi$. 
Theories with rather large Yukawa couplings to first generation quarks, $\kappa_d = 0.1$ and $\xi = 0.1$ are only tuned at the ten-percent level.

%%%%%%%%%%%%%%%%%%%%%%%%%%%%%%%%%%%%%%%%%%%%%%%%%%%%%%%%
\subsection{Flavor misalignment and constraints from radiatively induced FCNCs}
\label{ss.radiative_constraints}
%%%%%%%%%%%%%%%%%%%%%%%%%%%%%%%%%%%%%%%%%%%%%%%%%%%%%%%%
The corrections from RGE running can be separated in two types. 
First, 
the RGE terms that arise from Higgs anomalous dimensions and/or from gauge interactions preserve flavor alignment to long distances,
and they only lead to universal multiplicative rescalings of the 2HDM Yukawas. 
The second kind of corrections are the RGE evolution terms coming from the renormalization of the Yukawa three-point function,
which break flavor alignment and the SFV Ansatz. 
For instance, 
consider the up-type SFV 2HDM. 
In this case, 
at the SFV scale given by the singlet condensates in the UV completion of section \sref{uv_completion},
there exists a flavor basis in which the down-quark Yukawa matrices of the first- and second-doublet are both diagonal, 
so they are flavor-aligned. 
In this flavor basis, 
RGE evolution below that scale induces off-diagonal elements in both Yukawa matrices due to the terms proportional to $ \lambda^u \lambda^{u\dagger}= V^T Y_{u}^2 V^* $ in the beta function \eqref{betadown}.
If the couplings of the second-doublet to down quarks are large, 
the leading contributions to these off-diagonal elements at the electroweak scale are of the order
\begin{equation}
\frac{1}{16 \pi^2}  \big(V^T Y_{u}^2 V^*\big) K^d   \log\bigg(\frac{\Lambda_{\textrm{UV}}}{\Lambda_\textrm{EW}}\bigg)  \quad ,
\label{eq:misalignment1}
\end{equation} 
where $K^d$ is the real-diagonal matrix controlling the couplings of the second-doublet to down-quarks, c.f. \eqref{YukawasEFT2p}.
Naively, 
the flavor-misaligned terms are of the order \eqref{misalignment1}.
This expectation is incorrect since in order to calculate the misaligned terms, 
it is necessary to diagonalize the first-doublet Yukawa at the electroweak scale.
To see the effect of this diagonalization, 
consider the simpler two-family case, 
and with a second doublet coupling only to down-quarks  $\lambda^{d\dagger}_2=\textrm{diag}(\kappa_d,0)$.
In this scenario, 
the first-doublet Yukawa matrix at the electroweak scale is schematically of the form
\begin{equation}
\lambda^{d\dagger}_1
\sim
{\def\arraystretch{1}\tabcolsep=10pt
\Bigg(\begin{array}{cc} 
y_d
&  
0
\\  
\frac{1}{16 \pi^2}  y_c^2 \kappa_d \, V_{22} V_{21}  \,  \log\Big(\frac{\Lambda_{\textrm{UV}}}{\Lambda_\textrm{EW}}\Big)
 &
y_s
\end{array}\Bigg) \quad ,
}
\label{eq:misalignment2}
\end{equation}
where $y_d$ and $y_s$ are of the order of the SM down- and strange-Yukawas in our leading order estimate. 
If the off-diagonal corrections are small, 
the rotation angle needed to diagonalize the matrix \eqref{misalignment2} is of the order
\begin{equation}
\theta 
\sim 
\frac{1}{y_s-y_d}
\frac{1}{16 \pi^2}  y_c^2 \kappa_d  \, V_{22} V_{21}   \log\Big(\frac{\Lambda_{\textrm{UV}}}{\Lambda_\textrm{EW}}\Big)
\quad .
\end{equation}
Applying a rotation by an angle $\theta$ in the second-Higgs doublet Yukawa $\lambda^{d\dagger}_2=\textrm{diag}(\kappa_d,0)$ induces on it off diagonal elements of the order 
\begin{equation}
\frac{1}{y_s-y_d}
\frac{1}{16 \pi^2}  y_c^2 \kappa_d^2  \, V_{22} V_{21}   \log\Big(\frac{\Lambda_{\textrm{UV}}}{\Lambda_\textrm{EW}}\Big)
\quad ,
\end{equation}
which are larger than the naive expectation in \eqref{misalignment1} by a factor $\kappa_d/(y_s-y_d) \sim \kappa_d/y_s$. 
For $\kappa_d \sim 0.1$, 
this is an enhancement of the misaligned terms of order $\sim 10^2$.
This phenomenon is an elementary characteristic of matrix diagonalization
and is referred to as level repulsion, as it is most severe when two eigenvalues of a matrix are similar.
Level repulsion breaks the naive estimate of a polynomial flavor spurion expansion,
and is also relevant for other types of 2HDMs as the ones studied in \cite{Gori:2017qwg}.
Level repulsion is most significant for the misalignment of elements in the first two generations,
since repulsion between third-generation and the lighter quarks is only an effect of order $1/y_b$, 
instead of $1/y_s$.

Flavor misalignment leads to radiatively induced FCNCs in processes  mediated by the neutral Higgses.
While for $\Delta F=2$ processes these effects arise formally at two-loops since they require two insertions of one loop misaligned couplings, 
they are enhanced by large logarithms and level repulsion, 
and are potentially as large as the one loop FCNCs from charged Higgs boxes computed in \sref{flavor}.
To estimate these effects, 
we wrote a numerical code to calculate the flavor misaligned elements of the two-doublet Yukawa matrices in the up-type SFV 2HDM.
In the code, 
the Ansatz for the SFV Yukawas \eqref{YukawasEFT2p} is imposed at a high scale, 
corresponding to the scale of the SFV UV completion.
We take this scale to be $\Lambda_{\textrm{UV}} =  100\tev$.
We then evolve the Yukawas to the electroweak scale using the beta functions \eqref{betadown} and \eqref{betaup}.
In the infrared, 
we impose the known quark masses and CKM matrix elements to fix the first-doublet Yukawas. 
We then iterate between the UV and EW scales until we obtain consistency with both the SFV Ansatz boundary conditions at the UV, 
and the measured quark sector parameters at the EW scale. 

\begin{figure}[h]
\centering
\centering
\subfloat{{\includegraphics[width=0.47\linewidth]{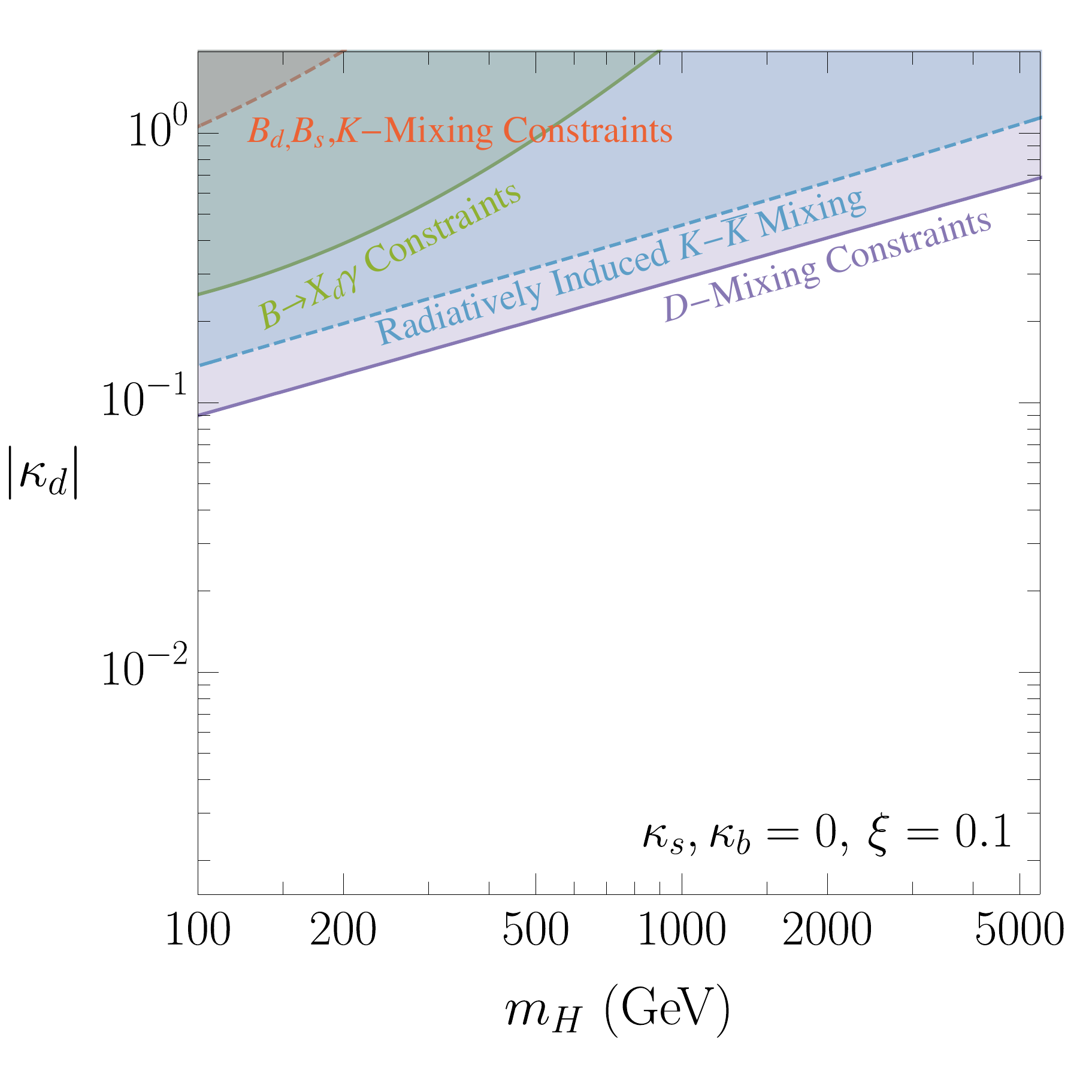} }}
\quad
\subfloat{{\includegraphics[width=0.47\linewidth]{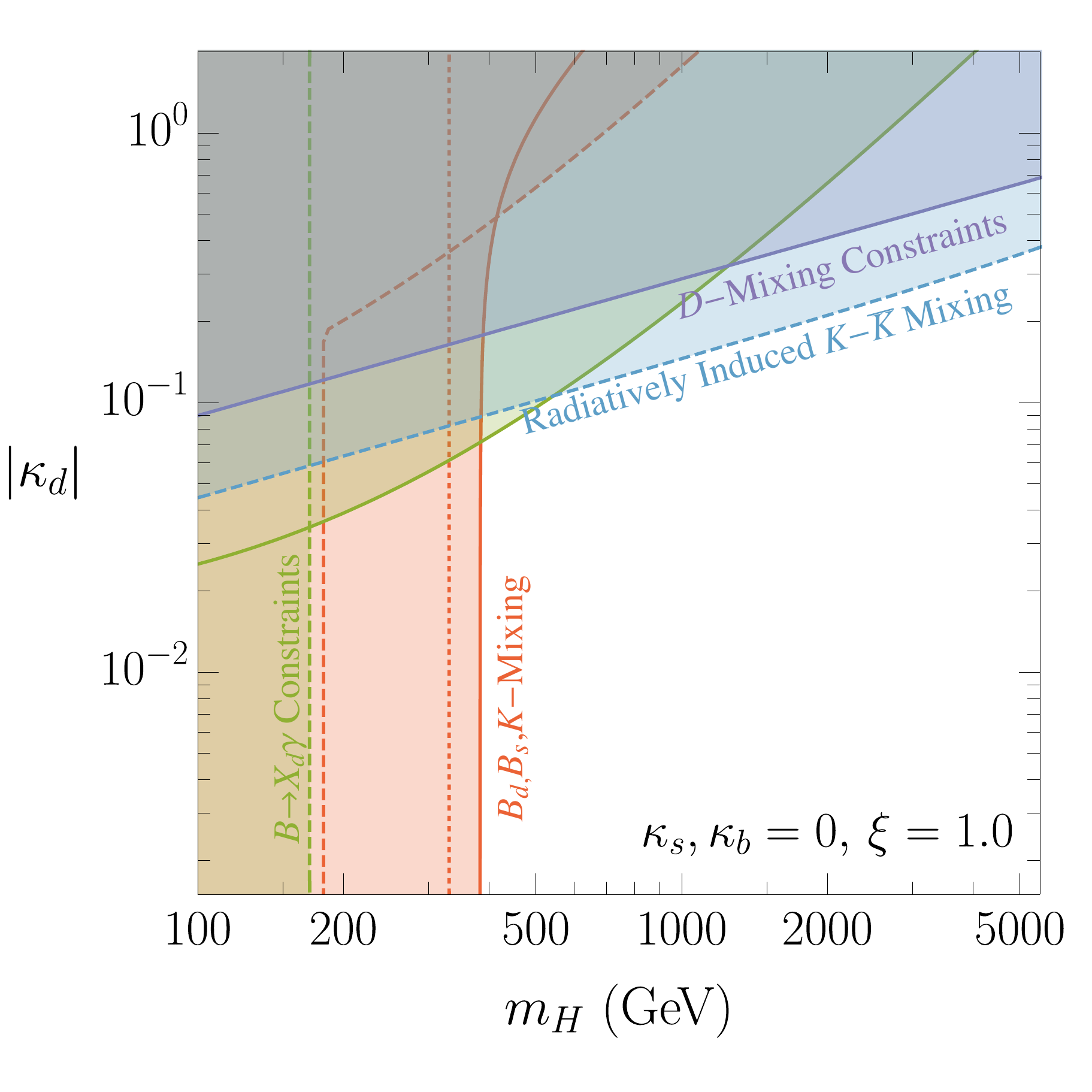} }}
\caption{
Constraints from $K-\bar{K}$ mixing arising from radiatively-induced off-diagonal Yukawa couplings, alongside other flavor bounds in the $\kappa_d$ vs. $m_H$ plane, assuming $\xi = 0.1$ (left), and $\xi = 1.0$ (right).
}
\label{fig:flavor_radiative_kappad}
\end{figure}

\begin{figure}[h]
\centering
\centering
\subfloat{{\includegraphics[width=0.47\linewidth]{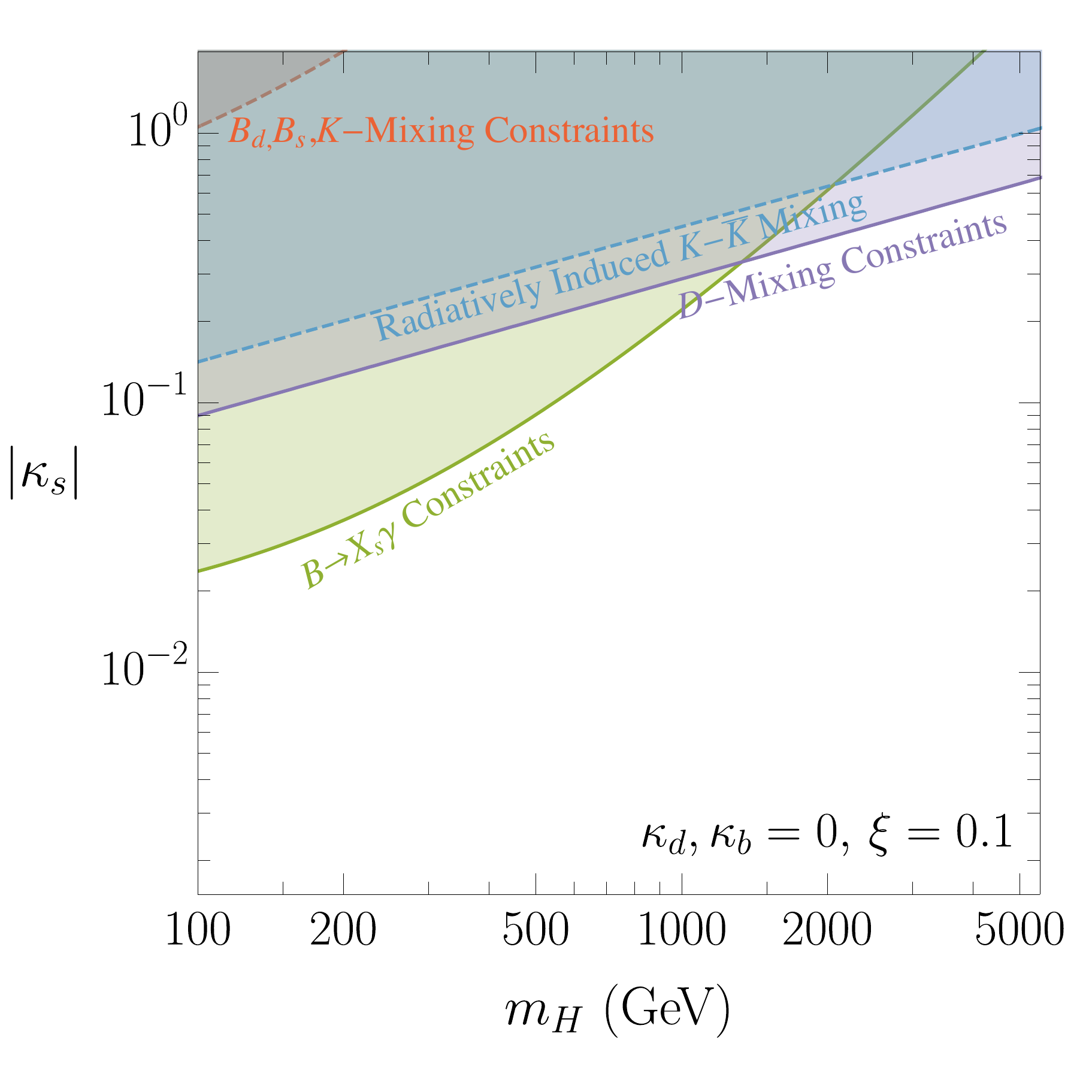} }}
\quad
\subfloat{{\includegraphics[width=0.47\linewidth]{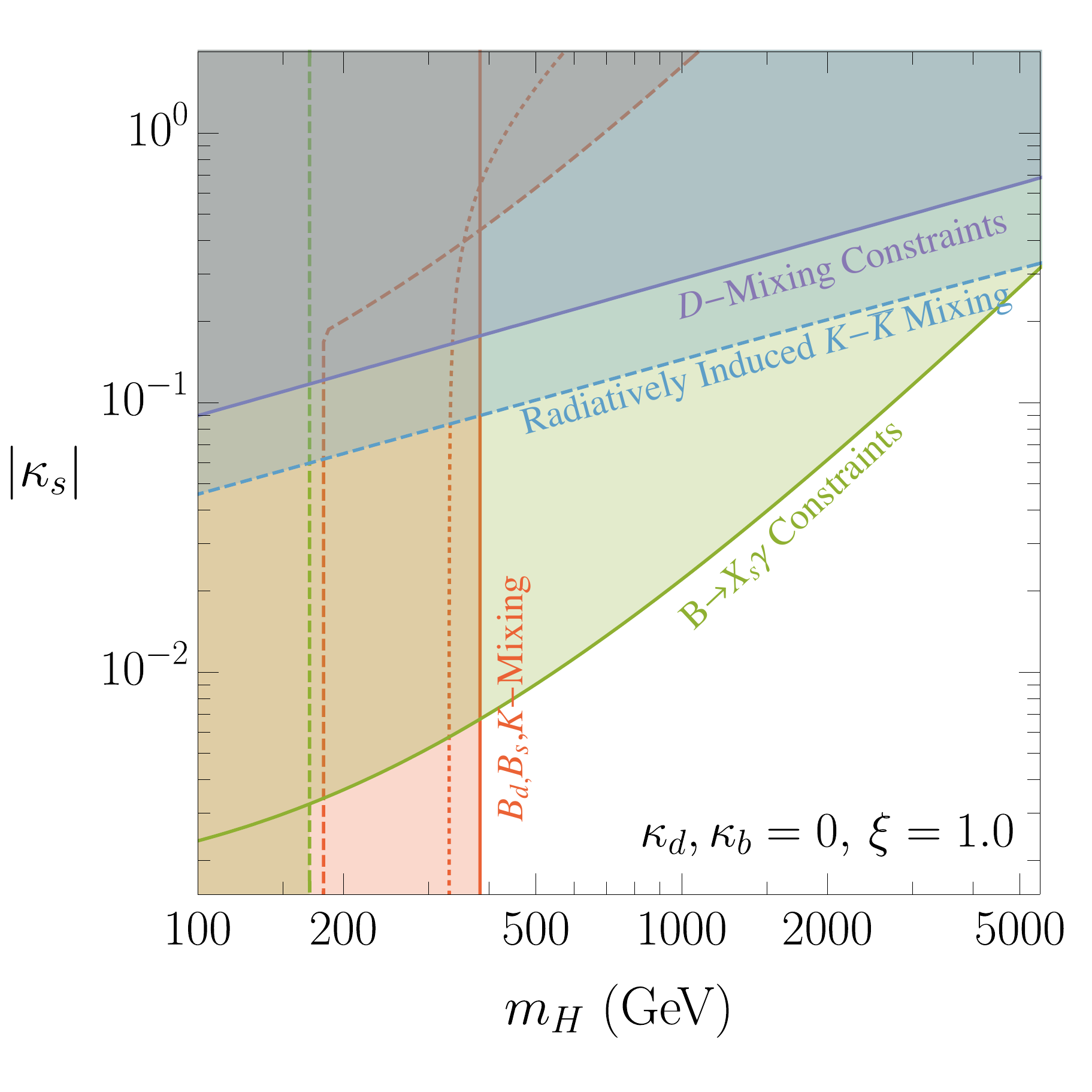} }}
\caption{
As in \figref{flavor_radiative_kappad}, but in the $\kappa_s$ vs. $m_H$ plane.}
\label{fig:flavor_radiative_kappas}
\end{figure}

Using our code, 
we find the second-doublet Yukawa matrices at the EW scale,
including their misaligned elements. 
Because of the factor $1/(y_s-y_d)$ from level repulsion, 
the largest misaligned elements are found in the first two generation Yukawas,
leading to significant effects in $K-\bar{K}$ mixing.
We find that these effects are the most constraining from all the RGE induced FCNCs. 
We present the corresponding limits from radiatively induced $K-\bar{K}$ mixing in the $\kappa_{d, s} - m_H$ in Figures \ref{fig:flavor_radiative_kappad} and \ref{fig:flavor_radiative_kappas}, 
in blue,
for both $\xi=0.1$ (left panel) and $\xi=1$ (right panel).
In the figures, 
we also show the direct constraints from charged Higgs boxes discussed in \sref{flavor}.
From the figures, 
we see that in most regions of parameter space, 
the flavor constraints from the charged Higgs boxes discussed in the body of this paper are dominant, 
with an exception being the constraints on $\kappa_d$ for $\xi=1$ and $m_H \gtrsim 500 \, \textrm{GeV}$, 
where bounds from $K-\bar{K}$ mixing dominate. 
In all cases, 
we see that the inclusion of flavor misalignment due to the RGE does not lead to constraints on our up-type SFV 2HDM much beyond the ones already discussed in \sref{flavor}.

%%%%%%%%%%%%%%%%%%%%%%%%%%%%%%%%%%%%%%%%%%%%%%%%%%%%%%%%
%%%%%%%%%%%%%%%%%%%%%%%%%%%%%%%%%%%%%%%%%%%%%%%%%%%%%%%%
%% SECTION BEGINS
%%%%%%%%%%%%%%%%%%%%%%%%%%%%%%%%%%%%%%%%%%%%%%%%%%%%%%%%
%%%%%%%%%%%%%%%%%%%%%%%%%%%%%%%%%%%%%%%%%%%%%%%%%%%%%%%%
\section{Comparison to other models}
\label{a.comparison}

In this appendix we compare the SFV 2HDM with other types of two doublet theories available in the literature. 
The different types of 2HDM and a summary of their Yukawa structure are given in table \ref{t:comparison}. 

 %%%%%%%%%%%%%%%%%%%%%%%%%
\begin{table}[htbp!]
\centering
\begin{tabular}{c | cc}
							& up-type 						& down-type \\ \hline
MFV				& polynomial of SM Yukawas 				& polynomial of SM Yukawas \\
gFC						& non-universally flavor aligned		& non-universally flavor aligned	 \\
NFC	(types I-IV)			& real proportional				& real proportional \\
Aligned 2HDM	& complex proportional 	& complex proportional  \\
up-type SFV			& real proportional				&non-universally flavor aligned	 \\
down-type SFV		& non-universally flavor aligned	 			& real proportional
\end{tabular}
\caption{Summary of the second doublet Yukawa structure for different 2HDMs. 
In each column we indicate the relation between the up- and down-type quark Yukawas for the second Higgs doublet and the SM Yukawa matrices. 
Non-universally flavor aligned stands for Yukawas that are flavor-aligned with the SM Yukawas, 
as in \eqref{flavor_basis_conditions},
without sharing the SM Yukawa hierarchies. 
Real (complex) proportional stands for proportionality to the corresponding up or down SM Yukawa matrix, with one up- and one down-type real (complex) proportionality coefficient.}
\label{t:comparison}
\end{table}
 %%%%%%%%%%%%%%%%%%%%%%%%%

\begin{itemize}
\item {\em Minimal Flavor Violating Theories}~\cite{DAmbrosio:2002vsn}:
the MFV Ansatz requires that the only spurions breaking the SM flavor group are the SM Yukawa matrices. 
This implies that, 
at leading order in an expansion in the SM Yukawa matrices,
the couplings of the second doublet to quarks are given by the SM Yukawas up to proportionality factors.
As a consequence, the second doublet couplings to all quarks maintain the SM Yukawa hierarchies. 
In MFV theories the only $CP$-violating phase at the perturbative level is the CKM phase. 
The SFV 2HDM trivially reduces to an MFV 2HDM truncated to the lowest order in the Yukawa expansion (at some boundary energy scale) when the matrices $K^{u,d}$ in \eqref{uptypeSFV} and \eqref{downtypeSFV} are proportional to the SM Yukawa coupling matrices $Y^{u,d}$.

\item {\em Aligned 2HDM}~\cite{Pich:2009sp, Pich:2010ic}: 
an extension of the MFV hypothesis in the 2HDM is the so-called Aligned 2HDM,
not to be confused with the more general idea of flavor alignment.
In this model, 
the Yukawa matrices for the second Higgs doublet are required to be proportional to the SM mass matrices at some boundary energy scale, 
but with the proportionality factor allowed to be arbitrary and complex. 
This type of 2HDM is a particular case of general MFV \cite{Kagan:2009bn} and allows for new $CP$ violating phases,
but otherwise retains the overall features of MFV. 

\item {\em Types I-IV 2HDMs}~\cite{Glashow:1976nt}: also referred to as ``Natural Flavor Conserving" (NFC) 2HDMs, or 2HDMs with Glashow-Weinberg conditions. 
These theories are obtained by imposing discrete symmetries on the two-Higgs doublets, 
which enforce proportionality of the second-doublet Yukawas with the Standard Model Yukawas.
As a consequence, 
the second-doublet couplings to up- and down-type quarks maintain the same hierarchies as in the SM, 
and the phenomenology is similar to the one of MFV models, 
with some important differences in flavor observables pointed out in \cite{Buras:2000dm}.

\item {\em General Flavor Conserving (gFC) 2HDMs}~\cite{Penuelas:2017ikk, Botella:2018gzy}: 
gFC is synonymous with flavor alignment in a 2HDM.
SFV is a subset of flavor aligned theories,
and differs from flavor alignment in its most general form in three respects. 
First, new generation-specific couplings only to \textit{either} up- \textit{or} down-type quarks are allowed in SFV,
while in generic flavor aligned theories it is possible to have generation-specific couplings to both types of quarks simultaneously.
Second, 
while in flavor aligned 2HDMs new $CP$-violating phases are allowed,
in SFV the only $CP$-violating phase at the perturbative level is the CKM phase.
Finally, 
while in SFV flavor alignment arises from a technically natural UV mechanism (see \sref{uv_completion}), 
there is no know mechanism to impose flavor alignment in its most general form, 
so these theories are usually strongly tuned.

2HDMs with flavor alignment are a particular example of Aligned Flavor Violation (AFV),
which corresponds to a systematic spurion definition of flavor alignment for generic BSM theories ~\cite{Egana-Ugrinovic:2018znw}. 

\item {\em Flavorful 2HDMs}~\cite{Das:1995df, Blechman:2010cs, Altmannshofer:2015esa, Ghosh:2015gpa, Botella:2016krk, Altmannshofer:2016zrn, Altmannshofer:2017uvs}: there are also models in the literature where the SM-like Higgs is responsible only for the masses of third-generation quarks, while the second doublet couples primarily to the first- and second-generation fermions. These models are known as ``Flavorful 2HDMs". Differently from the rest of the models in this list, such models are only free from tree level FCNCs in the first- and second-generations. 

\end{itemize}

Of all the above 2HDMs, the SFV 2HDM stands out as the only type of 2HDM which allows both for novel hierarchies in the couplings of the second doublet to the different SM quark generations \textit{and} 
is motivated by an UV completion.

%%%%%%%%%%%%%%%%%%%%%%%%%%%%%%%%%%%%%%%%%%%%%%%%%%%%%%%%
%%%%%%%%%%%%%%%%%%%%%%%%%%%%%%%%%%%%%%%%%%%%%%%%%%%%%%%%
%% SECTION BEGINS
%%%%%%%%%%%%%%%%%%%%%%%%%%%%%%%%%%%%%%%%%%%%%%%%%%%%%%%%
%%%%%%%%%%%%%%%%%%%%%%%%%%%%%%%%%%%%%%%%%%%%%%%%%%%%%%%%
\section{Physical fermion couplings to the Higgs bosons}
\label{a.couplings}

In this appendix we summarize the couplings of the Higgs bosons to the SM fermions.
In \tref{yukawaup} we present the couplings in the up-type SFV 2HDM, while in \tref{yukawadown} we present the couplings in the down-type SFV 2HDM.

\begin{table} [h!]
\begin{center}
$
\begin{array}{|c|c|c|c|}
\hline
%%% NEUTRAL SCALAR %%
\lambda_{h u_i \bar{u}_j}
& 
\delta_{ij} Y^u_i \left[ \sin(\beta - \alpha) + \xi \cos(\beta - \alpha) \right]
&
\lambda_{H u_i \bar{u}_j}
&
\delta_{ij} Y^u_i \left[ -\cos(\beta - \alpha) + \xi \sin(\beta - \alpha) \right], 
\\
\lambda_{h d_i \bar{d}_j} 
&
\delta_{ij} \left[ Y^d_i \sin(\beta - \alpha) + K^d_i \cos(\beta - \alpha) \right]
&
\lambda_{H d_i \bar{d}_j} 
&
 \delta_{ij} \left[ -Y^d_i \cos(\beta - \alpha) + K^d_i \sin(\beta - \alpha) \right]
 \\
 \lambda_{h \ell_i \bar{\ell}_j} 
&
\delta_{ij} Y^\ell_i \left[  \sin(\beta - \alpha) + \xi^\ell \cos(\beta - \alpha) \right]
&
\lambda_{H \ell_i \bar{\ell}_j} 
&
 \delta_{ij} Y^\ell_i  \left[ -\cos(\beta - \alpha) + \xi^\ell\sin(\beta - \alpha) \right]\\
%%% PSEUDOSCALAR and CHARGED %%%
\lambda_{A u_i \bar{u}_j}
&
i
\xi
\delta_{ij} 
Y^u_i
&
\lambda_{H^+ d_i \bar{u}_j}
&
-
\big[  \xi \, V^T~\! Y^u ~\! \big]_{ij}
\\
\lambda_{A d_i \bar{d}_j}
&
-i
\delta_{ij} 
K^d_i
&
\lambda_{H^- u_i \bar{d}_j}
&
\big[
V^*
K^d
\big]_{ij}
\\
\lambda_{A \ell_i \bar{\ell}_j}
&
-i
\xi^\ell \delta_{ij}  Y^\ell_i
&
\lambda_{H^- \ell_i \bar{\ell}_j}
&
\big[
\xi^\ell
Y^\ell
\big]_{ij}
\\
\hline
\end{array}
$
\end{center}
\caption{
Couplings of the physical Higgs bosons to the left-chiral fermion mass eigenstates in the up-type SFV 2HDM. 
Couplings are defined with a negative sign in the Lagrangian, e.g., $\mathcal{L} \supset - \lambda_{h f \bar{f}} h \bar{f} f$.
The couplings to the fermions with right-handed chirality are trivially obtained by hermitian conjugation.
$Y^{u,d,\ell}$ are the SM Yukawa couplings \eqref{sm_yukawas_svd}, 
$V$ is the CKM matrix \eqref{VCKM},
while $K^{d}=\mathrm{diag}( \kappa_d,\, \kappa_s,\, \kappa_b )$ are three new real Yukawas coupling the Higgs bosons
to the SM quarks with arbitrary hierarchies across generations, \eqref{uptypeSFV}.
$\xi$ and $\xi^\ell$ are free real proportionality constants, and
$\cos(\beta-\alpha)$ is the Higgs alignment parameter \eqref{alignment_angle}.
Note that neutral Higgs bosons do not have flavor off-diagonal terms so there are no tree level FCNCs,
as expected from the discussion in \ssref{SFV2HDM}.
Note also that the only source of $CP$-violation in the Higgs couplings is due to the CKM-mediated interactions of the charged Higgs.
}
\label{t:yukawaup}
\end{table}

\begin{table} [h!]
\begin{center}
$
\begin{array}{|c|c|c|c|}
\hline
%%% NEUTRAL SCALAR %%
\lambda_{h u_i \bar{u}_j}
& 
\delta_{ij} \left[ Y^u_i \sin(\beta - \alpha) + K^u_i \cos(\beta - \alpha) \right] 
&
\lambda_{H u_i \bar{u}_j}
&
\delta_{ij} \left[ - Y^u_i \cos(\beta - \alpha) + K^u_i \sin(\beta - \alpha) \right]
\\
\lambda_{h d_i \bar{d}_j} 
&
\delta_{ij} Y^d_i \left[ \sin(\beta - \alpha) + \xi \cos(\beta - \alpha) \right]
&
\lambda_{H d_i \bar{d}_j} 
&
\delta_{ij} 
Y^d_{i}
\big[
-
\cos (\beta-\alpha) 
+
\xi
\sin (\beta-\alpha) 
\big]
 \\
 \lambda_{h \ell_i \bar{\ell}_j} 
&
\delta_{ij} Y^\ell_i \left[  \sin(\beta - \alpha) + \xi^\ell \cos(\beta - \alpha) \right]
&
\lambda_{H \ell_i \bar{\ell}_j} 
&
 \delta_{ij} Y^\ell_i  \left[ -\cos(\beta - \alpha) + \xi^\ell\sin(\beta - \alpha) \right]\\
%%% PSEUDOSCALAR and CHARGED %%%
\lambda_{A u_i \bar{u}_j}
&
i
\delta_{ij} 
K^u_i
&
\lambda_{H^+ d_i \bar{u}_j}
&
-
\big[  V^T~\! K^u ~\! \big]_{ij}
\\
\lambda_{A d_i \bar{d}_j}
&
-i
\xi
\delta_{ij} 
Y^d_i
&
\lambda_{H^- u_i \bar{d}_j}
&
\big[
\xi
V^*
Y^d
\big]_{ij}
\\
\lambda_{A \ell_i \bar{\ell}_j}
&
-i
\xi^\ell \delta_{ij}  Y^\ell_i
&
\lambda_{H^- \ell_i \bar{\ell}_j}
&
\big[
\xi^\ell
Y^\ell
\big]_{ij}
\\
\hline
\end{array}
$
\end{center}
\caption{
Couplings of the physical Higgs bosons to the left-chiral fermion mass eigenstates in the down-type SFV 2HDM.
Couplings are defined with a negative sign in the Lagrangian, e.g., $\mathcal{L} \supset - \lambda_{h f \bar{f}} h \bar{f} f$. 
The couplings to the fermions with right-handed chirality are trivially obtained by hermitian conjugation.
$Y^{u,d,\ell}$ are the SM Yukawa couplings \eqref{sm_yukawas_svd}, 
$V$ is the CKM matrix \eqref{VCKM},
while $K^{u}=\mathrm{diag}( \kappa_u\, \kappa_c,\, \kappa_t )$ are three new real Yukawas coupling the Higgs bosons
to the SM up quarks with arbitrary hierarchies across generations, \eqref{downtypeSFV}.
$\xi$ and $\xi^\ell$ are free real proportionality constants.
$\cos(\beta-\alpha)$ is the Higgs alignment parameter \eqref{alignment_angle}.
Note that the leptonic couplings in the down-type SFV 2HDM in this table are the same than for the up-type SFV 2HDM in
\tref{yukawaup}: both types of 2HDMs only differ by their quark Yukawas. 
See also the notes on FCNCs and $CP$-violation in \tref{yukawaup}.
}
\label{t:yukawadown}
\end{table}

\FloatBarrier

%%%%%%%%%%%%%%%%%%%%%%%%%%%%%%%%%%%%%%%%%%%%%%%%%%%%%%%%
%%%%%%%%%%%%%%%%%%%%%%%%%%%%%%%%%%%%%%%%%%%%%%%%%%%%%%%%
%% SECTION BEGINS
%%%%%%%%%%%%%%%%%%%%%%%%%%%%%%%%%%%%%%%%%%%%%%%%%%%%%%%%
%%%%%%%%%%%%%%%%%%%%%%%%%%%%%%%%%%%%%%%%%%%%%%%%%%%%%%%%
\section{Loop functions}
\label{a.loop_functions}

Here we define the Loop functions used in our computations of $B \to X_{s,d}\, \gamma$ transitions and neutral meson mixing constraints.
The functions appearing in \eqref{bsgamma} are:
%boe$
\begin{equation}
\begin{aligned}
	C^0_{7,XY}(x) & = \frac{x}{12} \left[ \frac{ -5x^2 + 8x - 3 + (6x - 4)\log x}{(x-1)^3}\right], \\
	C^0_{8,XY}(x) & = \frac{x}{4} \left[ \frac{-x^2 + 4x - 3 - 2\log x}{(x-1)^3}\right], \\
	C^0_{7,YY}(x) & = \frac{x}{72} \left[ \frac{-8x^3 + 3x^2 + 12x - 7 + (18x^2 - 12x)\log x}{(x-1)^4}\right], \\
	C^0_{8,YY}(x) & = \frac{x}{24} \left[ \frac{-x^3 + 6 x^2 - 3x - 2  -6x \log x}{(x-1)^4}\right],
\end{aligned}
\end{equation}
%eoe%
while the box functions appearing in \eqref{SFVboxcoefficients1} and \eqref{SFVboxcoefficients2} are given by
%boe%
\begin{equation}
\begin{aligned}
D_0(m_1^2, m_2^2, m_3^2, m_4^2) & = \frac{m_1^2 \log m_1^2}{(m_4^2 - m_1^2)(m_3^2 - m_1^2) (m_2^2 - m_1^2)} \\ 
	& + ( 1 \leftrightarrow 2) + (1 \leftrightarrow 3) + (1 \leftrightarrow 4), \\
D_2(m_1^2, m_2^2, m_3^2, m_4^2) & = \frac{m_1^4 \log m_1^2}{(m_4^2 - m_1^2)(m_3^2 - m_1^2) (m_2^2 - m_1^2)} \\
	& + ( 1 \leftrightarrow 2) + (1 \leftrightarrow 3) + (1 \leftrightarrow 4). \\
\end{aligned}
\end{equation}
%eoe%
Note that our definition of $D_2$ is the same as in ref.~\cite{Crivellin:2013wna}, but differs from that in ref.~\cite{Altmannshofer:2007cs} by a factor of 4.

%%%%%%%%%%%%%%%%%%%%%%%%%%%%%%%%%%%%%%%%%%%%%%%%%%%%%%%%
%%%%%%%%%%%%%%%%%%%%%%%%%%%%%%%%%%%%%%%%%%%%%%%%%%%%%%%%

\bibliographystyle{utphys}
\bibliography{sfv_2hdm.bib}

\providecommand{\href}[2]{#2}\begingroup\raggedright\begin{thebibliography}{100}

\bibitem{DAmbrosio:2002vsn}
G.~D'Ambrosio, G.~F. Giudice, G.~Isidori, and A.~Strumia, ``{Minimal flavor
  violation: An Effective field theory approach},''
  \href{http://dx.doi.org/10.1016/S0550-3213(02)00836-2}{{\em Nucl. Phys.} {\bf
  B645} (2002)  155--187},
\href{http://arxiv.org/abs/hep-ph/0207036}{{\tt arXiv:hep-ph/0207036
  [hep-ph]}}.
%%CITATION = HEP-PH/0207036;%%.

\bibitem{Egana-Ugrinovic:2018znw}
D.~Egana-Ugrinovic, S.~Homiller, and P.~Meade, ``{Aligned and Spontaneous
  Flavor Violation},''
  \href{http://dx.doi.org/10.1103/PhysRevLett.123.031802}{{\em Phys. Rev.
  Lett.} {\bf 123} (2019) no.~3, 031802},
\href{http://arxiv.org/abs/1811.00017}{{\tt arXiv:1811.00017 [hep-ph]}}.
%%CITATION = ARXIV:1811.00017;%%.

\bibitem{Pich:2009sp}
A.~Pich and P.~Tuzon, ``{Yukawa Alignment in the Two-Higgs-Doublet Model},''
  \href{http://dx.doi.org/10.1103/PhysRevD.80.091702}{{\em Phys. Rev.} {\bf
  D80} (2009)  091702},
\href{http://arxiv.org/abs/0908.1554}{{\tt arXiv:0908.1554 [hep-ph]}}.
%%CITATION = ARXIV:0908.1554;%%.

\bibitem{Pich:2010ic}
A.~Pich, ``{Flavour constraints on multi-Higgs-doublet models: Yukawa
  alignment},'' \href{http://dx.doi.org/10.1016/j.nuclphysbps.2010.12.030}{{\em
  Nucl. Phys. Proc. Suppl.} {\bf 209} (2010)  182--187},
\href{http://arxiv.org/abs/1010.5217}{{\tt arXiv:1010.5217 [hep-ph]}}.
%%CITATION = ARXIV:1010.5217;%%.

\bibitem{Kagan:2009bn}
A.~L. Kagan, G.~Perez, T.~Volansky, and J.~Zupan, ``{General Minimal Flavor
  Violation},'' \href{http://dx.doi.org/10.1103/PhysRevD.80.076002}{{\em Phys.
  Rev.} {\bf D80} (2009)  076002},
\href{http://arxiv.org/abs/0903.1794}{{\tt arXiv:0903.1794 [hep-ph]}}.
%%CITATION = ARXIV:0903.1794;%%.

\bibitem{Nir:1993mx}
Y.~Nir and N.~Seiberg, ``{Should squarks be degenerate?},''
  \href{http://dx.doi.org/10.1016/0370-2693(93)90942-B}{{\em Phys. Lett.} {\bf
  B309} (1993)  337--343},
\href{http://arxiv.org/abs/hep-ph/9304307}{{\tt arXiv:hep-ph/9304307
  [hep-ph]}}.
%%CITATION = HEP-PH/9304307;%%.

\bibitem{Duarte-Campderros:2018ouv}
J.~Duarte-Campderros, G.~Perez, M.~Schlaffer, and A.~Soffer, ``{Probing the
  strange Higgs coupling at lepton colliders using light-jet flavor tagging},''
\href{http://arxiv.org/abs/1811.09636}{{\tt arXiv:1811.09636 [hep-ph]}}.
%%CITATION = ARXIV:1811.09636;%%.

\bibitem{Fraser:2018ieu}
K.~Fraser and M.~D. Schwartz, ``{Jet Charge and Machine Learning},''
  \href{http://dx.doi.org/10.1007/JHEP10(2018)093}{{\em JHEP} {\bf 10} (2018)
  093},
\href{http://arxiv.org/abs/1803.08066}{{\tt arXiv:1803.08066 [hep-ph]}}.
%%CITATION = ARXIV:1803.08066;%%.

\bibitem{Gatto:1978dy}
R.~Gatto, G.~Morchio, and F.~Strocchi, ``{Natural Flavor Conservation in the
  Neutral Currents and the Determination of the Cabibbo Angle},''
\href{http://dx.doi.org/10.1016/0370-2693(79)90213-2}{{\em Phys. Lett.} {\bf
  80B} (1979)  265--268}.
%%CITATION = PHLTA,80B,265;%%.

\bibitem{Gatto:1979mr}
R.~Gatto, G.~Morchio, G.~Sartori, and F.~Strocchi, ``{Natural Flavor
  Conservation in Higgs Induced Neutral Currents and the Quark Mixing
  Angles},''
\href{http://dx.doi.org/10.1016/0550-3213(80)90399-5}{{\em Nucl. Phys.} {\bf
  B163} (1980)  221--253}.
%%CITATION = NUPHA,B163,221;%%.

\bibitem{Sartori:1979gt}
G.~Sartori, ``{Discrete Symmetries, Natural Flavor Conservation and Weak Mixing
  Angles},''
\href{http://dx.doi.org/10.1016/0370-2693(79)90749-4}{{\em Phys. Lett.} {\bf
  82B} (1979)  255--259}.
%%CITATION = PHLTA,82B,255;%%.

\bibitem{Penuelas:2017ikk}
A.~Pe\~nuelas and A.~Pich, ``{Flavour alignment in multi-Higgs-doublet
  models},'' \href{http://dx.doi.org/10.1007/JHEP12(2017)084}{{\em JHEP} {\bf
  12} (2017)  084},
\href{http://arxiv.org/abs/1710.02040}{{\tt arXiv:1710.02040 [hep-ph]}}.
%%CITATION = ARXIV:1710.02040;%%.

\bibitem{Botella:2018gzy}
F.~J. Botella, F.~Cornet-Gomez, and M.~Nebot, ``{Flavour Conservation in Two
  Higgs Doublet Models},''
\href{http://arxiv.org/abs/1803.08521}{{\tt arXiv:1803.08521 [hep-ph]}}.
%%CITATION = ARXIV:1803.08521;%%.

\bibitem{Rodejohann:2019izm}
W.~Rodejohann and U.~Salda\~na Salazar, ``{Multi-Higgs-Doublet Models and
  Singular Alignment},'' \href{http://dx.doi.org/10.1007/JHEP07(2019)036}{{\em
  JHEP} {\bf 07} (2019)  036},
\href{http://arxiv.org/abs/1903.00983}{{\tt arXiv:1903.00983 [hep-ph]}}.
%%CITATION = ARXIV:1903.00983;%%.

\bibitem{Bodwin:2013gca}
G.~T. Bodwin, F.~Petriello, S.~Stoynev, and M.~Velasco, ``{Higgs boson decays
  to quarkonia and the $H\bar{c}c$ coupling},''
  \href{http://dx.doi.org/10.1103/PhysRevD.88.053003}{{\em Phys. Rev.} {\bf
  D88} (2013) no.~5, 053003},
\href{http://arxiv.org/abs/1306.5770}{{\tt arXiv:1306.5770 [hep-ph]}}.
%%CITATION = ARXIV:1306.5770;%%.

\bibitem{Kagan:2014ila}
A.~L. Kagan, G.~Perez, F.~Petriello, Y.~Soreq, S.~Stoynev, and J.~Zupan,
  ``{Exclusive Window onto Higgs Yukawa Couplings},''
  \href{http://dx.doi.org/10.1103/PhysRevLett.114.101802}{{\em Phys. Rev.
  Lett.} {\bf 114} (2015) no.~10, 101802},
\href{http://arxiv.org/abs/1406.1722}{{\tt arXiv:1406.1722 [hep-ph]}}.
%%CITATION = ARXIV:1406.1722;%%.

\bibitem{Perez:2015lra}
G.~Perez, Y.~Soreq, E.~Stamou, and K.~Tobioka, ``{Prospects for measuring the
  Higgs boson coupling to light quarks},''
  \href{http://dx.doi.org/10.1103/PhysRevD.93.013001}{{\em Phys. Rev.} {\bf
  D93} (2016) no.~1, 013001},
\href{http://arxiv.org/abs/1505.06689}{{\tt arXiv:1505.06689 [hep-ph]}}.
%%CITATION = ARXIV:1505.06689;%%.

\bibitem{Zhou:2015wra}
Y.~Zhou, ``{Constraining the Higgs boson coupling to light quarks in the H?ZZ
  final states},'' \href{http://dx.doi.org/10.1103/PhysRevD.93.013019}{{\em
  Phys. Rev.} {\bf D93} (2016) no.~1, 013019},
\href{http://arxiv.org/abs/1505.06369}{{\tt arXiv:1505.06369 [hep-ph]}}.
%%CITATION = ARXIV:1505.06369;%%.

\bibitem{Brivio:2015fxa}
I.~Brivio, F.~Goertz, and G.~Isidori, ``{Probing the Charm Quark Yukawa
  Coupling in Higgs+Charm Production},''
  \href{http://dx.doi.org/10.1103/PhysRevLett.115.211801}{{\em Phys. Rev.
  Lett.} {\bf 115} (2015) no.~21, 211801},
\href{http://arxiv.org/abs/1507.02916}{{\tt arXiv:1507.02916 [hep-ph]}}.
%%CITATION = ARXIV:1507.02916;%%.

\bibitem{Delaunay:2016brc}
C.~Delaunay, R.~Ozeri, G.~Perez, and Y.~Soreq, ``{Probing Atomic Higgs-like
  Forces at the Precision Frontier},''
  \href{http://dx.doi.org/10.1103/PhysRevD.96.093001}{{\em Phys. Rev.} {\bf
  D96} (2017) no.~9, 093001},
\href{http://arxiv.org/abs/1601.05087}{{\tt arXiv:1601.05087 [hep-ph]}}.
%%CITATION = ARXIV:1601.05087;%%.

\bibitem{Bishara:2016jga}
F.~Bishara, U.~Haisch, P.~F. Monni, and E.~Re, ``{Constraining Light-Quark
  Yukawa Couplings from Higgs Distributions},''
  \href{http://dx.doi.org/10.1103/PhysRevLett.118.121801}{{\em Phys. Rev.
  Lett.} {\bf 118} (2017) no.~12, 121801},
\href{http://arxiv.org/abs/1606.09253}{{\tt arXiv:1606.09253 [hep-ph]}}.
%%CITATION = ARXIV:1606.09253;%%.

\bibitem{Soreq:2016rae}
Y.~Soreq, H.~X. Zhu, and J.~Zupan, ``{Light quark Yukawa couplings from Higgs
  kinematics},'' \href{http://dx.doi.org/10.1007/JHEP12(2016)045}{{\em JHEP}
  {\bf 12} (2016)  045},
\href{http://arxiv.org/abs/1606.09621}{{\tt arXiv:1606.09621 [hep-ph]}}.
%%CITATION = ARXIV:1606.09621;%%.

\bibitem{Yu:2016rvv}
F.~Yu, ``{Phenomenology of Enhanced Light Quark Yukawa Couplings and the $W^\pm
  h$ Charge Asymmetry},'' \href{http://dx.doi.org/10.1007/JHEP02(2017)083}{{\em
  JHEP} {\bf 02} (2017)  083},
\href{http://arxiv.org/abs/1609.06592}{{\tt arXiv:1609.06592 [hep-ph]}}.
%%CITATION = ARXIV:1609.06592;%%.

\bibitem{Aaboud:2017xnb}
{\bf ATLAS} Collaboration, M.~Aaboud {\em et al.}, ``{Search for exclusive
  Higgs and $Z$ boson decays to $\phi\gamma$ and $\rho\gamma$ with the ATLAS
  detector},'' \href{http://dx.doi.org/10.1007/JHEP07(2018)127}{{\em JHEP} {\bf
  07} (2018)  127},
\href{http://arxiv.org/abs/1712.02758}{{\tt arXiv:1712.02758 [hep-ex]}}.
%%CITATION = ARXIV:1712.02758;%%.

\bibitem{Alves:2017avw}
D.~S.~M. Alves and N.~Weiner, ``{A viable QCD axion in the MeV mass range},''
  \href{http://dx.doi.org/10.1007/JHEP07(2018)092}{{\em JHEP} {\bf 07} (2018)
  092},
\href{http://arxiv.org/abs/1710.03764}{{\tt arXiv:1710.03764 [hep-ph]}}.
%%CITATION = ARXIV:1710.03764;%%.

\bibitem{Coyle:2019hvs}
N.~M. Coyle, C.~E.~M. Wagner, and V.~Wei, ``{Bounding the Charm Yukawa},''
\href{http://arxiv.org/abs/1905.09360}{{\tt arXiv:1905.09360 [hep-ph]}}.
%%CITATION = ARXIV:1905.09360;%%.

\bibitem{Ferreira:2004yd}
P.~M. Ferreira, R.~Santos, and A.~Barroso, ``{Stability of the tree-level
  vacuum in two Higgs doublet models against charge or CP spontaneous
  violation},'' \href{http://dx.doi.org/10.1016/j.physletb.2004.10.022,
  10.1016/j.physletb.2005.09.074}{{\em Phys. Lett.} {\bf B603} (2004)
  219--229}, \href{http://arxiv.org/abs/hep-ph/0406231}{{\tt
  arXiv:hep-ph/0406231 [hep-ph]}}.
[Erratum: Phys. Lett.B629,114(2005)].
%%CITATION = HEP-PH/0406231;%%.

\bibitem{Georgi:1978ri}
H.~Georgi and D.~V. Nanopoulos, ``{Suppression of Flavor Changing Effects From
  Neutral Spinless Meson Exchange in Gauge Theories},''
\href{http://dx.doi.org/10.1016/0370-2693(79)90433-7}{{\em Phys. Lett.} {\bf
  82B} (1979)  95--96}.
%%CITATION = PHLTA,82B,95;%%.

\bibitem{Botella:1994cs}
F.~J. Botella and J.~P. Silva, ``{Jarlskog - like invariants for theories with
  scalars and fermions},''
  \href{http://dx.doi.org/10.1103/PhysRevD.51.3870}{{\em Phys. Rev.} {\bf D51}
  (1995)  3870--3875},
\href{http://arxiv.org/abs/hep-ph/9411288}{{\tt arXiv:hep-ph/9411288
  [hep-ph]}}.
%%CITATION = HEP-PH/9411288;%%.

\bibitem{Egana-Ugrinovic:2015vgy}
D.~Egana-Ugrinovic and S.~Thomas, ``{Effective Theory of Higgs Sector Vacuum
  States},''
\href{http://arxiv.org/abs/1512.00144}{{\tt arXiv:1512.00144 [hep-ph]}}.
%%CITATION = ARXIV:1512.00144;%%.

\bibitem{Grimus:1986mh}
W.~Grimus and G.~Ecker, ``{On the Simultaneous Diagonalizability of
  Matrices},''
\href{http://dx.doi.org/10.1088/0305-4470/19/18/036}{{\em J. Phys.} {\bf A19}
  (1986)  3917}.
%%CITATION = JPAGA,A19,3917;%%.

\bibitem{Glashow:1976nt}
S.~L. Glashow and S.~Weinberg, ``{Natural Conservation Laws for Neutral
  Currents},''
\href{http://dx.doi.org/10.1103/PhysRevD.15.1958}{{\em Phys. Rev.} {\bf D15}
  (1977)  1958}.
%%CITATION = PHRVA,D15,1958;%%.

\bibitem{Gunion:2002zf}
J.~F. Gunion and H.~E. Haber, ``{The CP conserving two Higgs doublet model: The
  Approach to the decoupling limit},''
  \href{http://dx.doi.org/10.1103/PhysRevD.67.075019}{{\em Phys. Rev.} {\bf
  D67} (2003)  075019},
\href{http://arxiv.org/abs/hep-ph/0207010}{{\tt arXiv:hep-ph/0207010
  [hep-ph]}}.
%%CITATION = HEP-PH/0207010;%%.

\bibitem{Crivellin:2013wna}
A.~Crivellin, A.~Kokulu, and C.~Greub, ``{Flavor-phenomenology of
  two-Higgs-doublet models with generic Yukawa structure},''
  \href{http://dx.doi.org/10.1103/PhysRevD.87.094031}{{\em Phys. Rev.} {\bf
  D87} (2013) no.~9, 094031},
\href{http://arxiv.org/abs/1303.5877}{{\tt arXiv:1303.5877 [hep-ph]}}.
%%CITATION = ARXIV:1303.5877;%%.

\bibitem{Craig:2013hca}
N.~Craig, J.~Galloway, and S.~Thomas, ``{Searching for Signs of the Second
  Higgs Doublet},''
\href{http://arxiv.org/abs/1305.2424}{{\tt arXiv:1305.2424 [hep-ph]}}.
%%CITATION = ARXIV:1305.2424;%%.

\bibitem{Hou:1987kf}
W.-S. Hou and R.~S. Willey, ``{Effects of Charged Higgs Bosons on the Processes
  $b \to s\gamma$, $b\to s g^*$, and $b\to s \ell^+\ell^-$},''
\href{http://dx.doi.org/10.1016/0370-2693(88)91870-9}{{\em Phys. Lett.} {\bf
  B202} (1988)  591--595}.
%%CITATION = PHLTA,B202,591;%%.

\bibitem{Borzumati:1998tg}
F.~Borzumati and C.~Greub, ``{2HDMs predictions for $\bar{B} \to X_s \gamma$ in
  NLO QCD},'' \href{http://dx.doi.org/10.1103/PhysRevD.58.074004}{{\em Phys.
  Rev.} {\bf D58} (1998)  074004},
\href{http://arxiv.org/abs/hep-ph/9802391}{{\tt arXiv:hep-ph/9802391
  [hep-ph]}}.
%%CITATION = HEP-PH/9802391;%%.

\bibitem{Crivellin:2011ba}
A.~Crivellin and L.~Mercolli, ``{$B \to X_d \gamma$ and constraints on new
  physics},'' \href{http://dx.doi.org/10.1103/PhysRevD.84.114005}{{\em Phys.
  Rev.} {\bf D84} (2011)  114005},
\href{http://arxiv.org/abs/1106.5499}{{\tt arXiv:1106.5499 [hep-ph]}}.
%%CITATION = ARXIV:1106.5499;%%.

\bibitem{Capdevila:2017bsm}
B.~Capdevila, A.~Crivellin, S.~Descotes-Genon, J.~Matias, and J.~Virto,
  ``{Patterns of New Physics in $b\to s\ell^+\ell^-$ transitions in the light
  of recent data},'' \href{http://dx.doi.org/10.1007/JHEP01(2018)093}{{\em
  JHEP} {\bf 01} (2018)  093},
\href{http://arxiv.org/abs/1704.05340}{{\tt arXiv:1704.05340 [hep-ph]}}.
%%CITATION = ARXIV:1704.05340;%%.

\bibitem{deBoer:2017que}
S.~de~Boer and G.~Hiller, ``{Rare radiative charm decays within the standard
  model and beyond},'' \href{http://dx.doi.org/10.1007/JHEP08(2017)091}{{\em
  JHEP} {\bf 08} (2017)  091},
\href{http://arxiv.org/abs/1701.06392}{{\tt arXiv:1701.06392 [hep-ph]}}.
%%CITATION = ARXIV:1701.06392;%%.

\bibitem{Cerri:2018ypt}
A.~Cerri {\em et al.}, ``{Opportunities in Flavour Physics at the HL-LHC and
  HE-LHC},''
\href{http://arxiv.org/abs/1812.07638}{{\tt arXiv:1812.07638 [hep-ph]}}.
%%CITATION = ARXIV:1812.07638;%%.

\bibitem{Bona:2007vi}
{\bf UTfit} Collaboration, M.~Bona {\em et al.}, ``{Model-independent
  constraints on $\Delta F=2$ operators and the scale of new physics},''
  \href{http://dx.doi.org/10.1088/1126-6708/2008/03/049}{{\em JHEP} {\bf 03}
  (2008)  049},
\href{http://arxiv.org/abs/0707.0636}{{\tt arXiv:0707.0636 [hep-ph]}}.
%%CITATION = ARXIV:0707.0636;%%.

\bibitem{Bona:2016bvr}
{\bf UTfit} Collaboration, M.~Bona, ``{Unitarity Triangle analysis beyond the
  Standard Model from UTfit},''
{\em PoS} {\bf ICHEP2016} (2016)  149.
%%CITATION = POSCI,ICHEP2016,149;%%.

\bibitem{Aaij:2019jot}
{\bf LHCb} Collaboration, R.~Aaij {\em et al.}, ``{Measurement of the mass
  difference between neutral charm-meson eigenstates},''
\href{http://arxiv.org/abs/1903.03074}{{\tt arXiv:1903.03074 [hep-ex]}}.
%%CITATION = ARXIV:1903.03074;%%.

\bibitem{Bevan:2014tha}
{\bf UTfit} Collaboration, A.~J. Bevan {\em et al.}, ``{The UTfit collaboration
  average of D meson mixing data: Winter 2014},''
  \href{http://dx.doi.org/10.1007/JHEP03(2014)123}{{\em JHEP} {\bf 03} (2014)
  123},
\href{http://arxiv.org/abs/1402.1664}{{\tt arXiv:1402.1664 [hep-ph]}}.
%%CITATION = ARXIV:1402.1664;%%.

\bibitem{Bona:2017gut}
{\bf Utfit} Collaboration, M.~Bona and L.~Silvestrini, ``{Unitarity Triangle
  Analysis and D meson mixing in the Standard Model and Beyond},''
\href{http://dx.doi.org/10.22323/1.314.0205}{{\em PoS} {\bf EPS-HEP2017} (2017)
   205}.
%%CITATION = POSCI,EPS-HEP2017,205;%%.

\bibitem{Buras:2001mb}
A.~J. Buras, P.~H. Chankowski, J.~Rosiek, and L.~Slawianowska, ``{$\Delta$
  M($s$) / $\Delta$ M($d$), $\sin$ 2 Beta and the angle $\gamma$ in the
  presence of new $\Delta F=2$ operators},''
  \href{http://dx.doi.org/10.1016/S0550-3213(01)00517-X}{{\em Nucl. Phys.} {\bf
  B619} (2001)  434--466},
\href{http://arxiv.org/abs/hep-ph/0107048}{{\tt arXiv:hep-ph/0107048
  [hep-ph]}}.
%%CITATION = HEP-PH/0107048;%%.

\bibitem{Altmannshofer:2007cs}
W.~Altmannshofer, A.~J. Buras, and D.~Guadagnoli, ``{The MFV limit of the MSSM
  for low tan(beta): Meson mixings revisited},''
  \href{http://dx.doi.org/10.1088/1126-6708/2007/11/065}{{\em JHEP} {\bf 11}
  (2007)  065},
\href{http://arxiv.org/abs/hep-ph/0703200}{{\tt arXiv:hep-ph/0703200
  [hep-ph]}}.
%%CITATION = HEP-PH/0703200;%%.

\bibitem{Lenz:2010gu}
A.~Lenz, U.~Nierste, J.~Charles, S.~Descotes-Genon, A.~Jantsch, C.~Kaufhold,
  H.~Lacker, S.~Monteil, V.~Niess, and S.~T'Jampens, ``{Anatomy of New Physics
  in $B - \bar{B}$ mixing},''
  \href{http://dx.doi.org/10.1103/PhysRevD.83.036004}{{\em Phys. Rev.} {\bf
  D83} (2011)  036004},
\href{http://arxiv.org/abs/1008.1593}{{\tt arXiv:1008.1593 [hep-ph]}}.
%%CITATION = ARXIV:1008.1593;%%.

\bibitem{Becirevic:2001jj}
D.~Becirevic, M.~Ciuchini, E.~Franco, V.~Gimenez, G.~Martinelli, A.~Masiero,
  M.~Papinutto, J.~Reyes, and L.~Silvestrini, ``{$B_d - \bar{B}_d$ mixing and
  the $B_d \to J/\psi K_s$ asymmetry in general SUSY models},''
  \href{http://dx.doi.org/10.1016/S0550-3213(02)00291-2}{{\em Nucl. Phys.} {\bf
  B634} (2002)  105--119},
\href{http://arxiv.org/abs/hep-ph/0112303}{{\tt arXiv:hep-ph/0112303
  [hep-ph]}}.
%%CITATION = HEP-PH/0112303;%%.

\bibitem{Kagan:2009gb}
A.~L. Kagan and M.~D. Sokoloff, ``{On Indirect CP Violation and Implications
  for D0 - anti-D0 and B(s) - anti-B(s) mixing},''
  \href{http://dx.doi.org/10.1103/PhysRevD.80.076008}{{\em Phys. Rev.} {\bf
  D80} (2009)  076008},
\href{http://arxiv.org/abs/0907.3917}{{\tt arXiv:0907.3917 [hep-ph]}}.
%%CITATION = ARXIV:0907.3917;%%.

\bibitem{Bazavov:2017weg}
A.~Bazavov {\em et al.}, ``{Short-distance matrix elements for $D^0$-meson
  mixing for $N_f=2+1$ lattice QCD},''
  \href{http://dx.doi.org/10.1103/PhysRevD.97.034513}{{\em Phys. Rev.} {\bf
  D97} (2018) no.~3, 034513},
\href{http://arxiv.org/abs/1706.04622}{{\tt arXiv:1706.04622 [hep-lat]}}.
%%CITATION = ARXIV:1706.04622;%%.

\bibitem{Kou:2018nap}
{\bf Belle II} Collaboration, E.~Kou {\em et al.}, ``{The Belle II Physics
  Book},''
\href{http://arxiv.org/abs/1808.10567}{{\tt arXiv:1808.10567 [hep-ex]}}.
%%CITATION = ARXIV:1808.10567;%%.

\bibitem{DiazCruz:1992gg}
J.~L. Diaz-Cruz and O.~A. Sampayo, ``{Contribution of gluon fusion to the
  production of charged Higgs at hadron colliders},''
\href{http://dx.doi.org/10.1103/PhysRevD.50.6820}{{\em Phys. Rev.} {\bf D50}
  (1994)  6820--6823}.
%%CITATION = PHRVA,D50,6820;%%.

\bibitem{Alwall:2004xw}
J.~Alwall and J.~Rathsman, ``{Improved description of charged Higgs boson
  production at hadron colliders},''
  \href{http://dx.doi.org/10.1088/1126-6708/2004/12/050}{{\em JHEP} {\bf 12}
  (2004)  050},
\href{http://arxiv.org/abs/hep-ph/0409094}{{\tt arXiv:hep-ph/0409094
  [hep-ph]}}.
%%CITATION = HEP-PH/0409094;%%.

\bibitem{Craig:2015jba}
N.~Craig, F.~D'Eramo, P.~Draper, S.~Thomas, and H.~Zhang, ``{The Hunt for the
  Rest of the Higgs Bosons},''
  \href{http://dx.doi.org/10.1007/JHEP06(2015)137}{{\em JHEP} {\bf 06} (2015)
  137},
\href{http://arxiv.org/abs/1504.04630}{{\tt arXiv:1504.04630 [hep-ph]}}.
%%CITATION = ARXIV:1504.04630;%%.

\bibitem{Plehn:2002vy}
T.~Plehn, ``{Charged Higgs boson production in bottom gluon fusion},''
  \href{http://dx.doi.org/10.1103/PhysRevD.67.014018}{{\em Phys. Rev.} {\bf
  D67} (2003)  014018},
\href{http://arxiv.org/abs/hep-ph/0206121}{{\tt arXiv:hep-ph/0206121
  [hep-ph]}}.
%%CITATION = HEP-PH/0206121;%%.

\bibitem{Berger:2003sm}
E.~L. Berger, T.~Han, J.~Jiang, and T.~Plehn, ``{Associated production of a top
  quark and a charged Higgs boson},''
  \href{http://dx.doi.org/10.1103/PhysRevD.71.115012}{{\em Phys. Rev.} {\bf
  D71} (2005)  115012},
\href{http://arxiv.org/abs/hep-ph/0312286}{{\tt arXiv:hep-ph/0312286
  [hep-ph]}}.
%%CITATION = HEP-PH/0312286;%%.

\bibitem{Gunion:1986pe}
J.~F. Gunion, H.~E. Haber, F.~E. Paige, W.-K. Tung, and S.~S.~D. Willenbrock,
  ``{Neutral and Charged Higgs Detection: Heavy Quark Fusion, Top Quark Mass
  Dependence and Rare Decays},''
\href{http://dx.doi.org/10.1016/0550-3213(87)90600-6}{{\em Nucl. Phys.} {\bf
  B294} (1987)  621}.
%%CITATION = NUPHA,B294,621;%%.

\bibitem{Moretti:1996ra}
S.~Moretti and K.~Odagiri, ``{Production of charged Higgs bosons of the minimal
  supersymmetric standard model in b quark initiated processes at the large
  hadron collider},'' \href{http://dx.doi.org/10.1103/PhysRevD.55.5627}{{\em
  Phys. Rev.} {\bf D55} (1997)  5627--5635},
\href{http://arxiv.org/abs/hep-ph/9611374}{{\tt arXiv:hep-ph/9611374
  [hep-ph]}}.
%%CITATION = HEP-PH/9611374;%%.

\bibitem{He:1998ie}
H.-J. He and C.~P. Yuan, ``{New Method for Detecting Charged Scalars at
  Colliders},'' \href{http://dx.doi.org/10.1103/PhysRevLett.83.28}{{\em Phys.
  Rev. Lett.} {\bf 83} (1999)  28--31},
\href{http://arxiv.org/abs/hep-ph/9810367}{{\tt arXiv:hep-ph/9810367
  [hep-ph]}}.
%%CITATION = HEP-PH/9810367;%%.

\bibitem{Dittmaier:2007uw}
S.~Dittmaier, G.~Hiller, T.~Plehn, and M.~Spannowsky, ``{Charged-Higgs Collider
  Signals with or without Flavor},''
  \href{http://dx.doi.org/10.1103/PhysRevD.77.115001}{{\em Phys. Rev.} {\bf
  D77} (2008)  115001},
\href{http://arxiv.org/abs/0708.0940}{{\tt arXiv:0708.0940 [hep-ph]}}.
%%CITATION = ARXIV:0708.0940;%%.

\bibitem{Davidson:2005cw}
S.~Davidson and H.~E. Haber, ``{Basis-independent methods for the
  two-Higgs-doublet model},''
  \href{http://dx.doi.org/10.1103/PhysRevD.72.099902,
  10.1103/PhysRevD.72.035004}{{\em Phys. Rev.} {\bf D72} (2005)  035004},
  \href{http://arxiv.org/abs/hep-ph/0504050}{{\tt arXiv:hep-ph/0504050
  [hep-ph]}}.
[Erratum: Phys. Rev.D72,099902(2005)].
%%CITATION = HEP-PH/0504050;%%.

\bibitem{Albajar:1988rs}
{\bf UA1} Collaboration, C.~Albajar {\em et al.}, ``{Two Jet Mass Distributions
  at the CERN Proton - Anti-Proton Collider},''
\href{http://dx.doi.org/10.1016/0370-2693(88)91843-6}{{\em Phys. Lett.} {\bf
  B209} (1988)  127--134}.
%%CITATION = PHLTA,B209,127;%%.

\bibitem{Alitti:1990kw}
{\bf UA2} Collaboration, J.~Alitti {\em et al.}, ``{A Measurement of two jet
  decays of the $W$ and $Z$ bosons at the CERN $\bar{p} p$ collider},''
\href{http://dx.doi.org/10.1007/BF01570793}{{\em Z. Phys.} {\bf C49} (1991)
  17--28}.
%%CITATION = ZEPYA,C49,17;%%.

\bibitem{Alitti:1993pn}
{\bf UA2} Collaboration, J.~Alitti {\em et al.}, ``{A Search for new
  intermediate vector mesons and excited quarks decaying to two jets at the
  CERN $\bar{p} p$ collider},''
\href{http://dx.doi.org/10.1016/0550-3213(93)90395-6}{{\em Nucl. Phys.} {\bf
  B400} (1993)  3--24}.
%%CITATION = NUPHA,B400,3;%%.

\bibitem{Abe:1989gz}
{\bf CDF} Collaboration, F.~Abe {\em et al.}, ``{The Two jet invariant mass
  distribution at $\sqrt{s} = 1.8$ TeV},''
\href{http://dx.doi.org/10.1103/PhysRevD.41.1722}{{\em Phys. Rev.} {\bf D41}
  (1990)  1722--1725}.
%%CITATION = PHRVA,D41,1722;%%.

\bibitem{Abe:1993it}
{\bf CDF} Collaboration, F.~Abe {\em et al.}, ``{Search for quark
  compositeness, axigluons and heavy particles using the dijet invariant mass
  spectrum observed in $p\bar{p}$ collisions},''
\href{http://dx.doi.org/10.1103/PhysRevLett.71.2542}{{\em Phys. Rev. Lett.}
  {\bf 71} (1993)  2542--2546}.
%%CITATION = PRLTA,71,2542;%%.

\bibitem{Abe:1995jz}
{\bf CDF} Collaboration, F.~Abe {\em et al.}, ``{Search for new particles
  decaying to dijets in $p\bar{p}$ collisions at $\sqrt{s} = 1.8$ TeV},''
  \href{http://dx.doi.org/10.1103/PhysRevLett.74.3538}{{\em Phys. Rev. Lett.}
  {\bf 74} (1995)  3538--3543},
\href{http://arxiv.org/abs/hep-ex/9501001}{{\tt arXiv:hep-ex/9501001
  [hep-ex]}}.
%%CITATION = HEP-EX/9501001;%%.

\bibitem{Abe:1997hm}
{\bf CDF} Collaboration, F.~Abe {\em et al.}, ``{Search for new particles
  decaying to dijets at CDF},''
  \href{http://dx.doi.org/10.1103/PhysRevD.55.R5263}{{\em Phys. Rev.} {\bf D55}
  (1997)  R5263--R5268},
\href{http://arxiv.org/abs/hep-ex/9702004}{{\tt arXiv:hep-ex/9702004
  [hep-ex]}}.
%%CITATION = HEP-EX/9702004;%%.

\bibitem{Abazov:2003tj}
{\bf D0} Collaboration, V.~M. Abazov {\em et al.}, ``{Search for new particles
  in the two jet decay channel with the D0 detector},''
  \href{http://dx.doi.org/10.1103/PhysRevD.69.111101}{{\em Phys. Rev.} {\bf
  D69} (2004)  111101},
\href{http://arxiv.org/abs/hep-ex/0308033}{{\tt arXiv:hep-ex/0308033
  [hep-ex]}}.
%%CITATION = HEP-EX/0308033;%%.

\bibitem{Aaltonen:2008dn}
{\bf CDF} Collaboration, T.~Aaltonen {\em et al.}, ``{Search for new particles
  decaying into dijets in proton-antiproton collisions at s**(1/2) =
  1.96-TeV},'' \href{http://dx.doi.org/10.1103/PhysRevD.79.112002}{{\em Phys.
  Rev.} {\bf D79} (2009)  112002},
\href{http://arxiv.org/abs/0812.4036}{{\tt arXiv:0812.4036 [hep-ex]}}.
%%CITATION = ARXIV:0812.4036;%%.

\bibitem{Aad:2010ae}
{\bf ATLAS} Collaboration, G.~Aad {\em et al.}, ``{Search for New Particles in
  Two-Jet Final States in 7 TeV Proton-Proton Collisions with the ATLAS
  Detector at the LHC},''
  \href{http://dx.doi.org/10.1103/PhysRevLett.105.161801}{{\em Phys. Rev.
  Lett.} {\bf 105} (2010)  161801},
\href{http://arxiv.org/abs/1008.2461}{{\tt arXiv:1008.2461 [hep-ex]}}.
%%CITATION = ARXIV:1008.2461;%%.

\bibitem{Aad:2011aj}
{\bf ATLAS} Collaboration, G.~Aad {\em et al.}, ``{Search for New Physics in
  Dijet Mass and Angular Distributions in pp Collisions at $\sqrt{s} = 7$ TeV
  Measured with the ATLAS Detector},''
  \href{http://dx.doi.org/10.1088/1367-2630/13/5/053044}{{\em New J. Phys.}
  {\bf 13} (2011)  053044},
\href{http://arxiv.org/abs/1103.3864}{{\tt arXiv:1103.3864 [hep-ex]}}.
%%CITATION = ARXIV:1103.3864;%%.

\bibitem{Aad:2011fq}
{\bf ATLAS} Collaboration, G.~Aad {\em et al.}, ``{Search for New Physics in
  the Dijet Mass Distribution using 1 fb$^{-1}$ of $pp$ Collision Data at
  $\sqrt{s}=$7 TeV collected by the ATLAS Detector},''
  \href{http://dx.doi.org/10.1016/j.physletb.2012.01.035}{{\em Phys. Lett.}
  {\bf B708} (2012)  37--54},
\href{http://arxiv.org/abs/1108.6311}{{\tt arXiv:1108.6311 [hep-ex]}}.
%%CITATION = ARXIV:1108.6311;%%.

\bibitem{ATLAS:2012pu}
{\bf ATLAS} Collaboration, G.~Aad {\em et al.}, ``{ATLAS search for new
  phenomena in dijet mass and angular distributions using $pp$ collisions at
  $\sqrt{s}=7$ TeV},'' \href{http://dx.doi.org/10.1007/JHEP01(2013)029}{{\em
  JHEP} {\bf 01} (2013)  029},
\href{http://arxiv.org/abs/1210.1718}{{\tt arXiv:1210.1718 [hep-ex]}}.
%%CITATION = ARXIV:1210.1718;%%.

\bibitem{Khachatryan:2010jd}
{\bf CMS} Collaboration, V.~Khachatryan {\em et al.}, ``{Search for Dijet
  Resonances in 7 TeV pp Collisions at CMS},''
  \href{http://dx.doi.org/10.1103/PhysRevLett.105.211801,
  10.1103/PhysRevLett.106.029902}{{\em Phys. Rev. Lett.} {\bf 105} (2010)
  211801},
\href{http://arxiv.org/abs/1010.0203}{{\tt arXiv:1010.0203 [hep-ex]}}.
%%CITATION = ARXIV:1010.0203;%%.

\bibitem{Chatrchyan:2011ns}
{\bf CMS} Collaboration, S.~Chatrchyan {\em et al.}, ``{Search for Resonances
  in the Dijet Mass Spectrum from 7 TeV pp Collisions at CMS},''
  \href{http://dx.doi.org/10.1016/j.physletb.2011.09.015}{{\em Phys. Lett.}
  {\bf B704} (2011)  123--142},
\href{http://arxiv.org/abs/1107.4771}{{\tt arXiv:1107.4771 [hep-ex]}}.
%%CITATION = ARXIV:1107.4771;%%.

\bibitem{CMS:2012yf}
{\bf CMS} Collaboration, S.~Chatrchyan {\em et al.}, ``{Search for narrow
  resonances and quantum black holes in inclusive and $b$-tagged dijet mass
  spectra from $pp$ collisions at $\sqrt{s}=7$ TeV},''
  \href{http://dx.doi.org/10.1007/JHEP01(2013)013}{{\em JHEP} {\bf 01} (2013)
  013},
\href{http://arxiv.org/abs/1210.2387}{{\tt arXiv:1210.2387 [hep-ex]}}.
%%CITATION = ARXIV:1210.2387;%%.

\bibitem{Aad:2014aqa}
{\bf ATLAS} Collaboration, G.~Aad {\em et al.}, ``{Search for new phenomena in
  the dijet mass distribution using $p-p$ collision data at $\sqrt{s}=8$ TeV
  with the ATLAS detector},''
  \href{http://dx.doi.org/10.1103/PhysRevD.91.052007}{{\em Phys. Rev.} {\bf
  D91} (2015) no.~5, 052007},
\href{http://arxiv.org/abs/1407.1376}{{\tt arXiv:1407.1376 [hep-ex]}}.
%%CITATION = ARXIV:1407.1376;%%.

\bibitem{Chatrchyan:2013qha}
{\bf CMS} Collaboration, S.~Chatrchyan {\em et al.}, ``{Search for Narrow
  Resonances Using the Dijet Mass Spectrum in $pp$ Collisions at $\sqrt{s}$=8
  TeV},'' \href{http://dx.doi.org/10.1103/PhysRevD.87.114015}{{\em Phys. Rev.}
  {\bf D87} (2013) no.~11, 114015},
\href{http://arxiv.org/abs/1302.4794}{{\tt arXiv:1302.4794 [hep-ex]}}.
%%CITATION = ARXIV:1302.4794;%%.

\bibitem{Khachatryan:2015sja}
{\bf CMS} Collaboration, V.~Khachatryan {\em et al.}, ``{Search for resonances
  and quantum black holes using dijet mass spectra in proton-proton collisions
  at $\sqrt{s} =$ 8 TeV},''
  \href{http://dx.doi.org/10.1103/PhysRevD.91.052009}{{\em Phys. Rev.} {\bf
  D91} (2015) no.~5, 052009},
\href{http://arxiv.org/abs/1501.04198}{{\tt arXiv:1501.04198 [hep-ex]}}.
%%CITATION = ARXIV:1501.04198;%%.

\bibitem{Khachatryan:2016ecr}
{\bf CMS} Collaboration, V.~Khachatryan {\em et al.}, ``{Search for narrow
  resonances in dijet final states at $\sqrt(s)=$ 8 TeV with the novel CMS
  technique of data scouting},''
  \href{http://dx.doi.org/10.1103/PhysRevLett.117.031802}{{\em Phys. Rev.
  Lett.} {\bf 117} (2016) no.~3, 031802},
\href{http://arxiv.org/abs/1604.08907}{{\tt arXiv:1604.08907 [hep-ex]}}.
%%CITATION = ARXIV:1604.08907;%%.

\bibitem{ATLAS:2015nsi}
{\bf ATLAS} Collaboration, G.~Aad {\em et al.}, ``{Search for new phenomena in
  dijet mass and angular distributions from $pp$ collisions at $\sqrt{s}=$ 13
  TeV with the ATLAS detector},''
  \href{http://dx.doi.org/10.1016/j.physletb.2016.01.032}{{\em Phys. Lett.}
  {\bf B754} (2016)  302--322},
\href{http://arxiv.org/abs/1512.01530}{{\tt arXiv:1512.01530 [hep-ex]}}.
%%CITATION = ARXIV:1512.01530;%%.

\bibitem{Aaboud:2017yvp}
{\bf ATLAS} Collaboration, M.~Aaboud {\em et al.}, ``{Search for new phenomena
  in dijet events using 37 fb$^{-1}$ of $pp$ collision data collected at
  $\sqrt{s}=$13 TeV with the ATLAS detector},''
  \href{http://dx.doi.org/10.1103/PhysRevD.96.052004}{{\em Phys. Rev.} {\bf
  D96} (2017) no.~5, 052004},
\href{http://arxiv.org/abs/1703.09127}{{\tt arXiv:1703.09127 [hep-ex]}}.
%%CITATION = ARXIV:1703.09127;%%.

\bibitem{Aaboud:2018zba}
{\bf ATLAS} Collaboration, M.~Aaboud {\em et al.}, ``{Search for light
  resonances decaying to boosted quark pairs and produced in association with a
  photon or a jet in proton-proton collisions at $\sqrt{s}=13$ TeV with the
  ATLAS detector},''
\href{http://arxiv.org/abs/1801.08769}{{\tt arXiv:1801.08769 [hep-ex]}}.
%%CITATION = ARXIV:1801.08769;%%.

\bibitem{Aaboud:2018fzt}
{\bf ATLAS} Collaboration, M.~Aaboud {\em et al.}, ``{Search for low-mass dijet
  resonances using trigger-level jets with the ATLAS detector in $pp$
  collisions at $\sqrt{s}=13$ TeV},''
\href{http://arxiv.org/abs/1804.03496}{{\tt arXiv:1804.03496 [hep-ex]}}.
%%CITATION = ARXIV:1804.03496;%%.

\bibitem{Khachatryan:2015dcf}
{\bf CMS} Collaboration, V.~Khachatryan {\em et al.}, ``{Search for narrow
  resonances decaying to dijets in proton-proton collisions at $\sqrt(s) =$ 13
  TeV},'' \href{http://dx.doi.org/10.1103/PhysRevLett.116.071801}{{\em Phys.
  Rev. Lett.} {\bf 116} (2016) no.~7, 071801},
\href{http://arxiv.org/abs/1512.01224}{{\tt arXiv:1512.01224 [hep-ex]}}.
%%CITATION = ARXIV:1512.01224;%%.

\bibitem{Sirunyan:2016iap}
{\bf CMS} Collaboration, A.~M. Sirunyan {\em et al.}, ``{Search for dijet
  resonances in proton?proton collisions at $\sqrt{s}$ = 13 TeV and constraints
  on dark matter and other models},''
  \href{http://dx.doi.org/10.1016/j.physletb.2017.09.029,
  10.1016/j.physletb.2017.02.012}{{\em Phys. Lett.} {\bf B769} (2017)
  520--542}, \href{http://arxiv.org/abs/1611.03568}{{\tt arXiv:1611.03568
  [hep-ex]}}.
[Erratum: Phys. Lett.B772,882(2017)].
%%CITATION = ARXIV:1611.03568;%%.

\bibitem{Sirunyan:2017dnz}
{\bf CMS} Collaboration, A.~M. Sirunyan {\em et al.}, ``{Search for Low Mass
  Vector Resonances Decaying to Quark-Antiquark Pairs in Proton-Proton
  Collisions at $\sqrt{s}=13\text{ }\text{ }\mathrm{TeV}$},''
  \href{http://dx.doi.org/10.1103/PhysRevLett.119.111802}{{\em Phys. Rev.
  Lett.} {\bf 119} (2017) no.~11, 111802},
\href{http://arxiv.org/abs/1705.10532}{{\tt arXiv:1705.10532 [hep-ex]}}.
%%CITATION = ARXIV:1705.10532;%%.

\bibitem{Sirunyan:2017nvi}
{\bf CMS} Collaboration, A.~M. Sirunyan {\em et al.}, ``{Search for low mass
  vector resonances decaying into quark-antiquark pairs in proton-proton
  collisions at $ \sqrt{s}=13 $ TeV},''
  \href{http://dx.doi.org/10.1007/JHEP01(2018)097}{{\em JHEP} {\bf 01} (2018)
  097},
\href{http://arxiv.org/abs/1710.00159}{{\tt arXiv:1710.00159 [hep-ex]}}.
%%CITATION = ARXIV:1710.00159;%%.

\bibitem{Sirunyan:2018xlo}
{\bf CMS} Collaboration, A.~M. Sirunyan {\em et al.}, ``{Search for narrow and
  broad dijet resonances in proton-proton collisions at $ \sqrt{s}=13 $ TeV and
  constraints on dark matter mediators and other new particles},''
  \href{http://dx.doi.org/10.1007/JHEP08(2018)130}{{\em JHEP} {\bf 08} (2018)
  130},
\href{http://arxiv.org/abs/1806.00843}{{\tt arXiv:1806.00843 [hep-ex]}}.
%%CITATION = ARXIV:1806.00843;%%.

\bibitem{CMS:2018wxx}
{\bf CMS} Collaboration, C.~Collaboration,
``{Searches for dijet resonances in pp collisions at $\sqrt{s}=13~\mathrm{TeV}$
  using the 2016 and 2017 datasets},''.
%%CITATION = CMS-PAS-EXO-17-026;%%.

\bibitem{Alwall:2014hca}
J.~Alwall, R.~Frederix, S.~Frixione, V.~Hirschi, F.~Maltoni, O.~Mattelaer,
  H.~S. Shao, T.~Stelzer, P.~Torrielli, and M.~Zaro, ``{The automated
  computation of tree-level and next-to-leading order differential cross
  sections, and their matching to parton shower simulations},''
  \href{http://dx.doi.org/10.1007/JHEP07(2014)079}{{\em JHEP} {\bf 07} (2014)
  079},
\href{http://arxiv.org/abs/1405.0301}{{\tt arXiv:1405.0301 [hep-ph]}}.
%%CITATION = ARXIV:1405.0301;%%.

\bibitem{Abe:1998uz}
{\bf CDF} Collaboration, F.~Abe {\em et al.}, ``{Search for New Particles
  Decaying to $b\bar{b}$ in $p\bar{p}$ Collisions at $\sqrt{s} = 1.8$ TeV},''
  \href{http://dx.doi.org/10.1103/PhysRevLett.82.2038}{{\em Phys. Rev. Lett.}
  {\bf 82} (1999)  2038--2043},
\href{http://arxiv.org/abs/hep-ex/9809022}{{\tt arXiv:hep-ex/9809022
  [hep-ex]}}.
%%CITATION = HEP-EX/9809022;%%.

\bibitem{Sirunyan:2018pas}
{\bf CMS} Collaboration, A.~M. Sirunyan {\em et al.}, ``{Search for narrow
  resonances in the b-tagged dijet mass spectrum in proton-proton collisions at
  $\sqrt{s} =$ 8 TeV},''
  \href{http://dx.doi.org/10.1103/PhysRevLett.120.201801}{{\em Phys. Rev.
  Lett.} {\bf 120} (2018) no.~20, 201801},
\href{http://arxiv.org/abs/1802.06149}{{\tt arXiv:1802.06149 [hep-ex]}}.
%%CITATION = ARXIV:1802.06149;%%.

\bibitem{Aaboud:2016nbq}
{\bf ATLAS} Collaboration, M.~Aaboud {\em et al.}, ``{Search for resonances in
  the mass distribution of jet pairs with one or two jets identified as
  $b$-jets in proton--proton collisions at $\sqrt{s}=13$ TeV with the ATLAS
  detector},'' \href{http://dx.doi.org/10.1016/j.physletb.2016.05.064}{{\em
  Phys. Lett.} {\bf B759} (2016)  229--246},
\href{http://arxiv.org/abs/1603.08791}{{\tt arXiv:1603.08791 [hep-ex]}}.
%%CITATION = ARXIV:1603.08791;%%.

\bibitem{Aaboud:2018tqo}
{\bf ATLAS} Collaboration, M.~Aaboud {\em et al.}, ``{Search for resonances in
  the mass distribution of jet pairs with one or two jets identified as
  $b$-jets in proton-proton collisions at $\sqrt{s}=13$ TeV with the ATLAS
  detector},'' \href{http://dx.doi.org/10.1103/PhysRevD.98.032016}{{\em Phys.
  Rev.} {\bf D98} (2018)  032016},
\href{http://arxiv.org/abs/1805.09299}{{\tt arXiv:1805.09299 [hep-ex]}}.
%%CITATION = ARXIV:1805.09299;%%.

\bibitem{Aaltonen:2012zh}
{\bf CDF, D0} Collaboration, T.~Aaltonen {\em et al.}, ``{Search for Neutral
  Higgs Bosons in Events with Multiple Bottom Quarks at the Tevatron},''
  \href{http://dx.doi.org/10.1103/PhysRevD.86.091101}{{\em Phys. Rev.} {\bf
  D86} (2012)  091101},
\href{http://arxiv.org/abs/1207.2757}{{\tt arXiv:1207.2757 [hep-ex]}}.
%%CITATION = ARXIV:1207.2757;%%.

\bibitem{Aaboud:2017yyg}
{\bf ATLAS} Collaboration, M.~Aaboud {\em et al.}, ``{Search for new phenomena
  in high-mass diphoton final states using 37 fb$^{-1}$ of proton--proton
  collisions collected at $\sqrt{s}=13$ TeV with the ATLAS detector},''
  \href{http://dx.doi.org/10.1016/j.physletb.2017.10.039}{{\em Phys. Lett.}
  {\bf B775} (2017)  105--125},
\href{http://arxiv.org/abs/1707.04147}{{\tt arXiv:1707.04147 [hep-ex]}}.
%%CITATION = ARXIV:1707.04147;%%.

\bibitem{Khachatryan:2014ira}
{\bf CMS} Collaboration, V.~Khachatryan {\em et al.}, ``{Observation of the
  diphoton decay of the Higgs boson and measurement of its properties},''
  \href{http://dx.doi.org/10.1140/epjc/s10052-014-3076-z}{{\em Eur. Phys. J.}
  {\bf C74} (2014) no.~10, 3076},
\href{http://arxiv.org/abs/1407.0558}{{\tt arXiv:1407.0558 [hep-ex]}}.
%%CITATION = ARXIV:1407.0558;%%.

\bibitem{Aad:2014ioa}
{\bf ATLAS} Collaboration, G.~Aad {\em et al.}, ``{Search for Scalar Diphoton
  Resonances in the Mass Range $65-600$ GeV with the ATLAS Detector in $pp$
  Collision Data at $\sqrt{s}$ = 8 $TeV$},''
  \href{http://dx.doi.org/10.1103/PhysRevLett.113.171801}{{\em Phys. Rev.
  Lett.} {\bf 113} (2014) no.~17, 171801},
\href{http://arxiv.org/abs/1407.6583}{{\tt arXiv:1407.6583 [hep-ex]}}.
%%CITATION = ARXIV:1407.6583;%%.

\bibitem{Khachatryan:2015qba}
{\bf CMS} Collaboration, V.~Khachatryan {\em et al.}, ``{Search for diphoton
  resonances in the mass range from 150 to 850 GeV in pp collisions at
  $\sqrt{s} =$ 8 TeV},''
  \href{http://dx.doi.org/10.1016/j.physletb.2015.09.062}{{\em Phys. Lett.}
  {\bf B750} (2015)  494--519},
\href{http://arxiv.org/abs/1506.02301}{{\tt arXiv:1506.02301 [hep-ex]}}.
%%CITATION = ARXIV:1506.02301;%%.

\bibitem{Mariotti:2017vtv}
A.~Mariotti, D.~Redigolo, F.~Sala, and K.~Tobioka, ``{New LHC bound on low-mass
  diphoton resonances},''
  \href{http://dx.doi.org/10.1016/j.physletb.2018.06.039}{{\em Phys. Lett.}
  {\bf B783} (2018)  13--18},
\href{http://arxiv.org/abs/1710.01743}{{\tt arXiv:1710.01743 [hep-ph]}}.
%%CITATION = ARXIV:1710.01743;%%.

\bibitem{Aad:2015fna}
{\bf ATLAS} Collaboration, G.~Aad {\em et al.}, ``{A search for $ t\overline{t}
  $ resonances using lepton-plus-jets events in proton-proton collisions at $
  \sqrt{s}=8 $ TeV with the ATLAS detector},''
  \href{http://dx.doi.org/10.1007/JHEP08(2015)148}{{\em JHEP} {\bf 08} (2015)
  148},
\href{http://arxiv.org/abs/1505.07018}{{\tt arXiv:1505.07018 [hep-ex]}}.
%%CITATION = ARXIV:1505.07018;%%.

\bibitem{Aaboud:2017hnm}
{\bf ATLAS} Collaboration, M.~Aaboud {\em et al.}, ``{Search for Heavy Higgs
  Bosons $A/H$ Decaying to a Top Quark Pair in $pp$ Collisions at
  $\sqrt{s}=8\text{ }\text{ }\mathrm{TeV}$ with the ATLAS Detector},''
  \href{http://dx.doi.org/10.1103/PhysRevLett.119.191803}{{\em Phys. Rev.
  Lett.} {\bf 119} (2017) no.~19, 191803},
\href{http://arxiv.org/abs/1707.06025}{{\tt arXiv:1707.06025 [hep-ex]}}.
%%CITATION = ARXIV:1707.06025;%%.

\bibitem{Aaboud:2018mjh}
{\bf ATLAS} Collaboration, M.~Aaboud {\em et al.}, ``{Search for heavy
  particles decaying into top-quark pairs using lepton-plus-jets events in
  proton?proton collisions at $\sqrt{s} = 13$ $\text {TeV}$ with the ATLAS
  detector},'' \href{http://dx.doi.org/10.1140/epjc/s10052-018-5995-6}{{\em
  Eur. Phys. J.} {\bf C78} (2018) no.~7, 565},
\href{http://arxiv.org/abs/1804.10823}{{\tt arXiv:1804.10823 [hep-ex]}}.
%%CITATION = ARXIV:1804.10823;%%.

\bibitem{Yu:2013wta}
F.~Yu, ``{Di-jet resonances at future hadron colliders: A Snowmass
  whitepaper},''
\href{http://arxiv.org/abs/1308.1077}{{\tt arXiv:1308.1077 [hep-ph]}}.
%%CITATION = ARXIV:1308.1077;%%.

\bibitem{Chekanov:2017pnx}
S.~V. Chekanov, J.~T. Childers, J.~Proudfoot, D.~Frizzell, and R.~Wang,
  ``{Precision searches in dijets at the HL-LHC and HE-LHC},''
  \href{http://dx.doi.org/10.1088/1748-0221/13/05/P05022}{{\em JINST} {\bf 13}
  (2018) no.~05, P05022},
\href{http://arxiv.org/abs/1710.09484}{{\tt arXiv:1710.09484 [hep-ex]}}.
%%CITATION = ARXIV:1710.09484;%%.

\bibitem{CidVidal:2018eel}
{\bf Working Group 3} Collaboration, X.~Cid~Vidal {\em et al.}, ``{Beyond the
  Standard Model Physics at the HL-LHC and HE-LHC},''
\href{http://arxiv.org/abs/1812.07831}{{\tt arXiv:1812.07831 [hep-ph]}}.
%%CITATION = ARXIV:1812.07831;%%.

\bibitem{Aad:2019uzh}
{\bf ATLAS} Collaboration, G.~Aad {\em et al.}, ``{Combination of searches for
  Higgs boson pairs in $pp$ collisions at $\sqrt{s} = $13 TeV with the ATLAS
  detector},''
\href{http://arxiv.org/abs/1906.02025}{{\tt arXiv:1906.02025 [hep-ex]}}.
%%CITATION = ARXIV:1906.02025;%%.

\bibitem{Sirunyan:2018two}
{\bf CMS} Collaboration, A.~M. Sirunyan {\em et al.}, ``{Combination of
  searches for Higgs boson pair production in proton-proton collisions at
  $\sqrt{s} = $ 13 TeV},''
  \href{http://dx.doi.org/10.1103/PhysRevLett.122.121803}{{\em Phys. Rev.
  Lett.} {\bf 122} (2019) no.~12, 121803},
\href{http://arxiv.org/abs/1811.09689}{{\tt arXiv:1811.09689 [hep-ex]}}.
%%CITATION = ARXIV:1811.09689;%%.

\bibitem{Egana-Ugrinovic2019xxx}
D.~Egana-Ugrinovic, S.~Homiller, and P.~Meade {\em To appear}  .

\bibitem{ATLAS-CONF-2019-005}
{\bf ATLAS Collaboration} Collaboration, ``{Combined measurements of Higgs
  boson production and decay using up to $80$ fb$^{-1}$ of proton--proton
  collision data at $\sqrt{s}=$ 13 TeV collected with the ATLAS experiment},''
  Tech. Rep. ATLAS-CONF-2019-005, CERN, Geneva, Mar, 2019.
\newblock \url{https://cds.cern.ch/record/2668375}.

\bibitem{Aaboud:2018urx}
{\bf ATLAS} Collaboration, M.~Aaboud {\em et al.}, ``{Observation of Higgs
  boson production in association with a top quark pair at the LHC with the
  ATLAS detector},''
  \href{http://dx.doi.org/10.1016/j.physletb.2018.07.035}{{\em Phys. Lett.}
  {\bf B784} (2018)  173--191},
\href{http://arxiv.org/abs/1806.00425}{{\tt arXiv:1806.00425 [hep-ex]}}.
%%CITATION = ARXIV:1806.00425;%%.

\bibitem{Aaboud:2018zhk}
{\bf ATLAS} Collaboration, M.~Aaboud {\em et al.}, ``{Observation of $H
  \rightarrow b\bar{b}$ decays and $VH$ production with the ATLAS detector},''
  \href{http://dx.doi.org/10.1016/j.physletb.2018.09.013}{{\em Phys. Lett.}
  {\bf B786} (2018)  59--86},
\href{http://arxiv.org/abs/1808.08238}{{\tt arXiv:1808.08238 [hep-ex]}}.
%%CITATION = ARXIV:1808.08238;%%.

\bibitem{Aaboud:2018pen}
{\bf ATLAS} Collaboration, M.~Aaboud {\em et al.}, ``{Cross-section
  measurements of the Higgs boson decaying into a pair of $\tau$-leptons in
  proton-proton collisions at $\sqrt{s}=13$ TeV with the ATLAS detector},''
  {\em Submitted to: Phys. Rev.} (2018)  ,
\href{http://arxiv.org/abs/1811.08856}{{\tt arXiv:1811.08856 [hep-ex]}}.
%%CITATION = ARXIV:1811.08856;%%.

\bibitem{Sirunyan:2018koj}
{\bf CMS} Collaboration, A.~M. Sirunyan {\em et al.}, ``{Combined measurements
  of Higgs boson couplings in proton-proton collisions at $\sqrt{s}=$ 13
  TeV},'' \href{http://dx.doi.org/10.1140/epjc/s10052-019-6909-y}{{\em Eur.
  Phys. J.} {\bf C79} (2019) no.~5, 421},
\href{http://arxiv.org/abs/1809.10733}{{\tt arXiv:1809.10733 [hep-ex]}}.
%%CITATION = ARXIV:1809.10733;%%.

\bibitem{Sirunyan:2017exp}
{\bf CMS} Collaboration, A.~M. Sirunyan {\em et al.}, ``{Measurements of
  properties of the Higgs boson decaying into the four-lepton final state in pp
  collisions at $ \sqrt{s}=13 $ TeV},''
  \href{http://dx.doi.org/10.1007/JHEP11(2017)047}{{\em JHEP} {\bf 11} (2017)
  047},
\href{http://arxiv.org/abs/1706.09936}{{\tt arXiv:1706.09936 [hep-ex]}}.
%%CITATION = ARXIV:1706.09936;%%.

\bibitem{Aaboud:2018fhh}
{\bf ATLAS} Collaboration, M.~Aaboud {\em et al.}, ``{Search for the Decay of
  the Higgs Boson to Charm Quarks with the ATLAS Experiment},''
  \href{http://dx.doi.org/10.1103/PhysRevLett.120.211802}{{\em Phys. Rev.
  Lett.} {\bf 120} (2018) no.~21, 211802},
\href{http://arxiv.org/abs/1802.04329}{{\tt arXiv:1802.04329 [hep-ex]}}.
%%CITATION = ARXIV:1802.04329;%%.

\bibitem{Perez:2015aoa}
G.~Perez, Y.~Soreq, E.~Stamou, and K.~Tobioka, ``{Constraining the charm Yukawa
  and Higgs-quark coupling universality},''
  \href{http://dx.doi.org/10.1103/PhysRevD.92.033016}{{\em Phys. Rev.} {\bf
  D92} (2015) no.~3, 033016},
\href{http://arxiv.org/abs/1503.00290}{{\tt arXiv:1503.00290 [hep-ph]}}.
%%CITATION = ARXIV:1503.00290;%%.

\bibitem{Nelson:1983zb}
A.~E. Nelson, ``{Naturally Weak CP Violation},''
\href{http://dx.doi.org/10.1016/0370-2693(84)92025-2}{{\em Phys. Lett.} {\bf
  136B} (1984)  387--391}.
%%CITATION = PHLTA,136B,387;%%.

\bibitem{Barr:1984qx}
S.~M. Barr, ``{Solving the Strong CP Problem Without the Peccei-Quinn
  Symmetry},''
\href{http://dx.doi.org/10.1103/PhysRevLett.53.329}{{\em Phys. Rev. Lett.} {\bf
  53} (1984)  329}.
%%CITATION = PRLTA,53,329;%%.

\bibitem{Barr:1984fh}
S.~M. Barr, ``{A Natural Class of Nonpeccei-quinn Models},''
\href{http://dx.doi.org/10.1103/PhysRevD.30.1805}{{\em Phys. Rev.} {\bf D30}
  (1984)  1805}.
%%CITATION = PHRVA,D30,1805;%%.

\bibitem{Bento:1991ez}
L.~Bento, G.~C. Branco, and P.~A. Parada, ``{A Minimal model with natural
  suppression of strong CP violation},''
\href{http://dx.doi.org/10.1016/0370-2693(91)90530-4}{{\em Phys. Lett.} {\bf
  B267} (1991)  95--99}.
%%CITATION = PHLTA,B267,95;%%.

\bibitem{Hiller:2001qg}
G.~Hiller and M.~Schmaltz, ``{Solving the Strong CP Problem with
  Supersymmetry},'' \href{http://dx.doi.org/10.1016/S0370-2693(01)00814-0}{{\em
  Phys. Lett.} {\bf B514} (2001)  263--268},
\href{http://arxiv.org/abs/hep-ph/0105254}{{\tt arXiv:hep-ph/0105254
  [hep-ph]}}.
%%CITATION = HEP-PH/0105254;%%.

\bibitem{Hiller:2002um}
G.~Hiller and M.~Schmaltz, ``{Strong weak CP hierarchy from nonrenormalization
  theorems},'' \href{http://dx.doi.org/10.1103/PhysRevD.65.096009}{{\em Phys.
  Rev.} {\bf D65} (2002)  096009},
\href{http://arxiv.org/abs/hep-ph/0201251}{{\tt arXiv:hep-ph/0201251
  [hep-ph]}}.
%%CITATION = HEP-PH/0201251;%%.

\bibitem{Davidson:2007si}
S.~Davidson, G.~Isidori, and S.~Uhlig, ``{Solving the flavour problem with
  hierarchical fermion wave functions},''
  \href{http://dx.doi.org/10.1016/j.physletb.2008.04.005}{{\em Phys. Lett.}
  {\bf B663} (2008)  73--79},
\href{http://arxiv.org/abs/0711.3376}{{\tt arXiv:0711.3376 [hep-ph]}}.
%%CITATION = ARXIV:0711.3376;%%.

\bibitem{Leurer:1993gy}
M.~Leurer, Y.~Nir, and N.~Seiberg, ``{Mass matrix models: The Sequel},''
  \href{http://dx.doi.org/10.1016/0550-3213(94)90074-4}{{\em Nucl. Phys.} {\bf
  B420} (1994)  468--504},
\href{http://arxiv.org/abs/hep-ph/9310320}{{\tt arXiv:hep-ph/9310320
  [hep-ph]}}.
%%CITATION = HEP-PH/9310320;%%.

\bibitem{Cacciapaglia:2007fw}
G.~Cacciapaglia, C.~Csaki, J.~Galloway, G.~Marandella, J.~Terning, and
  A.~Weiler, ``{A GIM Mechanism from Extra Dimensions},''
  \href{http://dx.doi.org/10.1088/1126-6708/2008/04/006}{{\em JHEP} {\bf 04}
  (2008)  006},
\href{http://arxiv.org/abs/0709.1714}{{\tt arXiv:0709.1714 [hep-ph]}}.
%%CITATION = ARXIV:0709.1714;%%.

\bibitem{Csaki:2008eh}
C.~Csaki, A.~Falkowski, and A.~Weiler, ``{A Simple Flavor Protection for RS},''
  \href{http://dx.doi.org/10.1103/PhysRevD.80.016001}{{\em Phys. Rev.} {\bf
  D80} (2009)  016001},
\href{http://arxiv.org/abs/0806.3757}{{\tt arXiv:0806.3757 [hep-ph]}}.
%%CITATION = ARXIV:0806.3757;%%.

\bibitem{Csaki:2009wc}
C.~Csaki, G.~Perez, Z.~Surujon, and A.~Weiler, ``{Flavor Alignment via Shining
  in RS},'' \href{http://dx.doi.org/10.1103/PhysRevD.81.075025}{{\em Phys.
  Rev.} {\bf D81} (2010)  075025},
\href{http://arxiv.org/abs/0907.0474}{{\tt arXiv:0907.0474 [hep-ph]}}.
%%CITATION = ARXIV:0907.0474;%%.

\bibitem{Cvetic:1997zd}
G.~Cvetic, S.~S. Hwang, and C.~S. Kim, ``{One loop renormalization group
  equations of the general framework with two Higgs doublets},''
  \href{http://dx.doi.org/10.1142/S0217751X99000385}{{\em Int. J. Mod. Phys.}
  {\bf A14} (1999)  769--798},
\href{http://arxiv.org/abs/hep-ph/9706323}{{\tt arXiv:hep-ph/9706323
  [hep-ph]}}.
%%CITATION = HEP-PH/9706323;%%.

\bibitem{Barbieri:1987fn}
R.~Barbieri and G.~F. Giudice, ``{Upper Bounds on Supersymmetric Particle
  Masses},''
\href{http://dx.doi.org/10.1016/0550-3213(88)90171-X}{{\em Nucl. Phys.} {\bf
  B306} (1988)  63--76}.
%%CITATION = NUPHA,B306,63;%%.

\bibitem{Gori:2017qwg}
S.~Gori, H.~E. Haber, and E.~Santos, ``{High scale flavor alignment in
  two-Higgs doublet models and its phenomenology},''
  \href{http://dx.doi.org/10.1007/JHEP06(2017)110}{{\em JHEP} {\bf 06} (2017)
  110},
\href{http://arxiv.org/abs/1703.05873}{{\tt arXiv:1703.05873 [hep-ph]}}.
%%CITATION = ARXIV:1703.05873;%%.

\bibitem{Buras:2000dm}
A.~J. Buras, P.~Gambino, M.~Gorbahn, S.~Jager, and L.~Silvestrini, ``{Universal
  unitarity triangle and physics beyond the standard model},''
  \href{http://dx.doi.org/10.1016/S0370-2693(01)00061-2}{{\em Phys. Lett.} {\bf
  B500} (2001)  161--167},
\href{http://arxiv.org/abs/hep-ph/0007085}{{\tt arXiv:hep-ph/0007085
  [hep-ph]}}.
%%CITATION = HEP-PH/0007085;%%.

\bibitem{Das:1995df}
A.~K. Das and C.~Kao, ``{A Two Higgs doublet model for the top quark},''
  \href{http://dx.doi.org/10.1016/0370-2693(96)00031-7}{{\em Phys. Lett.} {\bf
  B372} (1996)  106--112},
\href{http://arxiv.org/abs/hep-ph/9511329}{{\tt arXiv:hep-ph/9511329
  [hep-ph]}}.
%%CITATION = HEP-PH/9511329;%%.

\bibitem{Blechman:2010cs}
A.~E. Blechman, A.~A. Petrov, and G.~Yeghiyan, ``{The Flavor puzzle in
  multi-Higgs models},'' \href{http://dx.doi.org/10.1007/JHEP11(2010)075}{{\em
  JHEP} {\bf 11} (2010)  075},
\href{http://arxiv.org/abs/1009.1612}{{\tt arXiv:1009.1612 [hep-ph]}}.
%%CITATION = ARXIV:1009.1612;%%.

\bibitem{Altmannshofer:2015esa}
W.~Altmannshofer, S.~Gori, A.~L. Kagan, L.~Silvestrini, and J.~Zupan,
  ``{Uncovering Mass Generation Through Higgs Flavor Violation},''
  \href{http://dx.doi.org/10.1103/PhysRevD.93.031301}{{\em Phys. Rev.} {\bf
  D93} (2016) no.~3, 031301},
\href{http://arxiv.org/abs/1507.07927}{{\tt arXiv:1507.07927 [hep-ph]}}.
%%CITATION = ARXIV:1507.07927;%%.

\bibitem{Ghosh:2015gpa}
D.~Ghosh, R.~S. Gupta, and G.~Perez, ``{Is the Higgs Mechanism of Fermion Mass
  Generation a Fact? A Yukawa-less First-Two-Generation Model},''
  \href{http://dx.doi.org/10.1016/j.physletb.2016.02.059}{{\em Phys. Lett.}
  {\bf B755} (2016)  504--508},
\href{http://arxiv.org/abs/1508.01501}{{\tt arXiv:1508.01501 [hep-ph]}}.
%%CITATION = ARXIV:1508.01501;%%.

\bibitem{Botella:2016krk}
F.~J. Botella, G.~C. Branco, M.~N. Rebelo, and J.~I. Silva-Marcos, ``{What if
  the masses of the first two quark families are not generated by the standard
  model Higgs boson?},''
  \href{http://dx.doi.org/10.1103/PhysRevD.94.115031}{{\em Phys. Rev.} {\bf
  D94} (2016) no.~11, 115031},
\href{http://arxiv.org/abs/1602.08011}{{\tt arXiv:1602.08011 [hep-ph]}}.
%%CITATION = ARXIV:1602.08011;%%.

\bibitem{Altmannshofer:2016zrn}
W.~Altmannshofer, J.~Eby, S.~Gori, M.~Lotito, M.~Martone, and D.~Tuckler,
  ``{Collider Signatures of Flavorful Higgs Bosons},''
  \href{http://dx.doi.org/10.1103/PhysRevD.94.115032}{{\em Phys. Rev.} {\bf
  D94} (2016) no.~11, 115032},
\href{http://arxiv.org/abs/1610.02398}{{\tt arXiv:1610.02398 [hep-ph]}}.
%%CITATION = ARXIV:1610.02398;%%.

\bibitem{Altmannshofer:2017uvs}
W.~Altmannshofer, S.~Gori, D.~J. Robinson, and D.~Tuckler, ``{The Flavor-locked
  Flavorful Two Higgs Doublet Model},''
  \href{http://dx.doi.org/10.1007/JHEP03(2018)129}{{\em JHEP} {\bf 03} (2018)
  129},
\href{http://arxiv.org/abs/1712.01847}{{\tt arXiv:1712.01847 [hep-ph]}}.
%%CITATION = ARXIV:1712.01847;%%.

\end{thebibliography}\endgroup


\providecommand{\href}[2]{#2}\begingroup\raggedright\endgroup

\end{document}